\documentclass[11pt,a4paper]{article}
\usepackage{jcappub}
\usepackage{multirow}
\usepackage{graphicx}
\usepackage{amsmath,amsfonts,amsthm}
\usepackage{array}
\usepackage{tcolorbox}

\usepackage{mdframed}
\definecolor{light-gray}{gray}{0.98}

\usepackage[normalem]{ulem}

\usepackage{newcommands}

\DeclareGraphicsExtensions{.png,.pdf}

\subheader{IFT-UAM/CSIC-XXX, LUPM:22-006, ULB-TH/22-06}

\title{Analytical insight into dark matter subhalo boost factors for Sommerfeld-enhanced $s$- and $p$-wave $\gamma$-ray signals}
\author[a]{Gaétan Facchinetti,}
\author[b]{Martin Stref,}
\author[c]{Thomas Lacroix,}
\author[d]{Julien Lavalle,}
\author[c]{Judit Pérez-Romero,}
\author[e]{David Maurin,}
\author[c]{and Miguel A.~S\'{a}nchez-Conde}
\affiliation[a]{Service de Physique Th\'eorique, Universit\'e Libre de Bruxelles, Boulevard du Triomphe, CP225, 1050 Brussels, Belgium}
\affiliation[b]{Laboratoire d'Annecy-le-Vieux de
Physique Th\'eorique (LAPTh), Universit\'e Grenoble Alpes, Universit\'e Savoie Mont Blanc, CNRS, F-74000 Annecy, France}
\affiliation[c]{Instituto de F\'isica Te\'orica (IFT) UAM/CSIC, Universidad Aut\'onoma de Madrid, Calle Nicol\'as Cabrera, 13-15, 28049 Madrid, Spain}
\affiliation[d]{Laboratoire Univers et Particules de Montpellier (LUPM), Universit\'e de Montpellier \& CNRS, Place Eug\`ene Bataillon, 34095 Montpellier Cedex 05, France}
\affiliation[e]{Laboratoire de Physique Subatomique \& Cosmologie (LPSC), Universit\'e Grenoble Alpes, CNRS/IN2P3, 53 avenue des Martyrs, 38026 Grenoble, France}
\emailAdd{gaetan.facchinetti@ulb.be}
\emailAdd{martin.stref@univ-smb.fr}
\emailAdd{thomas.lacroix@uam.es}
\emailAdd{lavalle@in2p3.fr}
\emailAdd{judit.perez@uam.es}
\emailAdd{david.maurin@lpsc.in2p3.fr}
\emailAdd{miguel.sanchezconde@uam.es}

\abstract{
As searches for thermal and self-annihilating dark matter (DM) intensify, it becomes crucial to include as many relevant physical processes and ingredients as possible to refine signal predictions, in particular those which directly relate to the intimate properties of DM. We investigate the combined impact of DM subhalos and of the (velocity-dependent) Sommerfeld enhancement of the annihilation cross section. Both features are expected to play an important role in searches for thermal DM particle candidates with masses around or beyond TeV, or in scenarios with a light dark sector. We provide a detailed analytical description of the phenomena at play, and show how they scale with the subhalo masses and the main Sommerfeld parameters. We derive approximate analytical expressions that can be used to estimate the overall boost factors resulting from these combined effects, from which the intricate phenomenology can be better understood. DM subhalos lead to an increase of the Sommerfeld effect by several orders of magnitude (for both $s$- and $p$-wave annihilation processes), especially on resonances, which makes them critical to get sensible gamma-ray signal predictions for typical targets of different masses (from dwarf galaxies to galaxy clusters).
}

\arxivnumber{2203.16491}

\begin{document}
\maketitle

\section{Introduction}
\label{sec:intro}
Gamma-ray astronomy, and more generally multimessenger astronomy, provides powerful probes of thermally produced non-asymmetric particle dark matter (DM), in particular scenarios in which DM self-annihilation proceeds through $s$-wave processes\footnote{In the non-relativistic regime that prevails since before chemical decoupling, the DM annihilation cross section can usually be expanded in powers of $v^2=v_{\rm rel}^2/4$, which directly relates to an expansion in partial waves \cite{GriestEtAl1990}. Accordingly, we generically refer to $s$-wave processes as those giving a velocity-independent annihilation rate, while $p$-wave processes come with a $\langle v^2\rangle$-dependent annihilation rate.} \cite{Bergstroem2000,BringmannEtAl2012c,LavalleEtAl2012,FornasaEtAl2015} --- we generically refer to this kind of scenarios as the weakly-interacting massive particle (WIMP) paradigm \cite{Bergstroem2009,Feng2010,ArcadiEtAl2017,BottaroEtAl2022}. Searches in the local universe are nicely complemented by early-universe probes, for instance those deriving from analyses of the cosmic-microwave background (CMB) radiation \cite{Slatyer2016}. Current constraints disfavor DM particle masses below $\sim 50$~GeV with canonical cross sections annihilating into a variety of standard-model final states \cite{Slatyer2021}. Indirect searches, namely the searches for DM annihilation or decay signals in astrophysical probes, are sensitive to parts of the parameter space usually hidden to direct searches (and vice versa), \ie~searches for DM particle collisions onto nuclear targets in underground detectors, due to the different velocity dependencies arising when Feynman diagrams are rotated from annihilation to elastic collision --- for instance, a $p$-wave annihilation into quarks can usually be efficiently probed by direct searches. Hence, deepening the exploration on both fronts is the best way to validate or exclude the thermal DM scenario.

As observations and experiments have now entered the thermal DM parameter space, and as theoretical modeling improves, it becomes important to refine predictions in order to explore further non-trivial, though very interesting, corners of theory space. One of such corners implies particle models in which DM self-interacts through long-range forces, which may lead to what is called the Sommerfeld enhancement of the annihilation cross section \cite{HisanoEtAl2004,HisanoEtAl2005,CirelliEtAl2007,Arkani-HamedEtAl2009,Iengo2009,Slatyer2010,FengEtAl2010a,Cassel2010,BlumEtAl2016}. It is basically triggered when the interaction range becomes larger than the spread of a DM-particle pair wavefunction. This typically occurs when there is a large mass hierarchy between the DM particle and interaction mediators, which is rather generic for multi-TeV DM, but can also be present on more general grounds in case of relatively light dark sectors. Recent examples can be found in, \eg, refs.~\cite{HambyeEtAl2020a,BottaroEtAl2022}.

The Sommerfeld enhancement of the annihilation cross section is similar to gravitational focusing as it depends on inverse powers of the relative speed, $v_{\rm rel}^{-n}$, where $n$ is an integer that will be specified later. Since the DM dispersion velocity relates to the mass of the self-gravitating halo, one can naturally expect that DM subhalos, the tiniest DM structures expected in the universe \cite{Peebles1982,GreenEtAl2004,BerezinskyEtAl2014,ZavalaEtAl2019a}, could be the sites for the largest enhancements. Of course, as we will see, that enhancement may saturate below some characteristic velocity inherent to the properties of DM particles and self-interactions, but subhalos still play a very important role in setting the overall annihilation signal predictions. In this paper, we restrict to DM annihilation into gamma rays, and our calculations are made in such a way that they can be applied to a diversity of DM targets in gamma-ray astronomy, from dwarf galaxies to galaxy clusters.

Several references have actually already addressed the Sommerfeld enhancement in specific DM systems with subhalos, \eg~\cite{KuhlenEtAl2009,Bovy2009,LattanziEtAl2009,PieriEtAl2009,SlatyerEtAl2012,ZavalaEtAl2014a,BoddyEtAl2019,RunburgEtAl2021} (see also refs.~\cite{BoddyEtAl2017,BoddyEtAl2018,BoucherEtAl2021,AndoEtAl2021,BoardEtAl2021}). The goal of this paper is rather to expand upon these works and to provide a more complete and generic analytical understanding of the intricate processes at play, applicable to all targets and covering the full (though simplified) parameter space relevant to the Sommerfeld effect. In particular, we will see that the role of subhalos is critical both on Sommerfeld resonances, and in the case of very large mass hierarchy between the DM particle and the interaction mediator. In some cases, the exploratory power of different gamma-ray targets (e.g., dwarf galaxies vs. galaxy clusters) could even be inverted, which points to new interesting complementary ways to constrain the Sommerfeld parameter space.

The paper develops as follows. In \citesec{sec:glimpse}, we present our general reasoning in simple physical terms, which will pave the way to our more technical discussion in the following sections, and already unveil some of the main results. In \citesec{sec:v-dep-signals}, we introduce the velocity dependencies of the general problem in more technical terms. We start by characterizing the Sommerfeld enhancement in \citesec{ssec:sommerfeld}, and then introduce the velocity-dependent $J$-factor in \citesec{ssec:phase-space}, which defines the amplitude of the gamma-ray signal for a given DM halo target. We further recall how one can get reasonable description of the phase-space distribution functions in self-gravitating DM halos. In \citesec{sec:subhalo_boost}, we extensively discuss how subhalos enter the game and affect the overall predictions, before concluding in \citesec{sec:concl}. A companion paper \cite{LacroixEtAl2022} explores in details the consequences of Sommerfeld enhancement and subhalo boosts on specific gamma-ray targets from dwarf galaxies to galaxy clusters, based on the full numerical calculation.

\section{A glimpse of the main results}
\label{sec:glimpse}
Before digging into the technical aspects of the work, it is useful to summarize them in more simple terms. DM subhalos are well-known boosters of DM annihilation signals; see \eg~refs.~\cite{SilkEtAl1993,BergstroemEtAl1999a,LavalleEtAl2008} for their effects in indirect DM searches with different messengers, and \eg~ref.~\cite{AndoEtAl2019} for a review. For velocity-independent $s$-wave processes, this is a mere consequence of $\langle\rho^2\rangle\geq\langle\rho\rangle^2$, where $\rho$ is the local DM density in a target host halo, which turns to a full inequality thanks to DM inhomogeneities such as subhalos. In that case, given the mass dependence of the signal for one object, the overall contribution is simply obtained by convolving this mass-dependent signal with the subhalo mass function. The latter derives from structure formation theory, and can be approached from both analytical considerations and cosmological simulations. If the signal associated with a single halo of mass $m$ is scale invariant and proportional to $m^{\beta}$ (which, we will show, is a reasonable approximation), and if the subhalo mass function scales like  $m^{-\alpha}$, with both $\alpha,\beta>0$, then the signal integrated over the subhalo mass range is simply $\phi_{\rm sub}\propto m^{-\alpha_{\rm eff}}$, with an effective index $\alpha_{\rm eff} = \alpha-\beta-1$. One readily sees that depending on the sign of $\alpha_{\rm eff}$, the signal will be dominated either by the light mass boundary of the integral (\ie, ~many small objects), or the heavy one (\ie, ~fewer massive objects). The former case generically leads to a stronger subhalo {\em boost factor}, which is a measure of the ratio $\langle\rho^2\rangle_V/\langle\rho\rangle_V^2$ in the relevant volume $V$, and which characterizes the annihilation signal enhancement due to DM inhomogeneities. The amplitude of this boost factor is linked to that of $\alpha_{\rm eff}$, and to the mass hierarchy between the host halo mass and the minimal subhalo mass, the latter being linked to the interaction properties of DM. This holds for $s$-wave annihilation processes.

For $p$-wave processes, no subhalo enhancement is expected because the $p$-wave suppression factor, proportional to $\langle v^2\rangle$, is even more severe in subhalos inside which bound DM particles must have a smaller dispersion velocity not to escape. This argument makes it straightforward to guess that, in scenarios in which subhalos would represent a significant fraction of the total mass, the overall $p$-wave signal could actually even be further subhalo-suppressed. On the other hand, it is also obvious that bigger halos, with larger dispersion velocities, will lead to larger global annihilation rates.

The Sommerfeld enhancement of the annihilation cross section strongly affects the above statements, because it is itself velocity dependent, with a different dependence between $s$- and $p$-wave processes. Then two questions arise: (i) Since the Sommerfeld effect is local by nature, how does it scale at the level of a full object? (ii) How does it propagate over a population of objects? The main complication comes from the fact that the Sommerfeld effect behaves differently depending on whether the relative DM de Broglie wavelength is greater (saturation regime) or lower (Coulomb regime) than the DM self-interaction range, a transition which therefore depends on DM velocity. That specific transition is actually fixed by particle physics independently of any astrophysics, and can therefore be predicted rather accurately (at least in simplified particle DM models). Moreover, the reasoning made just above for Sommerfeld-free $p$-wave processes indicates a possible way: although (sub)halos are featured by spatially-dependent velocity distribution functions, which need to be integrated over to properly describe the Sommerfeld distortion of the annihilation cross section, one could still hope to capture the net effect from a typical velocity for each halo, which would then be related to its mass. If this typical velocity is a scale-invariant function of the halo mass, for instance $\overline{v}\propto m^\nu$, then we can apply the same recipe as above. The transition between the two Sommerfeld regimes occurs at a specific velocity, $\vsat$, entirely fixed by particle physics, which can itself be translated into a specific halo mass, which we denote $\mtsat$. If the Sommerfeld enhancement scales locally like $v^{-s_1}$ in one of its regime, and like $v^{-s_2}$ in the other one, then this converts into a global scaling like $\overline{v}^{-s_1}\propto m^{-\nu\,s_1}$, say for $m>\mtsat$ (Coulomb regime), and $\overline{v}^{-s_2}\propto m^{-\nu\,s_2}$ for $m<\mtsat$ (saturation regime) --- $s_1$ and $s_2$ may take different values for $s$- and $p$-wave processes. In the same vein, the velocity dependence associated with the $p$-wave ``bare'' cross section scales like $\overline{v}^2\propto m^{2\nu}$. To figure out whether subhalos are susceptible to increase the signal, one needs to determine the overall velocity dependence of the Sommerfeld-corrected cross section (including the $p$-wave suppression). Basically, if $s_i>0$ (or $s_i-2>0$ for $p$-wave processes), where $i\in\{1,2\}$, then the Sommerfeld-corrected cross section will be larger with decreasing subhalo mass, which will make them increase the signal.

By integrating this Sommerfeld-corrected cross section times the ``bare'' mass-dependent signal ($\propto m^\beta$) over a power-law subhalo mass function of index $\alpha$, and assuming that the transition mass $\mtsat$ lies within the subhalo mass range defined by $[\mmin,\mmax]$, then we can readily infer two different contributions: one scaling like $m^{-\alpha_1}$, with $\alpha_1=\alpha+s_1-\beta-1$, and the other one scaling like $m^{-\alpha_2}$, with $\alpha_2=\alpha+s_2-\beta-1$ (with an additional factor of $m^{2\nu}$ for $p$-wave processes, and the corresponding change in the associated $\alpha_1$ and $\alpha_2$). The most contributing boundary will be either $\mtsat$ or $\mmax$ in the first regime, and either $\mtsat$ or $\mmin$ in the second regime, depending on the signs of the $\alpha$'s.

As a first important insight, we see that for $p$-wave processes, the mass hierarchy induced by the $p$-wave suppression factor can be fully compensated in the Sommerfeld-corrected case because globally, that suppression factor will simply disappear. The consequence in terms of target hierarchy is expected to be quite significant, as we shall see in more details below.

This paper will enter the technical details of this general program which turns out to work reasonably well, and which allows to derive fully analytical results --- our main results for the Sommerfeld-enhanced subhalo contribution to the signal are given in Eqs.~\eqref{eq:jtot_somm}-\eqref{eq:JSsub}. These (i) can help understand the different scaling with the different parameters at play (Sommerfeld particle physics parameters vs. cosmological or astrophysical subhalo parameters), and (ii) provide decent quantitative approximations to the more involved numerical calculations used in the companion paper \cite{LacroixEtAl2022}.

\section{Velocity-dependent annihilation: theoretical ingredients}
\label{sec:v-dep-signals}
Calculations of Sommerfeld-enhanced signal predictions consist in scaling the velocity dependence of a single annihilation process up to an ensemble of particle annihilations proceeding over an entire halo. Here, we first introduce the velocity dependence of the Sommerfeld enhancement at the level of a pair of DM particles, before describing its integration over a DM halo along the line of sight.
\subsection{Dark matter annihilation and self-interaction: the Sommerfeld effect}
\label{ssec:sommerfeld}
\subsubsection{Conventional formulation}
\label{sssec:conventional}

We start by shortly introducing the most important features of the Sommerfeld enhancement to the DM annihilation cross section that we are going to use throughout this paper. A slightly more extended introduction can be found in \citeapp{app:sommerfeld}. We assume a simplified model in which DM particles $\chi$ can self-interact through multiple exchanges of a single light mediator $\phi$ of mass $m_{\phi}$, with a coupling $g_{\chi} = \sqrt{4\pi\alpha_{\chi}}$, where $\alpha_{\chi}$ plays the role of a dark fine structure constant. If the interaction range, $1/m_\phi$, is larger than the DM Bohr radius, $1/(\alpha_{\chi}m_\chi)$ (close to the Compton length $1/m_\chi$), then the wave function of the two-DM-particle system can be distorted. This rather generically leads to a non-perturbative enhancement of the annihilation cross section called the Sommerfeld effect (we restrict ourselves to attractive interactions), which can often be effectively described by means of a Yukawa potential. This enhancement depends on the relative velocity between the DM particles and saturates when the DM de Broglie wavelength roughly exceeds the interaction range. More detailed descriptions of this phenomenon can be found in, \eg,~refs.~\cite{HisanoEtAl2004,HisanoEtAl2005,Iengo2009,CirelliEtAl2007,Arkani-HamedEtAl2009,Cassel2010,Slatyer2010,FengEtAl2010a,BlumEtAl2016}.

The Sommerfeld enhancement factor $\hat{\cal S}$ allows one to correct for this effect and applies to the nominal annihilation cross section as follows \cite{Sommerfeld1931}:
\ben
\label{eq:enhanced_cs}
\sigma v_{\rm rel} = (\sigma v_{\rm rel})_{0} \times \hat{\cal S}\,,
\een
where the subscript 0 refers to the cross section as commonly computed from perturbation theory. Working in natural units ($\hbar=c=1)$ and expressing velocities in units of the speed of light $c$ from now on, it turns useful to introduce the following dimensionless parameters,
\ben
\label{eq:def_epsilon}
\epsilon_{v} &\equiv& \dfrac{v}{\alpha_{\chi}}\\
\epsilon_{\phi} &\equiv& \dfrac{m_{\phi}}{\alpha_{\chi} m_{\chi}}\nn\,,
\een
where $\epsilon_{\phi}$ roughly expresses the ratio of the Bohr radius of a pair of DM particles, $2/(\alpha_{\chi} m_{\chi})$, to the interaction range, $1/m_\phi$, with $\epsilon_{\phi}\lesssim 1$ indicating the possible onset of the Sommerfeld enhancement. On the other hand, the ratio $\epsilon_v/\epsilon_{\phi}$ roughly characterizes the ratio of the interaction range to the DM de Broglie wavelength. When this ratio gets $<1$, then the long-range interaction is seen as finite again by the quantum system and the Sommerfeld effects saturates. The DM particle speed $v$ featuring above stands for half the relative speed of the pair, $v_{\rm rel}/2$, and $c$ is the speed of light. Parameters $\epsilon_{\phi}$ and $\epsilon_v$ fully characterize the Sommerfeld parameter space in our simple model, and encode the relevant properties of particles and interactions in the dark sector.

Here we focus on $s$-wave and $p$-wave annihilation processes. We recall the more tractable expressions of the Sommerfeld enhancement factor for these two cases, which can be derived from the general analytical solution obtained for the Hulth\'en potential approximation---see \citeapp{app:sommerfeld}. For an $s$-wave annihilation process, the enhancement factor can be written in a simple form as \cite{Slatyer2010}
\ben
\label{eq:Sommerfeld_enhancement_s_wave}
\hat{\cal S}_{s}(v,\epsilon_\phi) &\underset{[\text{Yukawa}\to\text{Hulth\'en}]}{\simeq}&
\dfrac{\pi}{\epsilon_{v}} \dfrac{\sinh\left( \dfrac{2\pi\epsilon_{v}}{\epsilon_{\phi}^{*}} \right)}{\cosh \left( \dfrac{2\pi\epsilon_{v}}{\epsilon_{\phi}^{*}} \right) - \cos\left( 2\pi\sqrt{\dfrac{1}{\epsilon_{\phi}^{*}} - \dfrac{\epsilon_{v}^{2}}{\epsilon_{\phi}^{*2}}} \right)} \\
&\overset{[\text{Coulomb limit}]}{\underset{[\epsilon_{v} \gg \sqrt{\epsilon_{\phi}^*}]}{\longrightarrow}}& \ \dfrac{\pi/\epsilon_{v}}{1 - {\rm e}^{-\pi/\epsilon_{v}}}\,,\nn
\een
where
\ben
\label{eq:def_epsphistar}
\epsilon_{\phi}^{*} \equiv \pi^{2}\epsilon_{\phi}/6\,.
\een
\change{The first line in the expression of the Sommerfeld factor in \citeeq{eq:Sommerfeld_enhancement_s_wave}, \ie~the standard result for the Hulth\'{e}n potential in the literature, provides a good approximation to the exact result (Yukawa potential) all over the parameter space characterized by $\epsilon_v,\epsilon_\phi\lesssim 1$ (taking $\cos\to\cosh$ when $\epsilon_v>\sqrt{\epsilon_\phi^*}$)~\cite{Cassel2010,Slatyer2010}. On the other hand, when $\epsilon_{v} \gg \sqrt{\epsilon_{\phi}^*}$ (second line), it boils down to the result obtained assuming the Coulomb potential, $V_{\rm C}(r) = -\alpha_{\chi}/r$. The condition $\epsilon_{v} \gg \sqrt{\epsilon_{\phi}^*}$ is sufficient but not necessary to ensure that the Coulomb-limit expression in \citeeq{eq:Sommerfeld_enhancement_s_wave} is accurate; for example, in the intermediate case where $\epsilon_{\phi}^* < \epsilon_{v} < \sqrt{\epsilon_{\phi}^*}$ and $\epsilon_v,\epsilon_\phi^*\ll 1$, the enhancement is also well-approximated by the Coulomb-limit expression \cite{LattanziEtAl2009,Slatyer2010}.}

%
For a $p$-wave annihilation process, the enhancement factor reads instead
\ben
\label{eq:Sommerfeld_enhancement_p_wave}
\hat{\cal S}_{p}(v,\epsilon_\phi) = \dfrac{\left( \dfrac{1}{\epsilon_{\phi}^{*}} -1 \right)^{2} + 4 \dfrac{\epsilon_{v}^{2}}{\epsilon_{\phi}^{*2}}}{1 + 4 \dfrac{\epsilon_{v}^{2}}{\epsilon_{\phi}^{*2}}} \times \hat{\cal S}_{s}(v,\epsilon_\phi)\,.
\een

Different regimes arise according to the values of the dimensionless parameters $\epsilon_{v}$ and $\epsilon_{\phi}$:
\begin{itemize}
\item {\bf Large velocity, $\epsilon_{v} \gg 1$} or {\bf heavy mediator, $\epsilon_{\phi} \gg 1$}\\
  There is no enhancement in that case: $\hat{\cal S}_{s} \approx 1$ and $\hat{\cal S}_{p} \approx 1$.
\item {\bf Intermediate velocities, $\epsilon_{\phi} \ll \epsilon_{v} \ll 1$}\\
  Here, we have $\hat{\cal S}_{s} \approx \pi/\epsilon_{v} \propto 1/v$ and $\hat{\cal S}_{p} \approx \pi/(4\epsilon_{v}^{3}) \propto 1/v^{3}$. This contains the regime in which the Yukawa potential tends to a Coulomb potential (for $\epsilon_{v} \gg \sqrt{\epsilon_{\phi}}$) but spans a broader range of values of $\epsilon_{v}$. 
\item {\bf Small velocities, $\epsilon_{v} \ll \epsilon_{\phi} \ll 1$}\\
  This corresponds to the saturation regime of the Sommerfeld effect for which 
\ben
\label{eq:S_saturation}
\hat{\cal S}_{s}(v,\epsilon_\phi) &\approx& \dfrac{12}{\epsilon_{\phi}} \dfrac{1}{1 + \dfrac{2\pi^{2}\epsilon_{v}^{2}}{\epsilon_{\phi}^{*2}}-\cos\left(2\pi\sqrt{1/\epsilon_{\phi}^{*}} \right)} \\
{\rm and}\ \ \hat{\cal S}_{p}(v,\epsilon_\phi) &\approx& \left( \dfrac{1}{\epsilon_{\phi}^{*}} -1 \right)^{2} \hat{\cal S}_{s}(v,\epsilon_\phi)\,,\nn
\een
which is almost independent of the velocity of the DM particles,
except at a series of resonances for
\ben
\label{eq:eps_res}
\epsilon_{\phi} = \epsilon_\phi^{{\rm res},n}\equiv 6/(\pi^{2} n^{2})\,,
\een
with $n$ an integer, for which $\hat{\cal S}_{s} \approx 1/(n^{2}\epsilon_{v}^{2}) \propto \epsilon_\phi^{\rm res}/v^{2}$ and $\hat{\cal S}_{p} \approx \left( n^{2} -1 \right)^{2}/(n^{2}\epsilon_{v}^{2}) \propto 1/(\epsilon_\phi^{\rm res} v^{2})$.
\end{itemize}

These analytical formulations match with the full numerical results within 10\%, except on resonances where larger differences are found. This comes from the fact that the Hulth\'{e}n potential approximation slightly offsets the solution from the one obtained with the Yukawa potential~\cite{FengEtAl2010a,BoddyEtAl2017}. However, for the purpose of this work, the features of the solution, comprising resonances, are sufficiently well accounted for by the analytical solution.

We note that the Sommerfeld factor neglects bound-state decay in the low-velocity regime, which leads to nonphysically large enhancements on resonances, where DM bound states can form, that violate the partial-wave unitarity limit. Consequently, the factorization in \citeeq{eq:enhanced_cs} is expected to fail at vanishing velocities, $\epsilon_v\ll\epsilon_\phi$. Actually, DM bound states have a finite lifetime, which induces a saturation of the enhancement at $v \approx \alpha_{\chi}^{3} m_{\phi}/m_{\chi}$ \cite{HisanoEtAl2005,FengEtAl2010a,BlumEtAl2016}, corresponding to $v \approx \alpha_{\chi}^{4}$ at resonances. As a result, a slightly modified version of \citeeq{eq:enhanced_cs} holds, with the nonphysical divergences regularized by replacing $v$ by $v + \alpha_{\chi}^{4}$~\cite{FengEtAl2010a}. \change{We emphasize that this is only an approximate parametric regularization expected to capture reasonably well the relevant physical effects in the current study --- for a more detailed description and discussion, see ref.~\cite{BlumEtAl2016}.} If $\alpha_{\chi} \ll 1$, bound-state effects mostly restrict to resonances \cite{BoddyEtAl2017}. We can therefore consider a benchmark value of $\alpha_{\chi} = 10^{-2}$, though generalized Sommerfeld corrections can be easily rescaled simply by shifting the values of $\epsilon_\phi$ and $\epsilon_v$.

\subsubsection{A practical ansatz}
\label{sssec:ansatz}
The practical reasoning we develop in this part is general and will turn useful when expressing the Sommerfeld enhancement at the level of a full DM halo. The important aspect is to correctly describe the velocity dependence of the Sommerfeld enhancement. In contrast to our formal definition of the Sommerfeld factor in \citesec{sssec:conventional}, here we absorb the $v^2$ dependence of the $p$-wave annihilation cross section into our effective definition of the Sommerfeld enhancement factor. To be specific, we introduce the following effective Sommerfeld enhancement factor:
\ben
\label{eq:def_Seff}
{\cal S}(v,\epsilon_\phi) \cong \left(\frac{v}{\vmax}\right)^p\times\hat{\cal S}(v,\epsilon_\phi)\,,
\een
where $\hat{\cal S}$ is the exact Sommerfeld factor introduced in the previous paragraph, $v$ is still half of the relative speed of the pair of DM particles, $\vmax$ is a reference speed that will be defined later, and
\ben
\label{eq:def_p}
p =
\begin{cases}
  0\;\; & \text{for $s$-wave annihilation}\\
  2\;\; & \text{for $p$-wave annihilation}
\end{cases}\,.
\een
Accordingly, the $p$-wave annihilation cross section can be expressed as:
\ben
\label{eq:def_pwave_xs}
(\sigma\, v)_\text{$p$-wave} = \sigma_p^0 \left(2\,v\right)^2 \times\hat{\cal S}(v,\epsilon_\phi)  \simeq \sigma_p^0 \left(2\,\vmax\right)^2\times {\cal S}(v,\epsilon_\phi) \,,\nn
\een
where $\sigma_p^0$ is the amplitude of the $p$-wave cross section. The extra factor of 2 is due to the fact that $v=v_{\rm rel}/2$ in our convention. This alternative form implies that there is no longer any velocity dependence in the reference cross section associated with the $p$-wave case. It is fully transferred to the effective Sommerfeld factor. Consequently, the effective $p$-wave Sommerfeld factor now scales like $1/v$ in the Coulomb regime (exactly like in the $s$-wave case), exhibits no speed dependence on resonances, and scales like $v^2$ between resonances in the saturation regime, as we shall review below. This redefinition allows us to introduce a unique ansatz for both $s$- and $p$-wave annihilation processes, and will further make the analytical estimate of the Sommerfeld-corrected subhalo boost factor much simpler to derive and to understand.

We now introduce a simplifying ansatz that captures the main features of the effective Sommerfeld enhancement in asymptotic regimes at the level of local interactions of test particles in a (sub)halo. This ansatz provides a good approximation to the exact result. We can write it as follows (disregarding resonances for the moment):
\ben
\label{eq:S_ansatz_v}
{\cal S}_\text{no-res}(v,\epsilon_\phi) &=&
S_0\left(\frac{v}{\vmax}\right)^{-1}
\left[1 +
  S_1^{-\frac{\overline{s}_{v,c}}{(1+p)}} \left(\frac{v}{\vsat(\epsilon_\phi)}\right)^{-\overline{s}_{v,c}}
  \right]^{-\frac{(1+p)}{\overline{s}_{v,c}}}\\
\Rightarrow  {\cal S}_\text{no-res}(v,\epsilon_\phi)
& \overset{\rm limits}{\longrightarrow}&
\begin{cases}
  S_0\,\left(\frac{v}{\vmax}\right)^{-1} & \forall \vsat(\epsilon_\phi) \ll v \ll \vmax\\
  S_0\,S_1\,\left(\frac{\vmax}{\vsat(\epsilon_\phi)}\right)\left(\frac{v}{\vsat(\epsilon_\phi)}\right)^{p} \propto v^p\,\epsilon_\phi^{-(1+p)}\; &\forall v \ll \vsat(\epsilon_\phi)
\end{cases}
\,,\nn
\een
where $\vsat$, which will be defined later on, marks the transition between the Coulomb and the saturation regimes, and $S_0$ and $S_1$ are calibration constants which can be calculated explicitly:
\ben
\label{eq:Sconstants}
S_0 = (2\,\pi)^{-p}\,;\; S_1 = (6/\pi)\, (12/\pi^2)^p\,.
\een
The ansatz of \citeeq{eq:S_ansatz_v} is valid only when the Sommerfeld effect becomes effective, which corresponds to velocities $v\leqslant \vmax$, where $\vmax$ is defined just below. Two power-law indices appear, $p$ and $\overline{s}_{v,c}$, all positive definite ($p$ is defined in \citeeq{eq:def_p}). Index $\overline{s}_{v,c}$ has no asymptotic relevance, and simply indicates how fast one transits from the Coulomb regime to the saturation regime. The Sommerfeld power-law index in the Coulomb regime is explicitly set to -1 for both $s$- and $p$-wave annihilation, as a consequence of absorbing the velocity dependence of the cross section in the definition of the Sommerfeld correction for the latter. Parameter $\vmax$ stands for the velocity beyond which the Sommerfeld effect roughly turns off, and $\vsat(\epsilon_\phi)$, which does explicitly depend on $\epsilon_\phi$, is the velocity below which the Sommerfeld effect saturates and resonances may appear. That transition occurs when the interaction range becomes shorter than the DM de Broglie wavelength. In between $\vsat$ and $\vmax$, we are in the Coulomb regime (infinite interaction-range limit). These critical velocities can actually be related to the coupling strength $\alpha_{\chi}$ and to the reduced Bohr radius $\epsilon_\phi$, which both characterize the parameter space of our minimal Sommerfeld-enhancement setup. The appropriate definitions read:
\ben
\label{eq:vmax}
\begin{cases}
\vmax & \equiv \pi\,\alpha_\chi\\
\vsat (\epsilon_\phi) &\equiv \epsilon_\phi \,\frac{\vmax}{\pi} \\
\vunit &\equiv \alpha_{\chi}^4
\end{cases}\,.
\een
Here, we only consider situations in which the DM Bohr radius ($\propto$ Compton wavelength) is shorter than the interaction range ($\epsilon_\phi\lesssim 1$), a condition to trigger the Sommerfeld enhancement. The saturation velocity $\vsat$ delineates a transition in velocity dependence, fixed by $\epsilon_v=v/\alpha_X=\epsilon_\phi$, at which the DM self-interaction range and the de Broglie wavelength are similar, and below which the finite range of self-interactions becomes manifest again. Then, the Sommerfeld enhancement saturates and its velocity dependence is frozen, except on resonances. A resonance of order $n$ can efficiently pop up if $v<\vsat(\epsilon_\phi=\epsilon_{\phi}^{{\rm res},n})$, where the saturation velocity is evaluated at the corresponding resonant value of the reduced Bohr radius, $\epsilon_\phi^{{\rm res},n}$. Finally, parameter $\vunit$ is meant to account for the unitarity constraint on Sommerfeld resonances.

\begin{mdframed}[backgroundcolor=light-gray]
\paragraph{Properties of resonances:}
We highlight the discussion of resonances, which will lead to non-trivial features throughout the paper and be specific zones in parameter space of gigantic signal enhancements. In the same spirit as above, we can write a simplifying ansatz to describe the enhancement on resonances, which we deliberately separate from the non-resonant ansatz of \citeeq{eq:S_ansatz_v} for clarity:
\ben
    {\cal S}_{{\rm res},n}(v,\epsilon_\phi) &\overset{n\geqslant 1+\frac{p}{2}}{=}&
    S_0^{\rm res}\,\left(\frac{\vmax}{\vsat(\epsilon_\phi)}\right)
    \left(\frac{v}{\vsat(\epsilon_\phi)}\right)^{(p-2)}
    \left(1+\frac{\vunit}{v}\right)^{-2}\nn\\
    &&\times \theta \left(\vsat(\epsilon_\phi)-v\right)\,
    \delta_{\epsilon_\phi/\{\epsilon_\phi^{{\rm res},n}\}}\nn\\
    &\propto& \,\epsilon_\phi^{(1-p)}\,v^{(p-2)}\,,
    \label{eq:S_ansatz_res_v}
\een
where $\vunit$ has been defined in \citeeq{eq:vmax}, and saturates the amplitudes of resonant peaks when $v<\vunit$, which allows us to effectively prevent any violation of the unitarity constraint (see discussion at the very end of \citesec{sssec:conventional}). We have introduced 
\ben
\label{eq:def_D_measure}
\delta_{\epsilon_\phi/\{\epsilon_\phi^{{\rm res},n}\}}\equiv
\begin{cases}
1\;\; & {\rm if}\;\epsilon_\phi\in\{\epsilon_\phi^{{\rm res},n}\}\\
0 & {\rm otherwise}
\end{cases}
\een
where again $p=0\,/\,2$ for $s/p$-wave annihilation (and for which the first resonance is at $n=1\,/\,2$, respectively). The constant reads:
\ben
S_0^{\rm res} = (\pi/6)\,(6/\pi^3)^{p}\,.
\een

It is important to recall the generic features of resonances, which occur at $\epsilon_\phi \sim \epsilon_\phi^{{\rm res},n}$, and can be triggered only if $v\leqslant \vsat(\epsilon_\phi^{{\rm res},n})$ --- in the above equations, for simplicity, we adopt an extreme simplification by means of a discrete measure, which triggers resonances only when $\epsilon_\phi$ sits exactly on one of its resonant values (to avoid numerical discontinuities, this can be replaced by an extremely thin unnormalized Gaussian function, or even a Cauchy function if one fancies better capturing the actual shapes of resonances).

In the $s$-wave case, resonances are boosted at low velocity $\propto 1/(n\,v)^2\propto\epsilon_\phi/v^2$, with decreasing amplitudes for higher-order resonances (in fact, linearly with $\epsilon_\phi$ [or $\vsat$], as the latter jumps to smaller and smaller resonant values $\epsilon_\phi^{{\rm res},n}$)---see \citeeq{eq:S_saturation} and \citeeq{eq:eps_res}. Note also the unitarity limit that saturates peak amplitudes to $\overset{\sim}{\propto}\epsilon_\phi/(n\,\vunit)^2$ when $v<\vunit$, which will have some impact when inspecting the translation in terms of subhalo masses. As for the inter-resonance baseline (saturation regime), it scales like $\propto 1/\vsat\propto 1/\epsilon_\phi$ and remains velocity independent---see \citeeq{eq:S_ansatz_v} and \citeeq{eq:vmax}.

In contrast, as a consequence of absorbing the $v^2$ suppression factor in the ansatz, $p$-wave resonances $\propto n^2 \propto 1/\epsilon_\phi$ are velocity independent and have their amplitudes increasing for higher-order resonances, \ie~lower resonant values of $\epsilon_\phi$. That feature is actually very important because it implies that the annihilation signal is then only set by the full DM squared density on $p$-wave resonant peaks, except again when approaching the unitarity limit, $v\sim \vunit$. Indeed, at velocities lower than $\vunit$, the $p$-wave suppression re-appears as $\propto (v/\vunit)^2$, which bounds from below the phase-space distribution available to amplify resonances. On the other hand, the inter-resonance baseline remains fully velocity suppressed $\propto v^2/\vsat^3\propto v^2/\epsilon_\phi^3$. Therefore, the surge of resonances in the $p$-wave case is due to the relative suppression of the baseline. Actually, the amplitude ratio ${\cal R}$ between resonances and baseline scales exactly the same for both the $s$-wave and $p$-wave cases in this formulation, and reduces to:
\ben
\label{eq:ratio_res_to_base}
{\cal R}(v,\epsilon_\phi=\epsilon_\phi^{{\rm res},n}) = 
\left(\frac{\pi}{6}\right)^2\,\left(\frac{v}{\vsat}\right)^{-2}\left(1+\frac{\vunit}{v}\right)^{-2}
\overset{\sim}{\propto} (\epsilon_\phi/v)^2\,.
\een
\change{The dependence of the resonant amplitudes on the reduced Bohr radius $\epsilon_\phi$ is shown in \citefig{fig:ansatz}, while their dependence on $v$ is shown in \citefig{fig:v_dependence}, which will be discussed further below.}
\end{mdframed}

All this can be wrapped up in a more synthetic form,
\ben
\label{eq:S_ansatz_tot_v}
{\cal S}(v,\epsilon_\phi) &=& {\cal S}_\text{no-res}(v,\epsilon_\phi)\left(1-  \sum_{n=1+\frac{p}{2}}
\delta_{\epsilon_\phi/\{\epsilon_\phi^{{\rm res},n}\}}
\right)
+ \sum_{n=1+\frac{p}{2}}
{\cal S}_{{\rm res},n}(v,\epsilon_\phi)\\
&\propto& (v/v_0)^{-s_v}\,,\nn
\een
where the generic index $s_v$ takes different values according to the different Sommerfeld regimes:
\ben
\label{eq:sv_values}
s_v=
\begin{cases}
  1\;\; & \text{(Coulomb regime)}\\
  -p & \text{(non-resonant saturation regime)}\\
  (2-p) & \text{(resonances)}\;\longrightarrow -p \;\text{(if $v\lesssim \vunit$)}
\end{cases}
\,,
\een
where $p=0/2$ for $s/p$-wave annihilation. We stress that on resonances, the scaling of peak amplitudes becomes $\propto (v/\vunit)^p$ as soon as $v\lesssim \vunit$, as a consequence of the unitarity limit. This translates into a transition of $s_v$ from $(2-p)$ to $-p$ on resonances at the unitarity boundary.

\begin{figure}[t!]
\centering
\includegraphics[width=0.49\linewidth]{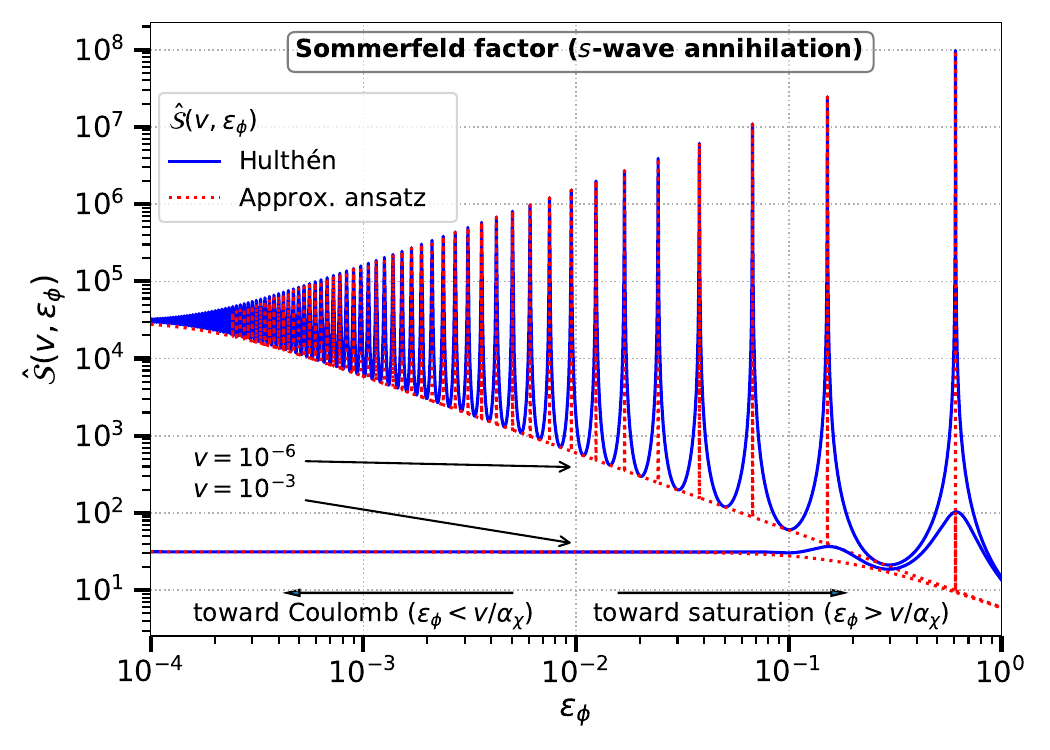} \hfill
\includegraphics[width=0.49\linewidth]{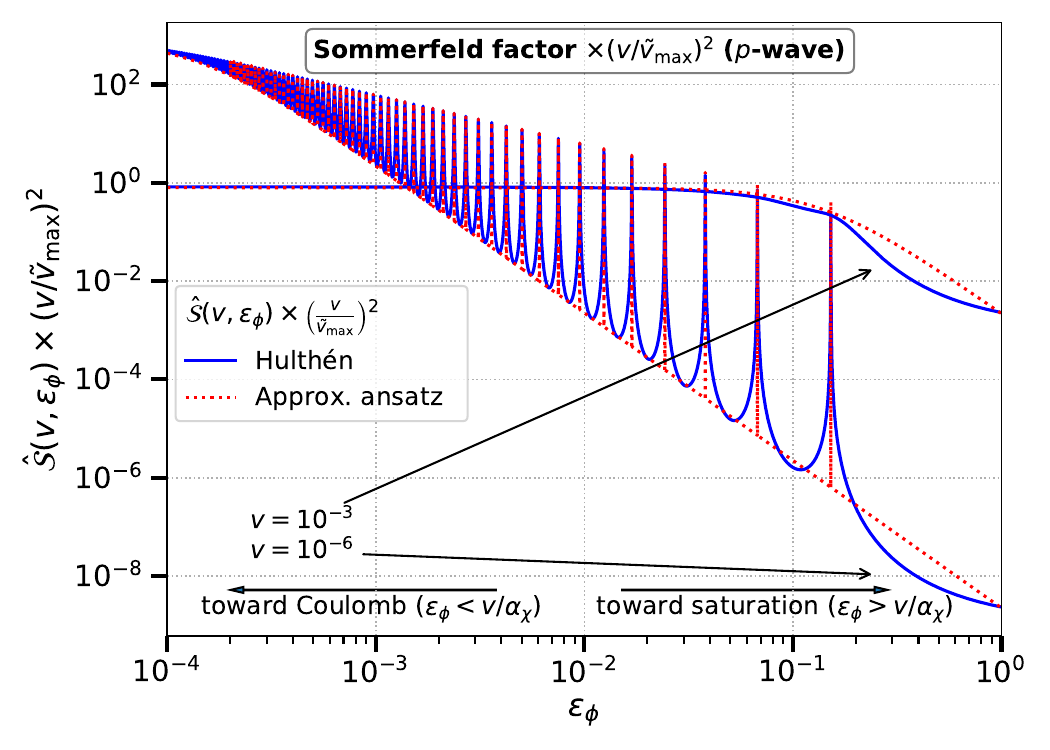}
\caption{{\bf Left panel:} Comparison between the Sommerfeld enhancement factor obtained for an $s$-wave annihilation process from \citeeq{eq:Sommerfeld_enhancement_s_wave} and the ansatz formulated in \citeeq{eq:S_ansatz_tot_v}, for two different values of the speed $v$. {\bf Right panel:} Same for a $p$-wave annihilation process, but then the actual Sommerfeld factor of \citeeq{eq:Sommerfeld_enhancement_p_wave} is multiplied by a factor of $(v/\vmax)^2$ to carry the full velocity dependence of the cross section.}
\label{fig:ansatz}
\end{figure}

\begin{figure}[t!]
\centering
\includegraphics[width=0.99\linewidth]{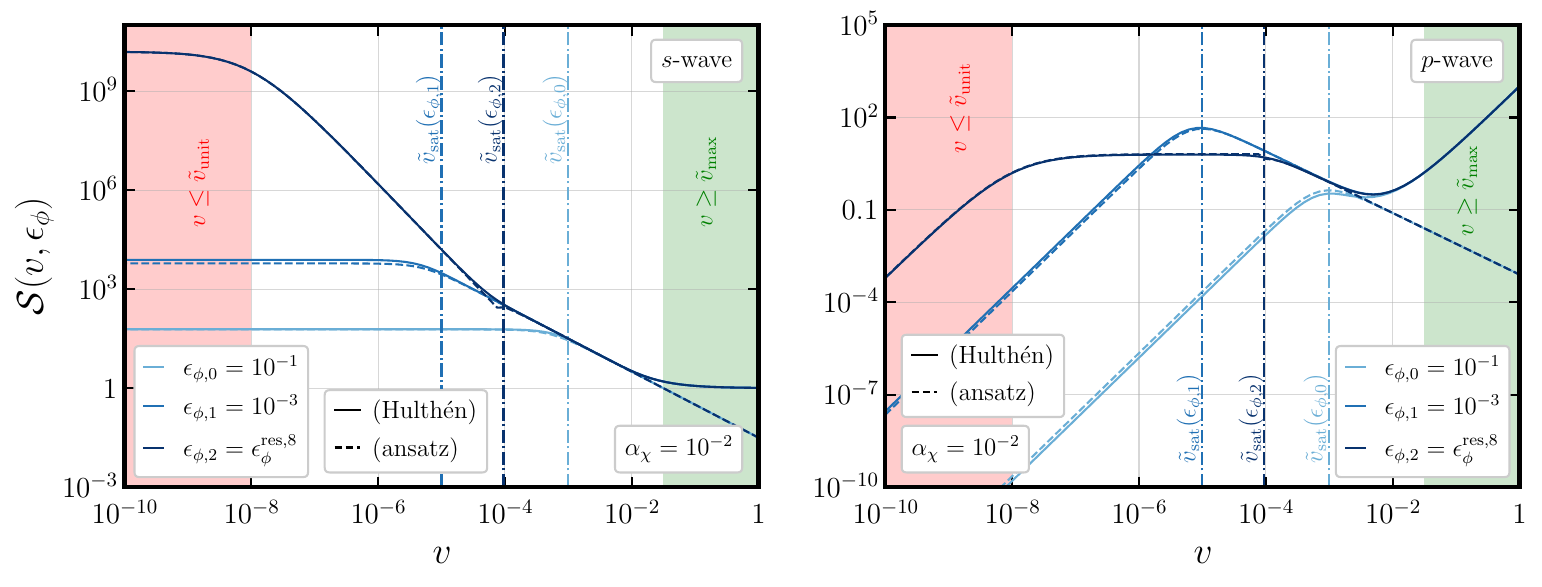}
\caption{\new{Effective Sommerfeld enhancement factor as a function of DM speed, for different values of the reduced Bohr radius $\epsilon_\phi$: a large value 0.1, a small value of $10^{-3}$, and an intermediate value of $\sim 10^{-2}$ sitting on the $n=8$ resonance. The enhancement factor is valid up to $\vmax$, and saturates below $\vunit$ on $s$-wave resonances (not on $p$-wave ones). Transition from Coulomb to saturation regimes occurs at $\vsat(\epsilon_\phi)$, reported as vertical dash-dotted lines. {\bf Left panel:} $s$-wave case. {\bf Right panel:} $p$-wave case.}}
\label{fig:v_dependence}
\end{figure}

Our general ansatz of \citeeq{eq:S_ansatz_tot_v} fully parameterizes the Sommerfeld enhancement factor at the level of local interactions in DM halos. It will serve as a basis to integrate the Sommerfeld effect over an entire (sub)halo. A comparison of this ansatz with the exact solution of the Sommerfeld enhancement factor is provided in \citefig{fig:ansatz} for both the $s$-wave and $p$-wave cases, assuming two values (high and low) of the relative DM speed. We see that this form closely matches with the exact result, except when $\epsilon_\phi$ approaches 1, as expected. In the $p$-wave case, the change of hierarchy in the Sommerfeld enhancement between the low and high velocity curves (with respect to the $s$-wave case)  is simply due to the fact that we have absorbed the $v^2$ suppression factor in the definition of the effective Sommerfeld factor. The virtue of this is that we directly see the true hierarchy of full cross sections as function of velocity from this effective definition. In particular, we see that even though there is a relative $p$-wave suppression of $10^{-6}$ between $v=10^{-3}$ and $v=10^{-6}$, the Sommerfeld-enhanced cross sections have similar amplitudes at $\epsilon_\phi\sim 1.5\times 10^{-3}$, with a net and increasing advantage to smaller velocities for smaller values of $\epsilon_\phi$. Already, this helps understand the fundamental role to be played by DM structures with small dispersion velocities in the following.

\new{To further illustrate the velocity dependency of the effective Sommerfeld factor, we explicitly show ${\cal S}$ as a function of DM speed in \citefig{fig:v_dependence}, for three different values of the reduced Bohr radius $\epsilon_\phi$: a relatively ``large'' value of 0.1, which implies a moderate hierarchy between the DM particle mass and that of the force carrier (moderate Sommerfeld enhancement); a small value of $10^{-3}$, hence a stronger hierarchy (significant enhancement); and an intermediate value of $\sim 10^{-2}$, but sitting exactly on the $n=8$ resonance (strong enhancement). The left (right) panel shows the dependence for an $s$-wave ($p$-wave) annihilation process. The saturation velocities $\vsat(\epsilon_\phi)$ associated with the different choices of $\epsilon_\phi$ are displayed as vertical dashed lines, which delineate the transition between the saturation (to the left thereof) and the Coulomb (to the right) regimes. This figure illustrates the reasonably good match between our Sommerfeld ansatz of \citeeq{eq:S_ansatz_tot_v} (dashed curves) and the exact formulation (solid curves). Following curves from right to left (decreasing velocity), for the $s$-wave (left panel), the enhancement $\propto 1/v$ in the Coulomb regime saturates as $v\leq \vsat$, except on the resonance for which it further increases $\propto 1/v^2$ down to the unitarity limit characterized by $\vunit(\alpha_\chi)$, at which it finally saturates. For the $p$-wave case (right panel), we actually see the product of the net Sommerfeld factor with the $p$-wave suppression factor $\propto v^2$ [\ie~the effective Sommerfeld factor as defined in \citeeq{eq:def_Seff}], which slightly delays the onset of the enhancement as $v$ decreases below $\vmax$. Then, as for the $s$-wave case, the Coulomb regime ($v>\vsat$) exhibits a $1/v$ scaling down to $\vsat$ (which hardly compensates for $p$-wave suppression for large reduced Bohr radii $\sim 0.1$, leading to a small net enhancement). Transitioning to the saturation regime, the effective Sommerfeld effect saturates to its maximal value for the $p$-wave case for $v\sim \vsat$, before the $p$-wave suppression factor takes over at velocities smaller than $\vsat$. On the resonance, however, the maximal saturation value is further maintained independent of the velocity all the way down to the unitary limit characterized by $\vunit$ (the actual Sommerfeld enhancement compensates for the $p$-wave suppression), below which $p$-wave suppression ends up taking over. All this explains the important role played by $\vsat(\epsilon_\phi)$ in the $p$-wave case, as well as the one of $\vunit$ on resonances. \citefig{fig:v_dependence} will later help better understand the mass-velocity dependencies at fixed values of $\epsilon_\phi$.}

\subsection{Gamma-ray signals: astrophysical factors and DM phase-space modeling}
\label{ssec:phase-space}

The DM-induced $\gamma$-ray flux integrated over a sky region of solid angle $\Delta \Omega$ about a target halo center reads~\cite{BergstroemEtAl1998a}\footnote{For easier comparison with the majority of previous works in the literature, we do not include in the definition of the $J$-factor the $1/(4\pi)$ factor that appears in the derivation of an intensity from a volume emissivity, and is a prefactor in \citeeq{eq:flux}. As a result, the $J$-factors given in this work are expressed in $\rm GeV^{2}\, cm^{-5}\, sr$.}
\ben
\label{eq:flux}
\dfrac{\dd\Phi_{\gamma}}{\dd E_{\gamma}} = \dfrac{1}{4 \pi} \dfrac{(\sigma v_{\rm rel})_{0}}{\eta m_{\chi}^{2}}\dfrac{\mathrm{d}N}{\dd E_{\gamma}} J_{\cal S} (\Delta \Omega)
\een
where $\dd N/\dd E_{\gamma}$ is the $\gamma$-ray spectrum per annihilation, and $\eta = 2$ for self-conjugate DM ($\eta = 4$ for non-self-conjugate DM). In the case of a velocity-dependent annihilation cross section that can be expressed as $(\sigma v_{\rm rel})_{0}\times {\cal S}(v)$, like in the effective formulation of the Sommerfeld enhancement above, the astrophysical factor $J_{\cal S}$ encodes the information on both the DM spatial and velocity distributions, and reads 
\ben
\label{eq:J_S_sp_wave}
J_{\cal S} (\Delta \Omega) = \int_{\Delta \Omega} \! \dd \Omega \, \int \! \dd s \, \int \! \dd ^{3}\vec{v}_{1} \, \int \! \dd ^{3}\vec{v}_{2} \, f(r(s,\Omega),\Vec{v}_{1}) \, f(r(s,\Omega),\Vec{v}_{2})\, {\cal S}\! \left(\dfrac{v_{\rm rel}}{2}\right)\,,
\een
where $\vec{v}_{\rm rel} = \vec{v}_{2} - \vec{v}_{1}$ is the relative velocity with $v_{\rm rel} = |\vec{v}_{\rm rel}|$, and $f(r,\vec{v})$ is the phase-space distribution function (PSDF) of the DM (assuming spherical symmetry). Here, the PSDF is normalized to the total mass of the gravitational system of interest, such that at halocentric radius $r$
\ben
\label{eq:rho_from_f}
\rho_{\chi}(r) = \int \! \dd ^{3}\vec{v}\, f(r,\Vec{v})\, .
\een
Note that if one trades the effective Sommerfeld factor ${\cal S}$ for its exact form $\hat{\cal S}$, one should add an additional factor of $(v/c)^2$ in the expression of the $J$-factor for $p$-wave annihilation --- see \citeeq{eq:def_Seff}. Our effective form allows to write a unique form for both $s$- and $p$-wave processes by absorbing the full velocity dependence in ${\cal S}$.
\citeeq{eq:J_S_sp_wave} is a generalization of the standard velocity-independent $J$-factor 
\ben
\label{eq:Jfactor_swave}
J(\Delta \Omega) = \int_{\Delta \Omega} \! \dd \Omega \, \int \! \dd s \, \rho_{\chi}^{2}(r(s,\Omega))\,,
\een
which is valid for $s$-wave annihilation without Sommerfeld enhancement.

Assuming spherical symmetry of the DM halo, the integral over solid angle becomes an integral over the angular distance $\theta$ from the center of the object, with $\dd \Omega = 2 \pi \sin \theta \, \dd\theta$ and $r(s,\Omega) \equiv r(s,\theta) = \sqrt{s^{2} + D^{2} - 2 s D \cos \theta}$, where $D$ is the distance of the observer to the center of the object. The integral is usually performed over an angular size $\theta_{\rm int}$ that depends on the target and the $\gamma$-ray detection technique. In this study, we will assume the distances of target halos to be sufficiently large to integrate the signals over angular extents exceeding the virial sizes of halos (point-like approximation).

In practice, \citeeq{eq:J_S_sp_wave} can be rewritten in terms of a $J$-factor for an effective squared density profile $\rho_{\chi,{\rm eff}}$ as
\ben
\label{eq:Jfactor_sommerfeld}
J_{\cal S} (\theta_{\rm int}) = 2 \pi \int_{0}^{\theta_{\rm int}} \! \dd \theta \, \sin \theta \int \! \dd s \, \rho_{\chi,{\rm eff}}^{2}(r(s,\theta))\,,
\een
assuming that the telescope points to the center of the target halo (this is easily generalized to any direction, see, e.g.,~\cite{KuhlenEtAl2008}), with a resolution angle of $\theta_{\rm int}$. Correspondingly, we introduce
\ben
\label{eq:def_rhoeff}
\rho_{\chi,{\rm eff}}^{2}(r) \equiv \left\langle {\cal S} \left(\dfrac{v_{\rm rel}}{2}\right) \right\rangle_v  (r)  \,\times\, \rho_{\chi}^{2}(r)\,,
\een
where $\langle\rangle_v$ denotes an average over the DM relative velocity distribution. The average of an observable $\mathcal{O}(v_{\rm rel})$ that depends on the relative velocity is conventionally given by 
\ben
\label{eq:average_O_vrel}
\left\langle \mathcal{O}(v_{\rm rel}) \right\rangle_v (r) = \int \! \dd ^{3}\vec{v}_{\rm rel}\, \mathcal{O}(v_{\rm rel})\, F_{\rm rel}(r,\vec{v}_{\rm rel})\,, 
\een
where the relative velocity distribution reads
\ben
F_{\rm rel}(r,\vec{v}_{\rm rel}) = \int \! \dd ^{3}\vec{v}_{\rm c}\, f_{\vec{v}}(r,\vec{v}_{1}) \, f_{\vec{v}}(r,\vec{v}_{2})\,,
\label{eq:relative_velocity_pdf}
\een
with $\vec{v}_{\rm c} = (\vec{v}_{1} + \vec{v}_{2})/2$ the center-of-mass velocity and $f_{\vec{v}}(r,\vec{v}) \equiv f(r,\vec{v})/\rho_{\chi}(r)$ the DM velocity distribution, defined as a probability density function (PDF), \ie~normalized to 1 over the relevant phase space. 

The accurate numerical results of this work are based on the Eddington inversion formalism \cite{Eddington1916a,BinneyEtAl2008}, which, assuming an isotropic velocity distribution for DM particles and a spherically symmetric halo in dynamical equilibrium, predicts the full DM PSDF $f(\vec{r},\vec{v})$. For a detailed discussion of the applicability of the Eddington inversion to different classes of DM halos, see ref.~\cite{LacroixEtAl2018}. Note that the predictive power of this formalism has been tested against cosmological simulations in ref.~\cite{LacroixEtAl2020}, and has been shown to predict the velocity moments of DM within $\sim 15$\% accuracy. Interestingly, such isotropic PSDF models have a similar predictive power as more elaborate models including anisotropy in the velocity field \cite{PetacEtAl2021}.

Starting from the PSDFs of DM halos (in a large range of masses) as predicted from the Eddington inversion method, we found that a very good estimate ($\leqslant 30$\% of error) of the averaged effective Sommerfeld factor could be obtained by picking the non-averaged Sommerfeld factor at some averaged values of the relative speed:
\ben
\left\langle {\cal S} \left(\dfrac{v_{\rm rel}}{2}\right) \right\rangle_v(r)
\simeq {\cal S} \left(\dfrac{\langle v_{\rm rel}^{\frac{2}{(p-1)}} \rangle^{\frac{(p-1)}{2}}(r)}{2}\right)\,,
\een
where the relative velocity moments $\langle v_{\rm rel}^{\pm n} \rangle_v(r)$ are calculated from the Eddington PSDF inferred for the considered halo. This is roughly valid for an extended range of halo masses, from very small subhalo to galaxy cluster masses. In the next section, we carefully inspect the impact of DM subhalos on the overall Sommerfeld-enhanced signal predictions.

\section{Subhalo boost factor for velocity-dependent annihilation}
\label{sec:subhalo_boost}
DM subhalos, which are a generic prediction of the theory of structure formation within CDM \cite{Peebles1982,BondEtAl1982,BlumenthalEtAl1984} and thus also characterize the WIMP class of models \cite{HofmannEtAl2001,GreenEtAl2004,Bertschinger2006a,BringmannEtAl2007a,BerezinskyEtAl2014}, are known to increase the $s$-wave annihilation rate, which is referred to as {\em subhalo boost factor} in the frame of indirect DM searches \cite{SilkEtAl1993,BergstroemEtAl1999a,UllioEtAl2002,LavalleEtAl2007,LavalleEtAl2008}. Predictions for subhalo boost factors have been mostly derived for vanilla $s$-wave annihilation processes, for which the annihilation rate is velocity-independent. The impact of subhalos is also expected to be important when the annihilation rate depends on (inverse powers) relative speed, but it is then slightly more difficult to calculate. Indeed, for the broad picture, since the internal average velocity dispersion of DM in subhalos decreases as their masses decrease, then the mass function of subhalos should translate into a non-trivial relative speed function. Since the Sommerfeld enhancement scales like powers of $1/v$, it is clear that the presence of small subhalos can significantly amplify predictions of the annihilation rate in target objects. We shall see below that $\epsilon_\phi$, the DM Bohr radius ($\sim$ Compton wavelength) in units of the interaction range, is actually the key parameter that determines the most relevant subhalo mass range. We shall also see that the related additional boost factor amounts to orders of magnitude. Before going into more details, we recall that the impact of subhalos was already studied in several references, \eg~\cite{KuhlenEtAl2009,Bovy2009,LattanziEtAl2009,PieriEtAl2009,vandenAarssenEtAl2012,SlatyerEtAl2012}, though with different perspectives.

Here we improve over past studies on several aspects. First, we rely on an analytical subhalo population model mostly built from constrained and controlled theoretical inputs, which self-consistently obeys the global kinematic and dynamical constraints on the host halo, and which includes subhalo tidal stripping. This means that given an observationally constrained global mass model for the host halo, we can self-consistently translate it into a halo model that comprises both a smooth distribution of DM and a substructure component. The bases and features of this model were proposed in ref.~\cite{StrefEtAl2017}, and further explored in, \eg,~refs.~\cite{CaloreEtAl2019,StrefEtAl2019,HuettenEtAl2019,FacchinettiEtAl2020}. This analytical subhalo population model can be easily applied to any host halo configuration. The complete model is used to get our more accurate numerical results on the Sommerfeld-enhanced subhalo contribution to $J$-factors, while further analytical approximations are used to derive fully analytical results. Second, similar to other recent studies (\eg~\cite{BoddyEtAl2017,BoddyEtAl2018,BoddyEtAl2019,BoucherEtAl2021}), we take advantage of the phase-space distribution studies performed in refs.~\cite{LacroixEtAl2018,LacroixEtAl2020}. The latter rigorously determine the regimes where the application of the Eddington inversion method \cite{Eddington1916a} can lead to a reliable description of the PSDF of DM in structures, which can be used to compute any velocity-dependent DM signal (see direct applications of these studies in, \eg,~ref.~\cite{BoudaudEtAl2019a} for $p$-wave annihilation, or in ref.~\cite{LopesEtAl2021} for DM capture by stars). Although hardly scalable to a full population of objects, the Eddington inversion applied to a reduced subhalo mass range can be used to calibrate analytical approximations and get accurate results.

In the following, we first give in \citesec{ssec:submodel} a description of the subhalo population model, before formalizing the general calculation of the induced boost factor in \citesec{ssec:numerical_boost}. Finally, we turn to an approximate analytical derivation of the boost factor in \citesec{ssec:analytical_boost}, which will allow us to make a detailed physical interpretation of the more accurate numerical results.

\subsection{Subhalo population model}
\label{ssec:submodel}

Here, we introduce the main properties of the subhalo population model proposed in ref.~\cite{StrefEtAl2017} (SL17 henceforth). The philosophy behind this model is to think of a DM halo as an assembly of smaller-scale pre-existing halos, consistently with the prescriptions of excursion set theory and merger-tree studies (see \cite{BondEtAl1991a,LaceyEtAl1993,vandenBosch2002} and, \eg,~\cite{JiangEtAl2016,IshiyamaEtAl2020a} for more recent approaches). Should these subhalos be hard spheres with negligible subhalo-subhalo encounter rate, they would simply track the global gravitational potential of the global host halo, which they are part of. However, they actually experience tidal mass loss and may even be disrupted in some cases. These phenomena depend on the time spent in the host and on their pericenter (deepest position in the host gravitational potential), and possibly encounters with stellar disks and individual stars in spiral galaxies. Sticking to a spherically symmetric description of both a smooth halo (which comprises both the originally diffuse DM and the DM stripped away from subhalos) and a subhalo population, one can write down a constrained {\em smoothed} mass density profile $\langle \rho_{\rm host}\rangle$, where $\langle\rangle$ denotes an average in spherical shells here, for the host in terms of two components:
\ben
\label{eq:rhohost}
\langle \rho_{\rm host}\rangle (R) = \rho_{\rm sm}(R) + \rho_{\rm sub}(R)\,,
\een
where $R$ is the distance to the host center, $\rho_{\rm sm}$ the smooth density profile and $\rho_{\rm sub}$ the averaged subhalo population density profile.\footnote{In practice, the smooth component is deduced from the subhalo population model and the global host profile, according to $\rho_{\rm sm}(R) = \langle\rho_{\rm host}\rangle(R) - \rho_{\rm sub}(R)$, and must obey the condition $\rho_{\rm sm}(R)\geqslant 0$.} The coarse-grained global host mass density profile $\langle\rho_{\rm host}\rangle$ is constrained from structure formation to be close to a Navarro-Frenk-White (NFW) profile \cite{Zhao1996,NavarroEtAl1996a,NavarroEtAl1997,Einasto1965,NavarroEtAl2004,MerrittEtAl2006,NavarroEtAl2010}, which is also consistent with observational constraints on different scales \cite{ReadEtAl2018,McMillan2017,EttoriEtAl2019}, pending ongoing debates about possible core-cusp issues \cite{deBlok2010,BullockEtAl2017,ZavalaEtAl2019a}. An important point is that $\rho_{\rm host}$ is also the one global density profile constrained by kinematic or dynamical studies of specific objects, should they be dwarf galaxies, galaxies, or galaxy clusters. A consistent subhalo population model should then be such that the sum of the smooth halo profile and the overall subhalo profile matches with observational constraints on the global host halo, whenever available.  

Rigorously, $\rho_{\rm sub}$ should be described as a discrete sum over all subhalos mapping all inhomogeneities, but the smoothed limit (\ie,~an average within spherical shells) can be considered to describe the overall subhalo density profile:
\ben
\label{eq:rhosub}
\rho_{\rm sub}(R) = \left\langle \sum_i^{N_{\rm tot}} \rho_i(|\vec{R}-\vec{r}_i|) \right\rangle_{|\vec{R}|}(R) = \int \dd m \,\frac{\dd n_{\rm sub}(R,m)}{\dd m}\,\langle m_{\rm t}\rangle_c (m,R)\,,
\een
where $\vec{r}_i$ is the position vector (center) of the $i^{\rm th}$ subhalo in the host frame and $\rho_i$ its spherical density profile, $m=m_{200}$ is the canonical virial mass a subhalo would have in a homogeneous background, $\langle m_{\rm t}\rangle_c(R)$ is the subhalo physical tidal mass $m_{\rm t}\leq m$ averaged over concentration at radius $R$, and $\dd n_{\rm sub}/\dd m$ is the differential number density of subhalos per unit mass, in the continuous limit. The implementation of tidal effects is hidden in the way the tidal mass $m_{\rm t}$ is predicted, given a fictitious virial mass $m$, a concentration $c$, and a prescription for the density profile, which will be specified later.

Therefore, designing a subhalo population model implies defining this continuous limit in terms of a subhalo number density consistent with \citeeq{eq:rhohost} while carrying imprints of initial cosmological conditions distorted by environmental effects (gravitational tides). Our model defines this number density in terms of PDFs describing the mass function $\dd{\cal P}_m(m)/\dd m$, the concentration function $\dd{\cal P}_c(c,m)/\dd c$, the {\em driving} spatial distribution $\dd\overline{\cal P}_V/\dd V$ (the meaning of {\em driving} is made clear in the appendix), and the total number of subhalos $N_{\rm tot}$ orbiting the host halo:
\ben
\label{eq:nsub}
\frac{\dd n_{\rm sub}(R,m)}{\dd m} = \frac{\dd^2 N_{\rm sub}}{\dd m\,\dd V} =  \frac{N_{\rm tot}}{K_{\rm tidal}}\frac{\dd \overline{\cal P}_V(R)}{\dd V} \int \dd c\, \frac{\dd^2{\cal P}_{c,m}(c,m,R)}{\dd c\,\dd m}\,.
\een
In this equation, $N_{\rm tot}$ is the total number of surviving subhalos, and $K_{\rm tidal}\leq 1$ is a normalization constant that ensures the whole PDF to be normalized to unity (said differently, it accounts for the fact that the nominal concentration and mass PDFs can be cut off by tidal effects). The concentration and mass PDFs are intricate as a result of tidal effects, which is explained in \citeapp{app:subhalo_model}. This comes from the fact that gravitational tides are more efficient in pruning less concentrated objects, which induces a selection of halos in concentration (hence on mass) depending on their averaged orbital distance to the host halo center. At this stage, it is therefore important to introduce two other parameters of the model: the minimal and maximal subhalo {\em virial} masses, $\mmin$ and $\mmax$, respectively (keeping in mind that the actual smallest masses in the population model can be much smaller than $\mmin$, due to tidal stripping). The former relates to the interaction properties of DM particles, and is in most cases fixed by the free-streaming length of DM at matter-radiation equality \cite{HofmannEtAl2001,GreenEtAl2004,BringmannEtAl2007a}---for WIMPs, it may take values in the range $10^{-12}$-$10^{-4}$~$\Msun$. The latter obviously depends on the host halo mass, and will be fixed to $\mmax=0.01\,M_{\rm host}$ throughout this work, similar to what is found in cosmological simulations \cite{DiemandEtAl2007a,SpringelEtAl2008,KlypinEtAl2011}.

While we include all the details in the full numerical calculations of $J$-factors, it is interesting to write down an approximation of the expected subhalo distribution as follows:
\ben
\label{eq:approx_dnsub}
\frac{\dd^2 n_{\rm sub}(R,m,c)}{\dd c\, \dd m} \approx
N_{\rm tot} \, \frac{\dd \overline{\cal P}_V(R)}{\dd V} \, \frac{\dd\overline{\cal P}_{m}(m)}{\dd m} \, \frac{\dd\overline{\cal P}_{c}(c)}{\dd c}\,.
\een
If subhalos were hard spheres insensitive to tides, this equation would give a decent description of the subhalo population, with concentration and mass PDFs being close to those of field subhalos. In such a hard-sphere approximation, the spatial distribution would be merely
\ben
\frac{\dd \overline{\cal P}_V(R)}{\dd V} = \frac{\langle\rho_{\rm host}\rangle(R)}{M_{\rm host}} \,.
\een
In fact, departures from the spatial matching between the total halo mass profile and the total subhalo mass profile are mostly observed in the inner parts of host halos in simulations, where tidal effects are strong \cite{DiemandEtAl2007a,SpringelEtAl2008}. Although the abundance of subhalos in the very central regions of host halos can hardly be measured reliably in simulations, due to resolution issues, this flattening of the subhalo spatial distribution in the central parts of host halos can actually still be predicted analytically by considering tidal disruption on top of tidal stripping, as detailed in \citeapp{app:subhalo_model}. The above approximation will still turn useful when trying to get analytical estimates of the Sommerfeld-enhanced subhalo boost factor.

\subsection{Subhalo boost factor: generalities}
\label{ssec:numerical_boost}
Here we establish the complete expressions that are used to perform generic computations of the subhalo boost factor --- see more detailed discussions in, e.g.,~\cite{BergstroemEtAl1998a,BergstroemEtAl1999a,KuhlenEtAl2008,PieriEtAl2011,CharbonnierEtAl2012,FornasaEtAl2015,StrefEtAl2017,CaloreEtAl2017,DiMauroEtAl2020,FacchinettiEtAl2020}. We have introduced the astrophysical $J$-factor in \citesec{ssec:phase-space}, which is proportional to the integral of the DM squared density profile $\rho^2_\chi$ along the line of sight, with the replacement $\rho^2_\chi\leftrightarrow \rho^2_{\chi,{\rm eff}}$ given in \citeeq{eq:def_rhoeff} to account for any velocity dependence in the annihilation signals. Neglecting subhalos amounts to setting $\rho^2_\chi(r)=\rho_{\rm host}^2(r)$, where the total DM profile of the host object is given in \citeeq{eq:rhohost} and includes both a smooth DM component and a subhalo population. A definition of the subhalo boost factor is straightforward and may readily be expressed in terms of the relevant $J$-factors:
\ben
\label{eq:def_boost}
{\cal B} \equiv \frac{J_{\rm tot}}{J_\text{smooth approx}}\,,
\een
where $J_{\rm tot}$ is the $J$-factor including rigorously both the smooth and subhalo contributions, while $J_\text{smooth approx}$ simply consider the contribution of the whole system after smoothing out all inhomogeneities. In this form, ${\cal B}$ is merely the multiplicative factor to apply to the smooth approximation of the $J$-factor to get the one accounting for subhalos. The fact that ${\cal B}\geqslant 1$ is a rather generic\footnote{Note that for nominal velocity-suppressed $p$-wave annihilation, we could actually have ${\cal B}\leqslant 1$.} consequence of that $\langle \rho_{\rm host}^2\rangle(r)\geqslant \langle \rho_{\rm host}\rangle^2(r)$ \cite{SilkEtAl1993}.

Using \citeeq{eq:Jfactor_swave}, we can already express the smooth approximation of the total $J$-factor as
\ben
J_\text{smooth approx} = \int_{\Delta \Omega} \! \dd \Omega \, \int \! \dd s \, \langle\rho_{\rm host}\rangle^{2}(R(s,\Omega))\,,
\een
which is the integral of the squared global density profile of the host halo along the line of sight, neglecting any inhomogeneous component. Assuming that subhalos contribute as point-like sources, the actual total $J$-factor should rather be expressed as
\ben
J_{\rm tot} = \int_{\Delta \Omega} \! \dd \Omega \, \int \! \dd s \, \langle \rho_{\rm host}^{2} \rangle (R(s,\Omega))\,,
\een
with
\ben
\label{eq:rho2undrl}
\langle\rho_{\rm host}^{2}\rangle(R) \cong \rho_{\rm sm}^2(R) + \underline{\rho_{\rm sub}}^{2}(R) + 2\, \rho_{\rm sm}(R)\,\rho_{\rm sub}(R)\,.
\een
We have introduced $\underline{\rho_{\rm sub}}^{2}\neq \rho_{\rm sub}^2$ to account for the fact that if subhalos contribute as point-like sources, their contribution to the annihilation flux is not proportional to their smooth mass density profile squared $\rho_{\rm sub}^2$, but rather to
\ben
\underline{\rho_{\rm sub}}^{2}(R) \equiv \rho_\circledast^2 \, \int \dd c\, \int \dd m\, \xi_{\rm t}(m,c,R)\,\frac{\dd^2n_{\rm sub}(c,m,R)}{\dd c\,\dd m}\,,
\een
where the subhalo number density in mass-concentration phase-space $\dd^2n_{\rm sub}/\dd c\,\dd m$ can be inferred from \citeeq{eq:nsub}, and is given an approximation in \citeeq{eq:approx_dnsub}. We have introduced the tidal annihilation volume $\xi_{\rm t}(m,c,R)$ for a subhalo of virial mass $m$, concentration $c$, and radial position $R$ in the host halo, defined as
\ben
\label{eq:def_xi}
\xi_{\rm t}(m,c,R) \equiv 4\,\pi \int_{r_{\rm t}(m,c,R)} \,\dd r \,r^2\,\left\{\frac{\rho(r,m,c,R)}{\rho_\circledast}\right\}^2\,.
\een
This is the integral of the inner subhalo density profile $\rho(r,m,c)$ performed over the assumed spherically symmetric subhalo tidal volume $\delta V_{\rm t}$ delineated by the tidal radius $r_{\rm t}(m,c,R)$, whose parametric dependencies are explicit. The constant parameter $\rho_\circledast$ is an arbitrary normalization density, which allows $\xi_{\rm t}$ to be interpreted as the effective volume a (sub)halo would need to reach the same annihilation rate as if it had a constant density of $\rho_\circledast$ \cite{LavalleEtAl2007,LavalleEtAl2008}.

Following \citeeq{eq:rho2undrl}, the total $J$-factor can be rewritten as
\ben
J_{\rm tot} = J_{\rm sm} + J_{\rm sub} + J_{\rm cross} \simeq J_{\rm sm} + J_{\rm sub} \,,
\een
where the definition of each term is now obvious, and where it is assumed that we can neglect the cross term to a very good approximation \cite{CharbonnierEtAl2012,StrefEtAl2017,FacchinettiEtAl2020}. We still include it, though, in our numerical calculations.

Therefore, one can fully compute the subhalo boost factor once $\dd^2n_{\rm sub}/\dd c\,\dd m$ and $\rho_{\rm sub}$ are determined (assuming an universal shape for the subhalo density profile). When a velocity dependence of the annihilation cross section is considered, the above expressions change only by the substitutions already introduced in \citesec{ssec:phase-space} (we specialize to the case of the Sommerfeld enhancement, though this statement is more general):
\ben
\begin{cases}
  \rho_{\rm sm}^2(R) &\longrightarrow \;\rho_{\rm sm,eff}^2(R)\equiv \left\langle {\cal S}\! \left(\dfrac{v_{\rm rel}}{2}\right) \right\rangle_v (R)  \,\times\, \rho_{\rm sm}^2(R)\,,\\
  \xi_{t}(m,c,R) &\longrightarrow\; \xi_{\rm t,eff}(m,c,R) \equiv 4\,\pi
  \displaystyle \int_0^{r_{\rm t}(m,c,R)} \,\dd r \,r^2\,\left\langle {\cal S}\! \left(\dfrac{v_{\rm rel}}{2}\right) \right\rangle_v (r)\,
   \left\{\frac{\rho(r,m,c)}{\rho_\circledast}\right\}^2\,.
\end{cases}\nn\\
\een
All full numerical calculations presented in this paper will be based on these equations. However, since the main goal is to get analytical insights of the results, we shall try to extract the simplest description that still allows to capture the correct orders of magnitude.

An additional simplification can be used if (i) the telescope is pointed to the center of the target host halo and (ii) if the smooth component dominates over the subhalo component there (which is expected as tidal stripping is very efficient in the central parts of host halos). In that case, we have $J_\text{smooth approx}\simeq J_{\rm sm}$ to an excellent approximation \cite{StrefEtAl2017}. Moreover, if the target host halo is sufficiently far away from the observer, at a distance $D\gg R_{\rm host}$, and appears (at least almost) as a point-like source, then we can further simplify the expressions of the $J$-factors, and thereby that of the subhalo boost factor. By defining
\ben
\label{eq:jpointlike}
J(m,c,D) \equiv \frac{\rho_\circledast^2\,\xi_{\rm t}(m,c)}{D^2}\,,
\een
where we now neglect tidal stripping and simply identify a (sub)halo with its conventional virial mass $m$ and concentration $c$, and where it is assumed that $\xi_{\rm t}$ is integrated up to the virial radius; then we get
\ben
\begin{cases}
  J_{\rm sm} &\simeq \; J_\text{smooth approx} = J_\text{host}\equiv J(M_{\rm host},c_{\rm host},D)\\
  J_{\rm sub} &\approx \; N_{\rm tot}\,\left\langle J(m,c,D)\right\rangle_{m,c}
\end{cases}\,,
\een
where from now on, $J_{\rm host}$ characterizes the smooth approximation of the $J$-factor for the host halo, and where $\langle\rangle_{m,c}$ denotes an average over mass and concentration phase space. For the latter, one can use the mass and concentration functions of field subhalos for decent order-of-magnitude estimates, because most of subhalos lie away from the central parts of the host halo, where tidal effects can be neglected. Consistently, the subhalo boost factor can be approximated by
\ben
{\cal B} \approx 1 + N_{\rm tot}\,\frac{\left\langle J(m,c,D)\right\rangle_{m,c}}{J_{\rm host}}\,,
\een
where it clearly appears that both the subhalo mass-concentration relation and the mass function will play decisive roles. To further account for any velocity dependence of the annihilation cross section, one has to trade $\xi_{\rm t}$ for $\xi_{\rm t,eff}$ in \citeeq{eq:jpointlike} [see \citeeq{eq:def_xieff}].

We are now equipped to investigate analytically how Sommerfeld effects act on the subhalo boost factor. In the next paragraph, we first review the Sommerfeld-free case before moving to the more complex and intricate velocity-dependent cases induced by the Sommerfeld enhancement.

\subsection{Subhalo boost factor: analytical insights}
\label{ssec:analytical_boost}
Here, we derive analytical approximations that will allow us to interpret our full results in terms of the driving physical parameters in the calculation, which remain to be determined.

Throughout this part, without so much loss of generality, we will assume that subhalos have NFW inner mass density profiles:
\ben
\label{eq:def_nfw}
\rho(x\equiv r/r_{\rm s}) = \rho_0\,\left\{f_{\rm nfw}(x)\equiv x^{-1}(1+x)^{-2}\right\}\,\theta(x_{\rm t}-x)\,,
\een
where we have introduced the shape function $f_{\rm nfw}(x)$, the dimensionless radius $x$, and tidal radius $x_{\rm t}$, and where the structural properties such as the scale radius $r_{\rm s}$ and scale density $\rho_0$ are conventionally fixed by the mass and the mass-concentration relation. We will neglect tidal effects in the following discussion, as they are not critical to develop a good physical understanding of the Sommerfeld effect (we do account for them in the full numerical calculations). Thus, we can first assume that whatever their positions in the host halo, subhalos keep their virial mass, hence $x_{\rm t}=x_{200}=c$. Rigorously, we should also take into account the fact that these structural properties are described by non-trivial PDFs when tidal effects are considered (we do so in the full numerical calculations). For simplicity here, we assume that both the mass function and the mass-concentration relation follow power laws:
\ben
\label{eq:fm_cm_approx}
\begin{cases}
  \frac{dN_{\rm sub}(m,M_{\rm host})}{dm} &\approx \frac{N_0(M_{\rm host})}{m_0} \,\left\{\mu\equiv \frac{m}{m_0}\right\}^{-\alpha}\\
  c(m) &\approx c_0\,\mu^{-\varepsilon}
\end{cases}
\een
where $m_0$ is an arbitrary reference mass, $N_0$ is the normalization of the number of subhalos, which depends on the host halo mass $M_{\rm host}$, and where we have introduced the dimensionless reduced mass $\mu$. That form of the mass function finds strong theoretical support, as discussed in \citeapp{app:subhalo_model} --- see \citeeq{eq:dNdm_approx}. We can further neglect the concentration PDF, and assume that all subhalos of a given mass $m$ have the same concentration $c(m)$. These are good approximations to the more precise description used in our full numerical treatment -- -our detailed subhalo population model is described in \citeapp{app:subhalo_model}, where we find that
\ben
\label{eq:num_mass_indices}
\begin{cases}
  \alpha & \approx 1.96\\
  \varepsilon &\approx 0.05
\end{cases}
\een
provide a decent matching to the numerical results over a significant subhalo mass range.\footnote{We find that $c_0\simeq 12.9$ for $m_0=10^{10}\,\Msun$ \cite{PieriEtAl2011} allows to get close to the parametric concentration function of field halos provided in ref.~\cite{Sanchez-CondeEtAl2014}, which characterizes the initial concentration function of subhalos in most of the mass range of interest in our study.}

Given \citeeq{eq:fm_cm_approx}, we can relate the total number of subhalos $N_{\rm tot}$ to the subhalo mass fraction in the host $f_{\rm sub}$ through the minimal and maximal reduced subhalo masses $\mu_{\rm min}$, and the averaged subhalo mass $\langle\mu\rangle_m$ as follows:
\ben
f_{\rm sub}\,\mu_{\rm host} = N_{\rm tot}\,\langle\mu\rangle_m
= \frac{N_0}{(2-\alpha)}\, \mu_{\rm max}^{2-\alpha}\left\{1-\left[\frac{\mu_{\rm min}}{\mu_{\rm max}}\right]^{2-\alpha}\right\}\,,
\een
where $\mu_{\rm host}\equiv M_{\rm host}/m_0$ is the reduced dimensionless host halo mass. In the limit $\mu_{\rm max}\gg \mu_{\rm min}$ and if $\alpha<2$, then we have
\ben
\begin{cases}
  N_{\rm tot} &\simeq \frac{N_0}{(\alpha-1)}\, \mu_{\rm min}^{1-\alpha}\simeq \frac{\gamma}{(\alpha-1)}\left\{\frac{\mumin}{\muhost}\right\}^{1-\alpha}\\
  \langle\mu\rangle_m & \simeq 10^{-2}\,\frac{(\alpha-1)}{(2-\alpha)}\, \mu_{\rm host} \,\left\{\frac{\mu_{\rm max}}{\mu_{\rm min}}\right\}^{1-\alpha}\\
  f_{\rm sub} &\simeq \frac{N_0}{(2-\alpha)}\, \frac{\mu_{\rm max}^{2-\alpha}}{\mu_{\rm host}}
  = \frac{\gamma}{(2-\alpha)}\,\left\{\frac{\mumax}{\mu_{\rm host}}\right\}^{2-\alpha}
  \simeq \frac{10^{2(\alpha-2)}\gamma}{(2-\alpha)}\sim 34\%
  \end{cases}\,.
\een
We have assumed $\mumax=10^{-2}\muhost$, and $N_0=\gamma\muhost^{\alpha-1}$, and except for the approximately universal subhalo mass fraction (before tidal stripping effects), the above values depend on $M_{\rm host}\leftrightarrow \mu_{\rm host}$ --- see details in \citeapp{eq:dNdm_approx}.
\subsubsection{Subhalo boost factor without Sommerfeld enhancement}
\label{sssec:boost_noSomm}
Let us consider first annihilation through an $s$-wave process and a subhalo of virial mass $m$, concentration $c$, and located at a radius $R$ in a host halo. The intrinsic annihilation luminosity can be expressed in terms of the effective annihilation volume $\xi_{\rm t}$ introduced in \citeeq{eq:def_xi}, that we can rewrite as
\ben
\label{eq:xi_ana}
\xi_{\rm t}(m,c,R) = \frac{4\pi}{3} r_{\rm s}^3 \frac{\rho_{0}^2}{\rho_\circledast^2} \left\{ \eta_{\rm t}\equiv 3\int_0^{x_{\rm t}}\dd x\,x^2\,f_{\rm nfw}^2(x) \right\}\,,
\een
where $r_{\rm s}$, $\rho_0 $, and $f_{\rm nfw}(x)$ were introduced in \citeeq{eq:def_nfw}. We see here that the luminosity is computed within the dimensionless {\em tidal} radius $x_{\rm t}$ of the subhalo, which, in principle, implies a spatial dependence of the luminosity even for a given virial mass. For an NFW profile, the tidal cut reads
\ben
\eta_{\rm t} =\eta_{\rm t}(m,c,R)= 1-(1+x_{\rm t}(m,c,R))^{-3}\lesssim 1\,.
\een
From the definition of the virial mass $m=m_{200}$, and assuming that the mass-concentration relation $c(m)=c_{200}(m)=r_{200}(m)/r_{\rm s}(m)$ obeys the power-law function in mass given in \citeeq{eq:fm_cm_approx}, we get
\ben
\label{eq:xit}
\xi_{\rm t}(m,c,R)\simeq \xi_{\rm t}(m) = \xi_{\rm t}(\mu=m/m_0) = \xi_0\,\mu^{1-3\varepsilon}\,,
\een
where
\ben
\label{eq:xi0}
\xi_0\equiv \frac{200\,\rho_{\rm c}\,m_0}{\rho_\circledast^2}\, \frac{c_0^3}{A^2}\,\eta_{\rm t}
\,,
\een
$\rho_{\rm c}$ being the critical density today, and
\ben
\label{eq:def_A}
A=A(c) = 3\left[\ln(c+1)-\frac{c}{(c+1)}\right]\approx \text{constant}\,{\cal O}(1-10)\,.
\een
As already mentioned above, we can at first order neglect the position of a subhalo in its host, characterized here by the radial coordinate $R$. This is because subhalos that will dominantly contribute to the $\gamma$-ray flux are typically those located beyond the scale radius of the host (often called ``field'' subhalos), for which tidal stripping effects are not so important. These field subhalos actually constitute the bulk of the subhalo population (in an NFW host halo, subhalos located beyond the scale radius represent $\gtrsim 90\%$ of the whole subhalo population). This holds true while the $\gamma$-ray flux is integrated over a volume bigger than the one encompassing the scale radius of the host (a more involved description is necessary for hosts much more extended than the angular resolution of the telescope, or for hosts that have experienced significant tidal stripping and have sizes of the order or less than their scale radii).

From the annihilation volume $\xi_{\rm t}$, we get an analytical expression for the point-like $J$-factor given in \citeeq{eq:jpointlike}:
\ben
\label{eq:jm}
J(m) = J(\mu)=J_0\times (2\,\overline{v}_0)^p\times \mu^{1-3\varepsilon+p\nu}\,,
\een
where
\ben
J_0\equiv \frac{\rho_\circledast^2 \,\xi_0}{D^2} = \frac{200\,\rho_{\rm c}\,m_0}{D^2}\, \frac{c_0^3}{A^2}\, \eta_{\rm t}\,,
\een
with $\xi_0$ given in \citeeq{eq:xi0}, and $\rho_\circledast$ the arbitrary reference mass density introduced in \citeeq{eq:def_xi}---it is then clear that neither $J(m)$ nor $J_0$ depend on $\rho_\circledast$, as shown explicitly in the above equations.

Note that the $p$-wave annihilation case is actually included in the previous two expressions, by setting $p=2$ (we remind that $p=0$ stands for the $s$-wave case). In fact, we have implicitly assumed that the radial profile of $\langle v^2\rangle_v(r)\,\rho^2(r)\approx \langle v^2\rangle_{v,V_{\rm h}}\,\rho^2(r)$, where $\langle v^2\rangle_{v,V_{\rm h}}$ is taken constant over the whole halo volume $V_{\rm h}$. Assuming $\langle v^2\rangle_{v,V_{\rm h}} = \overline{v}_0^2 \,\mu^{2\nu}\ll 1$ allows us to characterize the $p$-wave suppression factor in terms of halo mass. The spectral index $\nu$ will be specified later, and $\overline{v}_0$ is an arbitrary reference velocity associated with a halo of arbitrary reference mass $m_0$.

From this, we can predict the ratio of $J$-factors of two point-like halos of different masses, $m_1$ and $m_2$, and respectively located at distances $D_1$ and $D_2$ from the observer:
\ben
\label{eq:J_ratio}
\frac{J(m_1)}{J(m_2)} = \left\{\frac{D_2}{D_1}\right\}^2\,\left\{\frac{m_1}{m_2}\right\}^{1-3\varepsilon+p\,\nu}\,.
\een
This will be helpful to understand forthcoming results.

Let us now come back to our main working equation, \citeeq{eq:jm}. As we shall see just below, it will allow us to estimate the total contribution of subhalos to the $\gamma$-ray $J$-factor, assuming that the whole population of subhalos is contained within the field of view of the instrument. Indeed, this simply amounts to convolving \citeeq{eq:jm} with the subhalo mass function, which we take as the power law of index $\alpha$ introduced in \citeeq{eq:fm_cm_approx}. Then, the total $J$-factor associated with the contribution of the whole subhalo population reads:
\begin{tcolorbox}
\ben
\label{eq:JSfreesub}
J_{\rm sub} =  J_{{\rm sub},0} \,\mumin^{-\alpha_{\rm boost}}\left[1 - \left(\frac{\mumax}{\mumin}\right)^{-\alpha_{\rm boost}}\right]\,,
\een
\end{tcolorbox}
\noindent
where we have introduced the effective boost index $\alpha_{\rm boost}$. We have also used
\ben
J_{{\rm sub},0}(\muhost) \equiv \frac{N_0(\muhost)}{\alpha_{\rm boost}}\times J_0\times (2\,v_0)^p\,.
\een
The critical parameter in the above result is the effective boost index,
\ben
\label{eq:alpha_boost}
\alpha_{\rm boost} \equiv \alpha+3\varepsilon-2-p\nu\,,
\een
which fully characterizes the part of the mass function that sets the overall subhalo population luminosity. Indeed, three different regimes arise:
\ben
\label{eq:alpha_boost_sign}
\begin{cases}
\alpha_{\rm boost}&>0 \;\Longrightarrow \text{$m_{\rm min}$-dominated regime (strong boost)}\\
\alpha_{\rm boost}&=0 \;\Longrightarrow \text{democratic regime}\\
\alpha_{\rm boost}&<0 \;\Longrightarrow \text{$m_{\rm max}$-dominated regime (weak boost)\,.}
\end{cases}
\een
The positive sign convention has been chosen such that the boost is strong if $\alpha_{\rm boost}>0$, which means that the smallest, most numerous, and most concentrated subhalos carry the dominant contribution to the annihilation rate. The democratic regime corresponds to a logarithmic dependence in the subhalo masses, $\propto\ln(m_{\rm max}/m_{\rm min})$, in \citeeq{eq:JSfreesub}. The sign of $\alpha_{\rm boost}$ is therefore crucial here, as already known from past studies. From this very simple equation, since $\varepsilon\approx 0.05$, we understand that changing $\alpha$ from 1.9 to 2 amounts to going from an $m_{\rm max}$-dominated regime to an $m_{\rm min}$-dominated regime for an $s$-wave annihilation. In the latter case, the overall subhalo population luminosity becomes very sensitive to the subhalo minimal mass cutoff $\mmin$, as is well known. This is reinforced by the fact that the total number of subhalos $N_{\rm tot}\overset{\sim}{\propto} \mu_{\rm min}^{1-\alpha}$. We shall see later on that this effective power-law index $\alpha_{\rm boost}$ can also be expressed analytically in the Sommerfeld-enhanced case, which will allow us to use a reasoning very similar to the one presented here. Before moving to the Sommerfeld-enhanced case, let us just introduce an analytical expression for the Sommerfeld-free boost factor ${\cal B}$:
\ben
\label{eq:boost_Sfree}
{\cal B} -1 \simeq \frac{J_{\rm tot}^{\rm sub}}{J_{\rm host}} = 
\frac{J_{{\rm sub},0}}{J_{\rm host}}\,\mu_{\rm min}^{-\alpha_{\rm boost}}\left\{1- \left[ \frac{\mu_{\rm max}}{\mu_{\rm min}}\right]^{-\alpha_{\rm boost}}\right\}\,,
\een
where $J_{\rm host}$ is the $J$-factor calculated assuming a fully smooth density profile for the host halo ($\propto\int\dd s \, \langle \rho_{\rm host}\rangle^2$) --- we call this the smooth approximation. The "-1" on the left-hand side implicitly assumes that the contribution to the $J$-factor of the smooth part of the actual inhomogeneous host halo, $J_{\rm sm}$, equals the smooth approximation, but one should keep in mind that formally $J_{\rm host}\gtrsim J_{\rm sm}$. Given the analytical expressions introduced above, we finally get
\begin{tcolorbox}
\ben
\label{eq:boost_Sfree_approx}
    {\cal B} -1 &\simeq& \frac{N_0\,A^2_{\rm host}}{\alpha_{\rm boost}\,A_{\rm sub}^2\,\mu_{\rm host}^{1-3\varepsilon}}\,\mu_{\rm min}^{-\alpha_{\rm boost}}\left\{1- \left[ \frac{\mu_{\rm max}}{\mu_{\rm min}}\right]^{-\alpha_{\rm boost}}\right\}\\
   &\simeq& \frac{\gamma}{\alpha_{\rm boost}}\,\frac{A^2_{\rm host}}{A_{\rm sub}^2}\, \left\{\frac{\mu_{\rm min}}{\muhost}\right\}^{-\alpha_{\rm boost}}\left\{1- \left[ \frac{\mu_{\rm max}}{\mu_{\rm min}}\right]^{-\alpha_{\rm boost}}\right\} \,,\nn
\een
\end{tcolorbox}
\noindent
where $A_{\rm host}$ and $A_{\rm sub}$, introduced in \citeeq{eq:def_A}, are shown explicitly for definiteness because their ratio is not strictly 1. Note that the power-law dependence in the bracket on the right-hand-side becomes logarithmic, $\ln(\mumax/\mumin)$, when $\alpha_{\rm boost}=0$ (democratic regime, for which each decade of subhalo mass contributes the same signal). In the latest equation line above, we have traded $N_0$ for its dependence in $\muhost$ according to \citeeq{eq:dNdm_approx}, with $\gamma$ a constant predicted from a merger-tree calculations, which provides a very compact expression that depends only on the subhalo-to-host mass ratio and on the subhalo mass index.

From the numerical values introduced in \citeeq{eq:num_mass_indices}, we get $\alpha_{\rm boost}(p=0)\simeq 0.11>0$ for $s$-wave annihilation processes, hence a significant boost factor dominated by the contribution of the lightest subhalos to the annihilation rate (the last term in brackets in the right-hand-side of the above equation simplifies to 1):
\ben
\label{eq:boost_swave_approx}
    {\cal B}_\text{$s$-wave} -1 &\simeq& \frac{N_0\,A^2_{\rm host}}{\alpha_{\rm boost}\,A_{\rm sub}^2\,\mu_{\rm host}^{1-3\varepsilon}}\,\mu_{\rm min}^{-\alpha_{\rm boost}}\left\{1- \left[ \frac{\mu_{\rm max}}{\mu_{\rm min}}\right]^{-\alpha_{\rm boost}}\right\}\\
   &\simeq& \frac{\gamma}{\alpha_{\rm boost}}\,\frac{A^2_{\rm host}}{A_{\rm sub}^2}\, \left\{\frac{\mu_{\rm min}}{\muhost}\right\}^{-\alpha_{\rm boost}} \,.\nn
\een
In that case, the boost factor is fixed by the hierarchy between the host halo mass and the minimal subhalo mass, and modulated by the amplitude of the effective boost index $\alpha_{\rm boost}$.

There is no boost factor in the $p$-wave annihilation case because since the cross section is proportional to $v^2$ and the internal dispersion velocity decreases with the mass of a structure, the signal contributed by subhalos is strongly reduced with respect to that contributed by the host halo. Still under the assumption of $\langle v^2 \rangle \propto m^{2\,\nu}$, where $\nu$ will be evaluated later to be $\sim 1/3$, and that for a halo of density profile $\rho(r)$, $\langle v^2 \,\rho^2\rangle(r)\approx \langle v^2 \rangle \rho^2(r) \propto m^{2\,\nu} \,\rho^2(r)$, then it is easy to show from \citeeq{eq:JSfreesub} that
\ben
\label{eq:boost_pwave}
{\cal B}_\text{$p$-wave} -1 \approx - \frac{\gamma}{\alpha_{\rm boost}}\frac{A^2_{\rm host}}{A_{\rm sub}^2}\left\{\frac{\mumax}{\muhost}\right\}^{-\alpha_{\rm boost}} \approx - \frac{\gamma}{\alpha_{\rm boost}}\left\{\frac{1}{100}\right\}^{-\alpha_{\rm boost}}\ll 1\,.
\een
We have used the fact that $\mmax\simeq M_{\rm host}/100$, that the boost mass index for the $p$-wave annihilation $\alpha_{\rm boost}(p=2)=\alpha+3\varepsilon-2(1+\nu)\approx -0.56<0$ [see \citeeq{eq:JSfreesub}], and that $A^2_{\rm host}\approx A_{\rm sub}^2$. In that approximation, valid as long as $J_{\rm sm}\simeq J_\text{host}$, then clearly ${\cal B}_\text{$p$-wave}\simeq 1$. If $J_{\rm sm}< J_\text{host}$, which can be the case if the mass fraction in subhalos is significant within the scale radius of the host halo, then we could even have ${\cal B}_\text{$p$-wave}< 1$, which would imply that subhalos would no longer act as a boost factor, but rather as a damping factor to the signal. We will see just below that this picture changes radically when Sommerfeld effects kick in.

\subsubsection{Sommerfeld enhancement at the level of one (sub)halo.}
\label{sssec:S_for_one}

In this part, we initiate the derivation of an analytical expression for the subhalo boost factor further subject to Sommerfeld-enhancement effects. The derivation proceeds in three steps. We first develop an analytical understanding of the Sommerfeld effect at the level of a single structure (this paragraph). This is a crucial step before generalizing to a population of structures in the next paragraph, where we derive a full analytical expression for the Sommerfeld-enhanced $J$-factor associated with a subhalo population. Finally, we determine the overall boost factor by calculating the ratio between the Sommerfeld enhanced $J$-factor for the subhalo population and that of the host halo. This series of analytical developments is helpful to reach a clearer physical understanding of the intricate phenomena at play in terms of the specific particle physics model parameters, here characterized by the reduced DM Bohr radius, $\epsilon_\phi$. We recall that a Sommerfeld configuration is entirely fixed by $\epsilon_\phi$ and the coupling strength $\alpha_\chi$ in our simplified model. Decreasing $\epsilon_\phi$ roughly amounts to increasing the DM particle mass or the interaction coupling constants, or decreasing the mediator mass, assuming all of the other parameters are fixed.

We start by examining the overall Sommerfeld enhancement for one halo. The particle-velocity dependent ans\"atze introduced in \citesec{sssec:ansatz} suggest the possibility of formulating an effective Sommerfeld enhancement factor at the level of an entire (sub)halo. This can be done by picking the most representative value of the particle velocity in a DM structure, which depends on the structure mass (a mere consequence of the virial theorem for systems in dynamical equilibrium). If such a characteristic velocity in a (sub)halo can be estimated (\eg~from its averaged velocity dispersion), then one can effectively relate an average Sommerfeld enhancement to the (sub)halo mass. We can actually expect the characteristic velocity of a structure of mass $m$, $\overline{v}(m)$, to scale like
\ben
\label{eq:v_to_m_main}
\overline{v}(m)\sim \sqrt{\langle v^2\rangle}\sim\sqrt{\frac{G_{\rm N} \, m(r_{\rm c})}{r_{\rm c}}}\,,
\een
where $r_{\rm c}$ is some characteristic radius to be determined. For the sake of generality, considering that we can also relate that characteristic radius to the virial mass, we shall assume
\ben
\label{eq:v_to_m}
\overline{v} = \overline{v}_0\,\mu^\nu\,,
\een
where $\nu$ is the power-law index that relates the characteristic dimensionless velocity $\overline{v}/\overline{v}_0$ to the dimensionless (sub)halo mass $\mu=m/m_0$. Parameter $\overline{v}_0$ is the characteristic velocity associated with the arbitrary reference virial mass $m_0=m_0(r_{200,0})$ of an NFW halo, obeying the general relation
\ben
\label{eq:v_to_m_detail}
\overline{v}(m)\equiv \omega_0\,\sqrt{\frac{G_{\rm N} \, m(r_{\rm s})}{r_{\rm s}}}\,,
\een
where $\omega_0\sim 1$ is a tuning parameter, $r_{\rm s}$ is the scale radius associated with some halo of virial mass $m$, and $m(r_{\rm s})$ is the mass contained within $r_{\rm s}$. The speed parameter $\omega_0$ is meant to optimize the estimates of the speed moments relevant to the Sommerfeld enhancement over a given structure with a single value of $\overline{v}$, which should capture different regimes at the same time ($\propto \langle 1/v\rangle$ or $\langle 1/v^2\rangle$). By picking the subhalo characteristic mass and size at the scale radius of an NFW halo, it is easy to show that
\ben
\label{eq:nu_approx}
\nu\simeq \frac{1}{2}\left\{\frac{2}{3}-\varepsilon \right\}\approx \frac{1}{3}\,,
\een
where $\varepsilon$ is the power-law index of the approximate concentration-mass relation given in \citeeq{eq:fm_cm_approx}. More concretely, in numbers, this gives
\ben
\overline{v}  \simeq 6\times 10^{-6} \, \omega_0 \,\left\{\frac{m}{10^6\,\Msun}\right\}^{1/3}\,,
\een
with $\omega_0\sim 1$.

Even though $\overline{v}$ can be used as a characteristic velocity where most of the phase-space distribution is supposed to concentrate, one should still not forget that the speed of any DM particle bound to a (sub)halo can actually take any value between 0 and the escape speed. The Sommerfeld factor should therefore be integrated over the full available range---hence different parts of the (sub)halo phase-space distribution may contribute to different Sommerfeld regimes, not necessarily to a single regime. However, the characteristic velocity-mass relation written above is meant to reflect the typical velocity at which the bulk of annihilations in a (sub)halo of mass $m$ proceeds, which turns out to be a good approximation.

We can now opportunely reformulate the ansatz of \citeeq{eq:S_ansatz_v} by replacing the dependence on velocity $v$ by a dependence on the characteristic (sub)halo velocity $\overline{v}(m)$, and then by a dependence in (sub)halo mass $m$. This gives
\ben
\label{eq:S_ansatz_m}
\overline{\cal S}_\text{no-res}(m,\epsilon_\phi)&=& {\cal S}_\text{no-res}(\overline{v}(m),\epsilon_\phi)\\
&=& S_0\left(\frac{m}{\mtmax}\right)^{-\nu} 
\left[1 + S_1^{-\frac{\overline{s}_{v,c}}{(1+p)}}\left(\frac{m}{\mtsat}\right)^{-\nu \overline{s}_{v,c}} \right]^{-\frac{(1+p)}{\overline{s}_{v,c}}}\,,\nn
\een
with $m$ the (sub)halo mass, and the constants $S_0$ and $S_1$ given in \citeeq{eq:Sconstants}. This ansatz is essentially valid for $\overline{v}(m)  \leqslant \vmax$, or equivalently $m\leqslant \mtmax=m(\vmax)$, for which the Sommerfeld effect starts being operative. This maximal mass $\mtmax$ should not be confused with the maximal subhalo mass $\mmax$ in a given host halo; it is really the (sub)halo mass beyond which the characteristic velocity of DM is too large for the Sommerfeld enhancement to be turned on efficiently. The power-law indices have been introduced in \citeeq{eq:S_ansatz_v} up to a correction by the speed-to-mass index $\nu$, introduced in \citeeq{eq:v_to_m}, and evaluated in \citeeq{eq:nu_approx}. Switching from velocity to mass dependence, the power-law index in the Coulomb regime becomes $-\nu$. We have also introduced $\mtsat=\mtsat(\epsilon_\phi)=m(\vsat(\epsilon_\phi))$, the halo mass below which most of the halo phase-space volume is in the Sommerfeld saturation regime and resonances may appear. The different velocity dependencies in the different regimes are summarized below \citeeq{eq:S_ansatz_v}.

Similarly to the corresponding velocities, the transition masses introduced above can be expressed in terms of the main Sommerfeld parameters:
\ben
\label{eq:msat}
\begin{cases}
  \mtmax& = m_0\left( \frac{\vmax}{\overline{v}_0}\right)^{\frac{1}{\nu}}
  = m_0 \left( \frac{\pi\,\alpha_X}{\overline{v}_0}\right)^{\frac{1}{\nu}} \approx 9.6\times 10^{17}\,\Msun\left( \frac{\alpha_X}{0.01}\right)^{3}\\
  \mtsat(\epsilon_\phi) & = m_0\left( \frac{\vsat}{\overline{v}_0}\right)^{\frac{1}{\nu}} =
  m_0 \left( \frac{\alpha_X \,\epsilon_\phi}{\overline{v}_0}\right)^{\frac{1}{\nu}} \approx 7.6\times 10^9\,\Msun\left(\frac{\alpha_X}{0.01}\times\frac{\epsilon_\phi}{0.01}\right)^3\\
  \mtunit &= m_0\left( \frac{\vunit}{\overline{v}_0}\right)^{\frac{1}{\nu}} \approx 8\times 10^{-4}\,\Msun\left(\frac{\alpha_X}{0.01}\right)^{12}
\end{cases}\,.
\een
We stress that $\mtsat$ is a smooth function of $\epsilon_\phi$ even on resonances. This saturation mass defines a threshold in phase space: halos with masses below $\mtsat$ will have most of their phase-space distribution in the saturation regime. Resonant saturation masses are simply characterized by $\mtsat=\mtsat(\epsilon_\phi=\epsilon_\phi^{{\rm res},n})$. The power-law dependence of $\mtsat\propto \epsilon_\phi^{1/\nu}$ can be predicted from \citeeq{eq:nu_approx} to be close to $\mtsat\propto \epsilon_\phi^{3}$. This is actually recovered from a numerical calculation of \citeeq{eq:v_to_m_detail}, as shown in the bottom right panel of \citefig{fig:comparison_JSomm_subhalos}, which will be discussed more thoroughly later on. Finally, the numerical estimate of $\mtunit$ given above can make us anticipate the important role it will play in the determination of the resonant peak amplitudes, and then the intrinsic limit set in the potential of the latter to probe the minimal (sub)halo masses if $\mtunit>\mmin$. We stress that this mass boundary $\mtunit$ is extremely sensitive to the DM fine structure constant, \change{as it scales like $\sim \alpha_\chi^{12}$ in our approximate parametric regularization (but see the discussion at the end of \citesec{sssec:conventional})}.

We can get a similar formulation for the halo mass dependent Sommerfeld factor on resonances by inserting $v=\overline{v}(m)$ in \citeeq{eq:S_ansatz_res_v}:
\ben
\label{eq:S_ansatz_res_m}
\overline{\cal S}_{\text{res},n}(m,\epsilon_\phi)&=& {\cal S}_{\text{res},n}(\overline{v}(m),\epsilon_\phi)\\
&\overset{n\geqslant 1+\frac{p}{2}}{=}& S_0^{\rm res}\,\left(\frac{\mtsat(\epsilon_\phi)}{\mtmax}\right)^{-\nu}\,
\left(\frac{m}{\mtsat(\epsilon_\phi)}\right)^{-\nu(2-p)}\,\left(1+\frac{\mtunit^\nu}{m^\nu}\right)^{-2} \nn\\
&&\times \,\theta \left(\mtsat(\epsilon_\phi)-m\right)\,
    \delta_{\epsilon_\phi/\{\epsilon_\phi^{{\rm res},n}\}}\nn
\een
where $\delta_{\epsilon_\phi/\{\epsilon_\phi^{{\rm res},n}\}}$ was defined in \citeeq{eq:def_D_measure}, and where we see that a halo can efficiently trigger resonances provided its mass $m\ll \mtsat(\epsilon_\phi^{{\rm res},n})$.

\change{A full understanding of the mass dependence in the resonant regime actually follows from that of the velocity dependence discussed around \citeeq{eq:S_ansatz_res_v} and illustrated in \citefig{fig:v_dependence}, keeping in mind that $\overline{v}\propto m^\nu$---focus on dark blue curves in both panels. Indeed, beside the step function responsible for turning resonances on or off, the only direct dependence of the above resonant Sommerfeld factor on the halo mass $m$ shows up in the $s$-wave case, down to the unitarity limit characterized by $\mtunit$.} In contrast, the amplitudes of $p$-wave resonant peaks ($p=2$) do not, essentially, depend on $m$, which is reminiscent from the fact that the effective Sommerfeld enhancement (which includes the $v^2$ $p$-wave suppression factor as well) is velocity independent on $p$-wave peaks, as shown in \citeeq{eq:S_ansatz_res_v} and in the right panel of \citefig{fig:v_dependence}. The $p$-wave suppression factor re-appears once the unitarity bound is reached, and translates into a mass-dependent suppression factor of $(1+\mtunit/m)^{-2\nu}$ that becomes operative when $m\lesssim\mtunit$. Hence, the potential numerical error made by converting a local velocity into a global velocity is significantly reduced on $p$-wave resonances (except close to the step function threshold, $m\lesssim\mtsat$, where only part of the phase-space distribution lies in the saturation regime, or close to the unitarity bound).

All this allows us to translate the velocity-dependent ansatz of \citeeq{eq:S_ansatz_tot_v} in terms of a halo-mass-dependent and generic effective Sommerfeld factor,
\ben
\label{eq:S_ansatz_tot_m}
\overline{\cal S}(m,\epsilon_\phi) &=& \overline{\cal S}_\text{no-res}(m,\epsilon_\phi)\left(1-  \sum_{n=1+\frac{p}{2}}\delta_{\epsilon_\phi/\{\epsilon_\phi^{{\rm res},n}\}}\right) + \sum_{n=1+\frac{p}{2}}\overline{\cal S}_{{\rm res},n}(m,\epsilon_\phi)\\
&\propto& \mu^{-s_m}\,,\nn
\een
where $\overline{\cal S}_{\rm res}(m)$ is the transcript of ${\cal S}_{\rm res}(v)$ of \citeeq{eq:S_ansatz_res_v} in terms of mass $m$ (evaluated at $v=\overline{v}(m)$). At the level of a (sub)halo, the Sommerfeld enhancement can be written as power law in mass, whose effective index $s_m=\nu\,s_v$ can be readily inferred from the possible values of $s_v$ listed in \citeeq{eq:sv_values}:
\ben
\label{eq:sm_values}
s_m=
\begin{cases}
  \nu\;\; & \text{(Coulomb regime)}\\
  -\nu\,p & \text{(non-resonant saturation regime)}\\
  \nu\,(2-p) & \text{(resonances)}\;\longrightarrow -p\,\nu \;\text{(if $m\lesssim \mtunit$)}
\end{cases}
\,.
\een

We stress that the correspondence between the characteristic speed and the (sub)halo mass in the Sommerfeld factor has only a global meaning---we shall refer to $\overline{v}$ as the characteristic speed in a (sub)halo, and to $v$ as an arbitrary or local speed from now on. Indeed, as mentioned above, the DM speed $v$ in a subhalo can take any value between $\sim 0\ll \overline{v}$ and the escape speed $v_{\rm e}\gtrsim \overline{v}$. The whole (sub)halo lies in the Sommerfeld-enhancement regime typically when $v_{\rm e} \leq \vmax$. Then, the part of the phase-space distribution located between $\vsat(\epsilon_\phi)$ and $\vmax$ will essentially participate in the Coulomb enhancement ($\propto 1/v$), while the part of the phase-space distribution below $\vsat(\epsilon_\phi)$ will instead contribute in the saturation regime. In the latter case, the speed dependence saturates, except at resonances, which are triggered at special values of $\epsilon_\phi=\epsilon_\phi^{{\rm res},n}$ at which all of the phase-space distribution located below $\vsat$ participates in the enhancement ($\propto 1/v^2$ for $s$-wave annihilation). The very fact that different parts of the phase-space distribution of a (sub)halo feed different Sommerfeld regimes implies that the ansatz of \citeeq{eq:S_ansatz_tot_m} cannot lead to accurate predictions. However, we shall see below that it is still very powerful to capture the main phenomenological features of the intricate phenomena at play.

It is instructive to further inspect the relative amplitudes of resonant peaks when the Sommerfeld factor is applied over an entire halo. To do so, let us briefly convert the ratio of resonance-to-baseline enhancement in the saturation regime, introduced in \citeeq{eq:ratio_res_to_base}, in terms of an overall mass-dependent ratio:
\ben
\label{eq:eff_ratio_res_to_base}
\overline{\cal R}(\mu,\epsilon_\phi) = \left(\frac{\pi}{6}\right)^2\,\left\{\frac{\mu}{\mutsat}\right\}^{-2\,\nu} \,\left\{1+\left(\frac{\mutunit}{\mu}\right)^\nu\right\}^{-2}\overset{\sim}{\propto}\epsilon_\phi^2\, \mu^{-2/3}\,.
\een
Therefore, the relative amplitudes of peaks scale like $\overline{v}^{-2}(m)\leftrightarrow m^{-2/3}$ for both $s$- and $p$-wave processes, which means that resonances are more pronounced for less massive halos (though saturating when $m\lesssim \mtunit$), while still suppressed like $(\epsilon_\phi\sim \epsilon_\phi^{{\rm res},n})^2$ at higher and higher resonances, with respect to the saturation baseline.

We pursue by writing down the analytical expression obtained for the $J$-factor corrected for the Sommerfeld enhancement at the level of one structure of mass $m$ (or dimensionless reduced mass $\mu=m/m_0$), combining \citeeq{eq:jm} and \citeeq{eq:S_ansatz_tot_m}:
\ben
\label{eq:JS_halo}
J_{\cal S}(m,\epsilon_\phi) &=& J_{\cal S}(\mu,\epsilon_\phi) =  J(\mu)\times\left\{\frac{\mu}{\mutmax}\right\}^{-p\nu}\times\overline{\cal S}(\mu,\epsilon_\phi)\\
&=& J_0\times (2\,\vmax)^p\times \mu^{1-3\varepsilon}\,\overline{\cal S}(\mu,\epsilon_\phi)\nn\\
&=& J_{0,{\cal S}}(\epsilon_\phi)\,\mu^{1-3\varepsilon-s_m}\,,\nn
\een
which gives an implicit definition to the factor $J_{0,{\cal S}}(\epsilon_\phi)$, and where the Sommerfeld mass index $s_m$ was introduced in \citeeq{eq:sm_values}. Note that the factor $(\mu/\mutmax)^{-p\nu}=(\overline{v}/\vmax)^{-p}$ is simply there not to double count the $v^2$ dependence of the $p$-wave cross section that is included in the definition of the nominal $J$-factor $J(m)$ [see \citeeq{eq:jm}], which we have also conveniently absorbed in the definition of the effective Sommerfeld enhancement factor [see \citeeq{eq:def_Seff}].

Like in the Sommerfeld-free case, it is interesting to determine the ratio of $J$-factors for halos of different masses, say $m_1\leftrightarrow\mu_1$ and $m_2\leftrightarrow\mu_2$ (assumed here to be located at different distances, $D_1$ and $D_2$, from the observer):
\ben
\label{eq:JS_ratio}
\frac{J_{\cal S}(m_1,\epsilon_\phi)}{J_{\cal S}(m_2,\epsilon_\phi)} = \left\{\frac{D_2}{D_1}\right\}^2\left\{\frac{m_1}{m_2}\right\}^{1-3\varepsilon}\,\left\{\frac{m_1^{-s_{m_1}}}{m_2^{-s_{m_2}}}\right\}\,,
\een
where $s_{m_i}$ refers to the Sommerfeld mass index of the halo of index $i$. In the absence of Sommerfeld enhancement, $J_1/J_2\sim (m_1/m_2)^{(1-3\varepsilon+p\,\nu)}$ according to \citeeq{eq:J_ratio}, where the $\overline{v}^p(m)$ factor relevant to the $p$-wave case remains (contributing $\nu p$ in the power-law mass index, a contribution hidden in the definition of the index $s_m$ in the Sommerfeld-enhanced case).

Now, we determine the asymptotic expressions for the different Sommerfeld regimes, which will turn useful later because subhalos are not necessarily all in the same Sommerfeld regime, nor necessarily in the same Sommerfeld regime as the host halo itself. This gives:
\bi
\item {\bf Coulomb regime $(\mu\geqslant\mutsat(\epsilon_\phi))$:}
  \ben
\label{eq:JS_Coulomb}
  J_{\cal S}(\mu,\epsilon_\phi) & \overset{\mu\geqslant\mutsat}{\underset{\rm Coulomb}{\longrightarrow}} &
 J_0\,(2\,\vmax)^p\,S_0\,\mutmax^{1-3\varepsilon}\,\left\{\frac{\mu}{\mutmax}\right\}^{1-3\varepsilon-\nu} \\
 & \propto &\mu^{1-3\varepsilon-\nu}\,.\nn
 \een
\item {\bf Saturation $(\mu\leqslant\mutsat(\epsilon_\phi))$:}
  \ben
\label{eq:JS_Saturation}
 J_{\cal S}(\mu,\epsilon_\phi) & \overset{\mu\leqslant\mutsat}{\underset{\rm saturation}{\longrightarrow}} &
 J_0\,(2\,\vmax)^p\,S_0\,S_1\,\left\{\frac{\mutmax}{\mutsat}\right\}^{\nu}\mutsat^{1-3\varepsilon}\,\left\{\frac{\mu}{\mutsat}\right\}^{1-3\varepsilon+\nu p}\\
 &\propto& \epsilon_\phi^{-(1+p)}\,\mu^{1-3\varepsilon+\nu p}\,.\nn
\een
\item {\bf Resonances $(\mu\ll \mutsat(\epsilon_\phi^{{\rm res},n}))$:}
  \ben
  \label{eq:JS_Resonances}
 J_{\cal S}(\mu,\epsilon_\phi=\epsilon_\phi^{{\rm res},n}) & \overset{\mu\ll\mutsat}{\underset{\rm resonance/s}{\longrightarrow}} &
 J_0\,(2\,\vmax)^p\,S_0^{\rm res}\,\left\{\frac{\mutmax}{\mutsat}\right\}^{\nu}\,\mutsat^{1-3\varepsilon}\,\left\{\frac{\mu}{\mutsat}\right\}^{1-3\varepsilon-\nu(2-p)}\\
 &&\times \left\{1 + \left(\frac{\mutunit}{\mu}\right)^\nu \right\}^{-2}\nn\\
 & \propto &\epsilon_\phi^{(1-p)}\,\mu^{1-3\varepsilon-\nu(2-p)}\,.\nn 
 \een  
\ei

\begin{figure}[t!]
\centering
\includegraphics[width=0.99\linewidth]{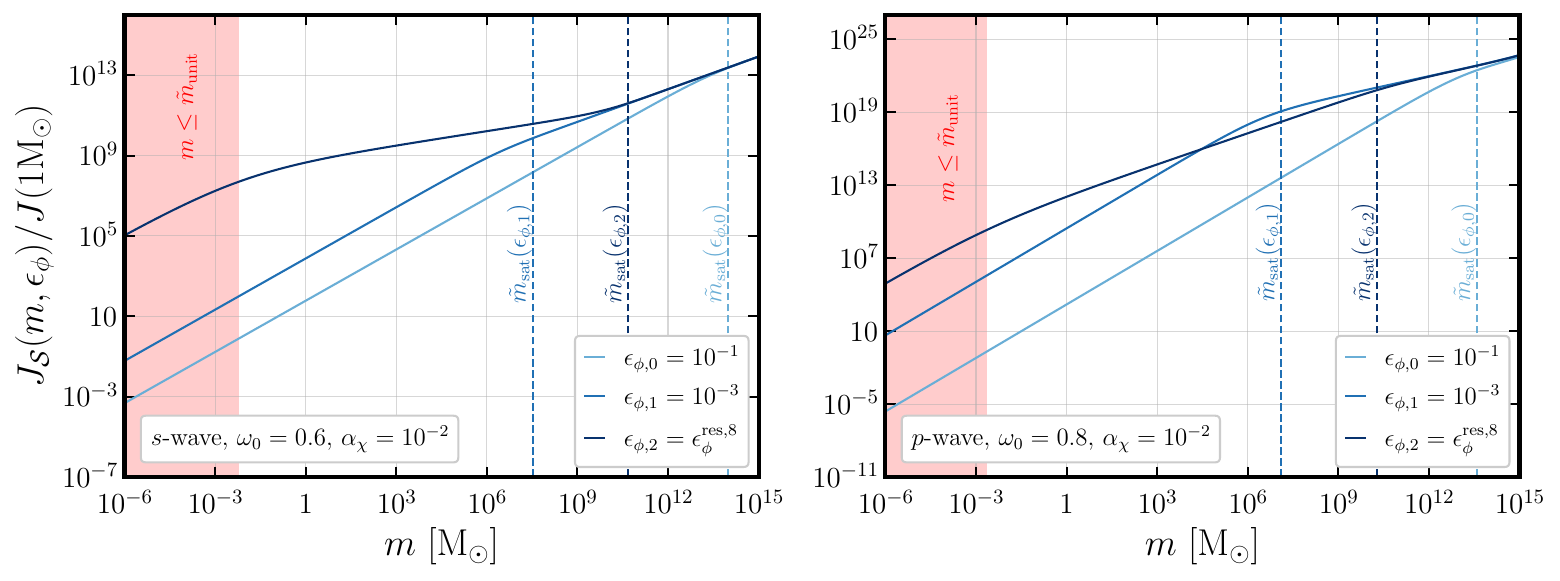}
\caption{\new{Sommerfeld-enhanced $J_{\cal S}$ factor as a function of halo mass, for different values of the reduced Bohr radius $\epsilon_\phi$: a large value 0.1, a small value of $10^{-3}$, and an intermediate value of $\sim 10^{-2}$ sitting on the $n=8$ resonance. This figure is somewhat a translation of the local effective Sommerfeld factor as a function of DM velocity shown in \citefig{fig:v_dependence} (times the nominal $J$ factor). Transition from Coulomb to saturation regimes occurs around $\mtsat(\epsilon_\phi)$, reported as vertical dashed lines. {\bf Left panel:} $s$-wave case. {\bf Right panel:} $p$-wave case.}}
\label{fig:JS_vs_m}
\end{figure}

\new{The full mass dependence (which derives from velocity dependence) of the overall Sommerfeld-enhanced $J_{\cal S}$ factor at the level of a single halo is shown in \citefig{fig:JS_vs_m}, where the power-law scalings derived just above are illustrated for different values of the reduced Bohr radius $\epsilon_\phi$, hence for different particle physics configurations. We actually took the same reference cases as in \citefig{fig:v_dependence}, $\epsilon_\phi=0.1$ (moderate enhancement), $10^{-3}$ (significant enhancement), and the $n=8$ resonance popping up at an intermediate value of $\epsilon_\phi=\epsilon_\phi^{\rm res, 8}\simeq 10^{-2}$ (strong enhancement). The corresponding saturation masses $\mtsat(\epsilon_\phi)$ are reported as dashed vertical lines, which mark the transition between the Coulomb regime domination of the phase-space volume ($m>\mtsat$) and the saturation regime domination ($m<\mtsat$). We see that in both the $s$- (left panel) and $p$-wave (right panel) cases, this transition is characterized by a change of logarithmic slope for $J_{\cal S}$, as analytically predicted above. The important point to note here is that the slope gets generically steeper below the saturation mass (stronger decrease with decreasing mass), except on the resonance, where the mass dependence is much shallower toward low masses down to the unitary mass $\mtunit$ (from both panels, we see that the resonant curves maintain a higher level of $J_{\cal S}$ factor compared to non-resonant curves, as $m$ decreases down to $\mtunit$). This will obviously have strong consequences when integrated over a subhalo population, whose power-law mass function will act as an extra weight in favor of low-mass subhalos.}

We are now equipped with all necessary analytical results to understand the Sommerfeld enhancement at the level of an entire halo. \change{A final characteristic ingredient to better identify the Sommerfeld regime a given a (sub)halo of mass $m$ should fall in is the value of $\epsilon_\phi$ for which that halo would transition from the Coulomb to the saturation regime}. Since the saturation mass $\mtsat$ is defined from $\epsilon_\phi$, we can conversely assign a reference value $\epsilon_\phi^{\rm sat}(m)$ to a halo of virial mass $m$ such that
\ben
\label{eq:def_eps_star}
\mtsat(\epsilon_\phi^{{\rm sat}}) &=& m \;\;\text{(definition of $\epsilon_\phi^{\rm sat}$)}\\
\Rightarrow \epsilon_\phi^{\rm sat}(m) &\simeq & 0.01 \times\left\{\frac{\alpha_X}{0.01}\right\}^{-1}\times\left\{\frac{m}{6\times 10^9\,\Msun}\right\}^\nu\overset{\sim}{\propto} m^{1/3}\nn\,.
\een
From this definition, we can have a better intuition of the Sommerfeld enhancement regime in which a (sub)halo sits: if $\epsilon_\phi>\epsilon_\phi^{\rm sat}(m)$ ($\epsilon_\phi<\epsilon_\phi^{\rm sat}(m)$), then most of the halo phase-space distribution is located in the saturation regime (Coulomb regime, respectively).

\change{It is also helpful to understand the dependence on $\epsilon_\phi$. At fixed values of the coupling strength $\alpha_\chi$, decreasing $\epsilon_\phi$ amounts to exploring different particle physics model configurations (increasing the DM particle mass, or equivalently decreasing the mediator mass). This also amounts to decreasing  $\mtsat(\epsilon_\phi)$ accordingly, hence moving the halo phase-space distribution from the saturation regime domination ($m<\mtsat$) to the Coulomb regime domination ($m>\mtsat$)}. We have the following behaviors for the {\em effective} Sommerfeld enhancement as a function of (sub)halo mass $m$:
\bi
\item $m > \mtsat(\epsilon_\phi) \leftrightarrow \epsilon_\phi^{\rm sat} > \epsilon_\phi $: The bulk of the phase-space volume is located in the Coulomb regime of the Sommerfeld factor, since $\vsat(\epsilon_\phi)<\overline{v}$, so the enhancement is $\propto(m/\mtmax)^{-\nu} \propto (\overline{v}/\vmax)^{-1}$. Hence, the enhancement factor does not depend on $\epsilon_\phi$, and it is fixed at a value $\propto \overline{v}^{-1}(m)$ even for decreasing (while non-resonant) $\epsilon_\phi$. This situation is typically encountered when $\epsilon_\phi$ is very small, or when the halo is very massive.
\item $m<\mtsat(\epsilon_\phi\neq \epsilon_\phi^{{\rm res},n} )\leftrightarrow \epsilon_\phi^{\rm sat}<\epsilon_\phi\neq \epsilon_\phi^{{\rm res},n}$: In this configuration, most of the halo phase space lies in the non-resonant saturation regime ($\overline{v}<\vsat(\epsilon_\phi)$), and the Sommerfeld factor scales at $\propto (\vmax/\vsat(\epsilon_\phi))^{-1}(v/\vsat(\epsilon_\phi))^{p} \propto (\mtmax/\mtsat(\epsilon_\phi))^{-\nu}(m/\mtsat(\epsilon_\phi))^{p \nu} \epsilon_\phi^{-(p+1)} $, independent of mass only in the $s$-wave case. The overall Sommerfeld factor is therefore entirely set by $\epsilon_\phi$ (and is $\propto 1/\epsilon_\phi$ or $1/\epsilon_\phi^3$ for $s$- or $p$-wave annihilation).
\item $m<\mtsat(\epsilon_\phi\sim \epsilon_\phi^{{\rm res},n})\leftrightarrow \epsilon_\phi^{\rm sat}<\epsilon_\phi\sim \epsilon_\phi^{{\rm res},n}$: Here, we sit on the $n^{\rm th}$ resonance, and since $\overline{v}(m)<\vsat^{{\rm res},n}$, the bulk of the phase-space volume participates in the enhancement, whose amplitude is maximized when $\vunit\lesssim\overline{v}\ll\vsat^{{\rm res},n}\leftrightarrow \mtunit\lesssim m\ll \mtsat^{{\rm res},n}$. The amplitude of the resonance peak {\em relative to} the baseline enhancement is larger for smaller halos, as predicted from \citeeq{eq:eff_ratio_res_to_base}. It is also suppressed like $\epsilon_\phi^2$ at higher and higher resonances, which, combined with the $1/\epsilon_\phi$ scaling of the baseline, explains why the amplitude of the series of peaks globally decreases linearly with $\epsilon_\phi$ as $\epsilon_\phi$ decreases.
\item $\mtsat(\epsilon_\phi\sim \epsilon_\phi^{{\rm res},n})<m \leftrightarrow \epsilon_\phi\sim \epsilon_\phi^{{\rm res},n}<\epsilon_\phi^{\rm sat}$: Here, we also sit on the $n^{\rm th}$ resonance, but only the lower tail of the phase-space volume participates in the enhancement because $\overline{v}>\vsat(\epsilon_\phi^{{\rm res},n})$. The remaining (higher) part of the phase-space volume is in the Coulomb regime. The amplitude of the resonance is therefore controlled by the reduced volume of available relevant phase space, and then suppressed if $\overline{v}\gg \vsat^{{\rm res},n}$. In that case, only the Coulomb enhancement is active, and actually saturates at the characteristic velocity of the (sub)halo $\propto (\overline{v}/\vmax)^{-1}\leftrightarrow (m/\tilde{m}_{\rm max})^{-\nu}$.
\ei

\begin{figure*}[t!]
\centering
\includegraphics[width=0.49\textwidth]{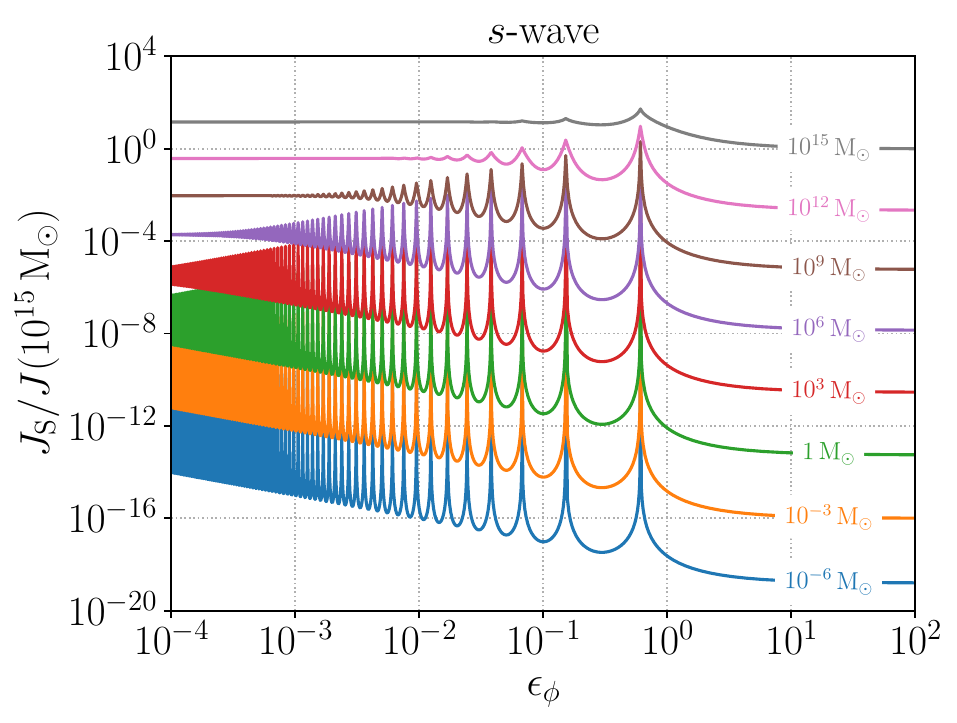} \hfill
\includegraphics[width=0.49\textwidth]{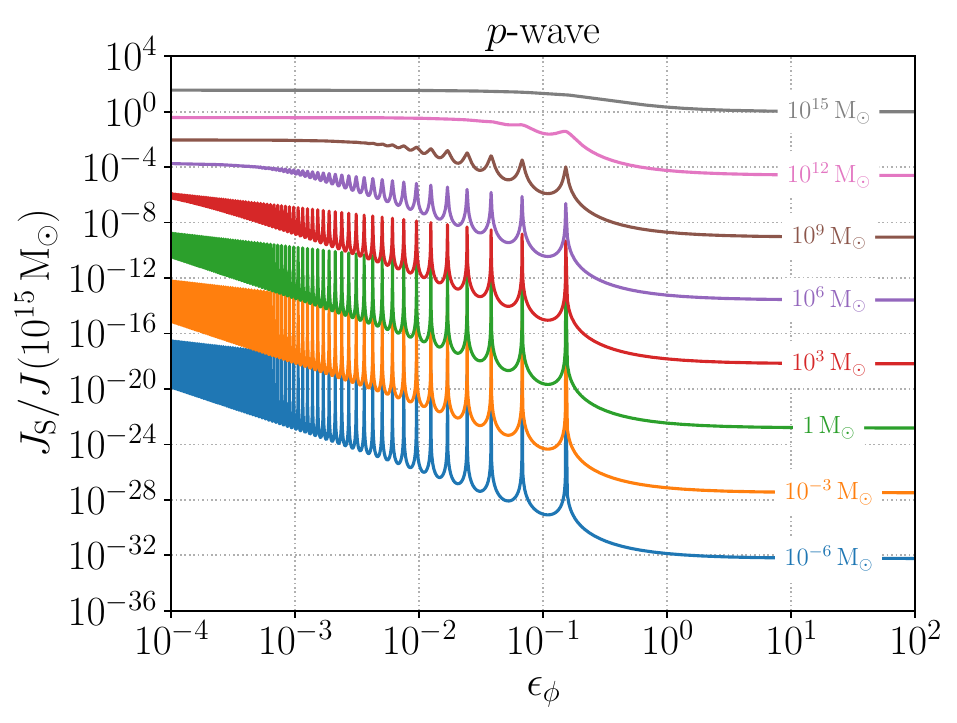} \hfill
\includegraphics[width=0.49\linewidth]{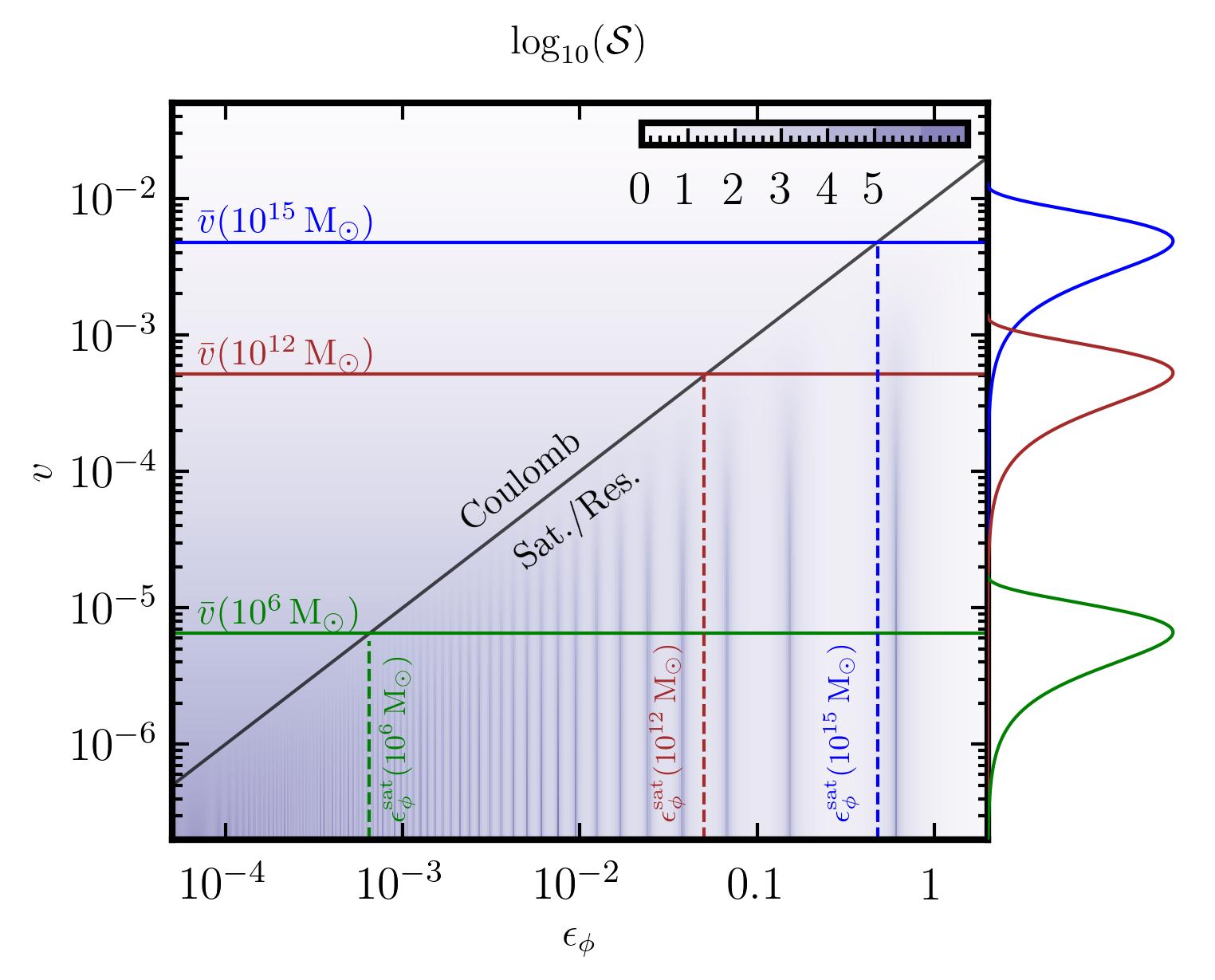} \hfill
\includegraphics[width=0.49\textwidth]{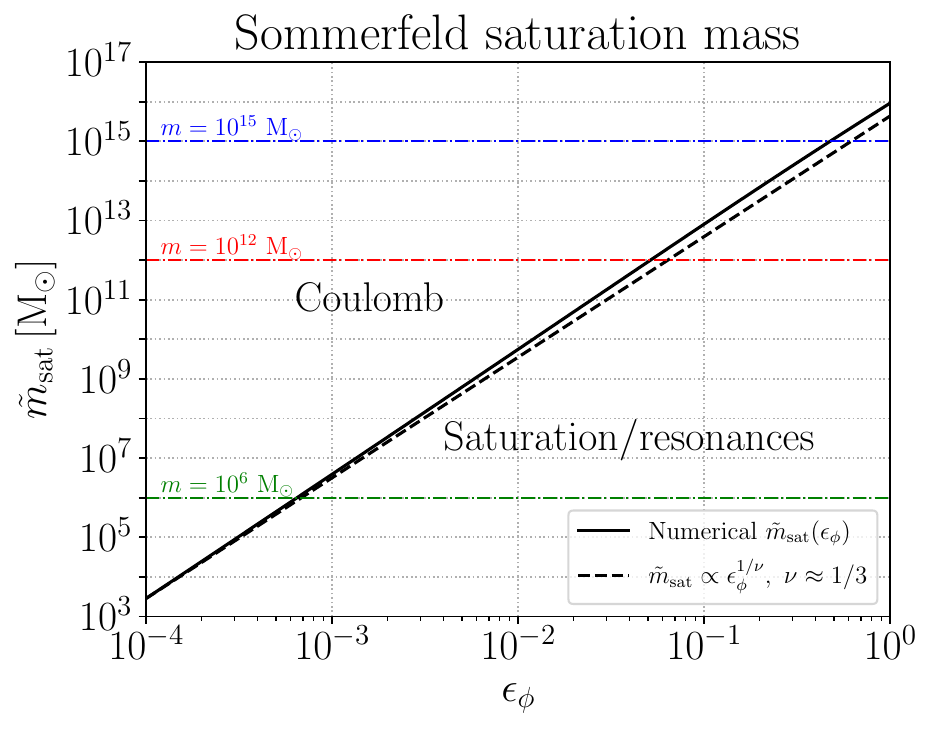}
\caption{Sommerfeld-enhanced $J_{\cal S}$-factors as a function of $\epsilon_\phi$ for DM halos of different masses located at the same distance, normalized to the Sommerfeld-free $J$-factor of a $10^{15}\Msun$ halo. {\bf Top left}: The $s$-wave annihilation case. {\bf Top right}: Same as left panel, but for a $p$-wave annihilation. {\bf Bottom left}: Saturation velocity as a function of $\epsilon_\phi$ (solid black curve), delineating the transition between the Coulomb and saturation regimes. The ($\log_{10}$ of the) Sommerfeld factor is represented as the third dimension (gray color scale), as a function of velocity $v$ and $\epsilon_\phi$. The characteristic speeds of $10^{15}$, $10^{12}$, and $10^{6}\,{\rm M}_\odot$ halos are indicated, with the corresponding (unnormalized) full Eddington velocity distribution taken at the scale radius, along the right vertical axis for illustration. {\bf Bottom right}: Saturation mass of the Sommerfeld effect as a function of $\epsilon_{\phi}$.}
\label{fig:comparison_JSomm_subhalos}
\end{figure*}

All this is illustrated in \citefig{fig:comparison_JSomm_subhalos}, where the top panels show the Sommerfeld-enhanced $J_{\cal S}$-factors for halos of different masses located at the same distance from the observer (normalized to the Sommerfeld-free $J$-factor of a reference halo of $10^{15}\Msun$). They are plotted as a function of the Bohr-to-interaction length ratio $\epsilon_\phi$. These enhanced $J_{\cal S}$-factors are calculated fully numerically assuming an $s$-wave ($p$-wave) annihilation in the top left (right) panel, with, from the top to bottom curves, predictions for halos of masses from $10^{15}\,\Msun$ (typical of galaxy clusters) down to $10^{-6}\,\Msun$ (typical of the cutoff mass in the matter power spectrum for WIMPs). This corresponds to characteristic speeds $\overline{v}$ spanning a range from $\sim 10^{-3}$ down to $\sim 10^{-9}$ (in natural units), hence of $\overline{\epsilon}_v=\overline{v}/\alpha_X$ of $\sim 10^{-1}$, down to $\sim 10^{-7}$.

As a practical toolkit to better understand these results, we also trace in the bottom left panel the key relation between the saturation velocity $\vsat$ and $\epsilon_\phi$ (solid black curve), with the Sommerfeld enhancement factor represented in a third dimension as a function of both the velocity $v$ and $\epsilon_\phi$ (gray color contrast code). In the same panel, we report the characteristic DM velocity $\overline{v}(m)$ for halos of masses $10^{15}$, $10^{12}$, and $10^{6}\,\Msun$, as inferred from \citeeq{eq:v_to_m_main}. To each of them, we associate along the right vertical axis the full (unnormalized) DM velocity distribution calculated from the Eddington inversion at the scale radii of these halos. This plot is particularly helpful to illustrate how the phase-space distribution of DM concentrates around $\overline{v}(m)$, though with a lower and a higher tail. The bottom right panel of \citefig{fig:comparison_JSomm_subhalos} further shows the scaling of the saturation mass $\mtsat$ with $\epsilon_\phi$, namely $\mtsat\propto\epsilon_\phi^{1/\nu} \approx \epsilon_\phi^{3}$, which follows from \citeeq{eq:msat} and from \citeeq{eq:nu_approx}. This approximation matches pretty well with the exact numerical result.

Let us now describe the top left panel of \citefig{fig:comparison_JSomm_subhalos}, which shows the Sommerfeld-enhanced $J_{\cal S}$-factors in the $s$-wave case for different halo masses. Large values of $\epsilon_\phi$ imply that most halos have the bulk of their phase-space volume in the saturation regime, as long as $m$ is smaller than $\mtsat(\epsilon_\phi)$ (equivalently $\overline{v}(m)<\vsat(\epsilon_\phi)$). The transition to the Coulomb regime occurs as $\epsilon_\phi$ decreases below a specific value, $\epsilon_\phi=\epsilon_\phi^{\rm sat}(m)$, defined in \citeeq{eq:def_eps_star} (equivalently $\vsat(\epsilon_\phi^{\rm sat})=\overline{v}(m)$). We can read values of $\epsilon_\phi^{\rm sat}(m)$ off the plot in the bottom right panel of \citefig{fig:comparison_JSomm_subhalos}, and can also use the bottom left panel to translate the halo masses in terms of characteristic velocities and velocity distributions. For illustration, let us focus on three specific virial halo masses, $10^{15}$, $10^{12}$, and $10^{6}\,{\rm M}_\odot$. All these masses have characteristic velocities smaller than $\sim \alpha_\chi = 0.01$, and are therefore subject to Sommerfeld enhancement ($m<\mtmax$). Let us follow their $J_{\cal S}$-curves on the top left panel from the right to left (large to small $\epsilon_\phi$ or $\mtsat$), while keeping in mind \citeeq{eq:JS_Coulomb}, which describes the Coulomb regime, \citeeq{eq:JS_Saturation} the saturation regime, and \citeeq{eq:JS_Resonances} resonances:
\bi
\item $m=10^{15}\,{\rm M}_\odot \leftrightarrow \overline{v} \sim 6\times 10^{-3}$: The transition from saturation to the Coulomb regime occurs at $\epsilon_\phi^{\rm sat}\sim 0.5$, a value just below the first resonance. For $\epsilon_\phi>\epsilon_\phi^{\rm sat}$, we are in the saturation regime ($m<\mtsat(\epsilon_\phi)$): the Sommerfeld factor saturates at $\propto (\mtsat/\mtmax)^{-\nu}\propto (\vsat/\vmax)^{-1}\propto 1 / \epsilon_\phi $, but takes a small value because $\mtsat\lesssim \mtmax$. When $\epsilon_\phi$ hits the first resonance, $\epsilon_\phi^{{\rm res},1}\simeq 2/3$, only a tiny part of the phase-space volume can participate in the enhancement $\propto 1/v^2$, because the bulk of the velocity distribution lies around $\overline{v}\sim \vsat^{{\rm res},1}$. Consequently, the amplitude of the first resonance is suppressed. The transition from the saturation regime to the Coulomb regime occurs when $\epsilon_\phi<\epsilon_\phi^{\rm sat}$, below which the bulk of the Sommerfeld boost becomes $\propto 1/\overline{v}(m)$. Once the whole phase-space distribution finds itself in the Coulomb regime, the Sommerfeld boost factor remains fixed at a constant value $\propto 1/\overline{v}$ determined by the characteristic velocity of the halo. For the same reason, higher-order resonances are suppressed (no phase-space volume left below $\vsat(\epsilon_\phi)$). Therefore, for further decreasing values of $\epsilon_\phi$, even though in the Coulomb regime, the Sommerfeld enhancement factor remains constant, fixed by the characteristic velocity $\propto 1/\overline{v}$: the $J$-factor stops evolving accordingly and remains flat.
\item $m=10^{12}\,{\rm M}_\odot \leftrightarrow \overline{v} \sim 6\times 10^{-4}$: The transition from the saturation to the Coulomb regime occurs at $\epsilon_\phi^{\rm sat}\sim 0.05$, which is located between the fourth ($\epsilon_\phi^{{\rm res},4}\simeq 1/24$) and third ($\epsilon_\phi^{{\rm res},3}\simeq 2/27$) resonances. For $\epsilon_\phi>\epsilon_\phi^{\rm sat}$, we are in the regime $\mtsat>m$, hence in the saturation regime for which the enhancement is $\propto 1/\epsilon_\phi$. When $\epsilon_\phi$ hits resonant values of order $n<3$, a significant part of the phase-space volume can participate in the $(\vsat/v)^2$ enhancement, which cannot exceed $\sim (\vsat/\overline{v})^2$ because most of the phase-space volume concentrates around $\overline{v}$. Therefore, even though the first resonances are turned on, their amplitudes are phase-space limited. When $\epsilon_\phi$ further decreases below $\epsilon_\phi^{\rm sat}$, the bulk of the phase-space volume switches to the Coulomb regime, but with a Sommerfeld factor asymptoting to a constant $\propto \vmax/\overline{v}$. There is no longer enough phase-space volume available below $\vsat(\epsilon_\phi)$ to trigger higher-order resonances, and the $J_{\cal S}$-factor stops evolving and remains flat.
\item $m=10^{6}\,{\rm M}_\odot \leftrightarrow \overline{v} \sim 6\times 10^{-6}$: The saturation-Coulomb transition occurs at $\epsilon_\phi^{\rm sat}\sim 5\times 10^{-4}$, which is located in the resonance forest. Like in the previous case, as long as $\epsilon_\phi>\epsilon_\phi^{\rm sat}$ (equivalently $\mtsat>m$), we are in the saturation regime, and the Sommerfeld factor is $\propto 1/\vsat\propto 1/\epsilon_\phi$. All resonances encountered by decreasing $\epsilon_\phi$ down to $\epsilon_\phi^{\rm sat}$ are turned on and have their amplitudes roughly set by $1/(n\,\overline{v})^2\propto \epsilon_\phi/\overline{v}^2$ --- the amplitudes decrease linearly with $\epsilon_\phi$ as the latter decreases. When $\epsilon_\phi$ becomes smaller than $\epsilon_\phi^{\rm sat}$, we switch to the Coulomb regime, and the enhancement is frozen to $\propto 1/\overline{v}$, and there is not enough phase-space volume available in the lower tail to turn the remaining resonances on. Hence, the Sommerfeld enhancement remains frozen and no longer evolves as $\epsilon_\phi$ keeps on decreasing below $\epsilon_\phi^{\rm sat}$.
\ei

The ratio of $J$-factors for two halos of masses $m_1$ and $m_2$ can easily be estimated from \citeeq{eq:J_ratio} and \citeeq{eq:JS_ratio}. In the absence of Sommerfeld enhancement, i.e. $\epsilon_\phi\gtrsim 1$, then $J_1/J_2\sim (m_1/m_2)^{1-3\,\varepsilon+p\,\nu}$. From this rough scaling relation, we can predict a factor of $\sim 3.5\times 10^2$ between each successive curve in the top left panel of \citefig{fig:comparison_JSomm_subhalos} for the $s$-wave case, and $\sim 3.5\times 10^{4}$ for the $p$-wave case in the top right panel. This is reasonably close to the exact results, $\sim 5\times 10^2$ and $4\times 10^4$, respectively. When the Sommerfeld enhancement kicks in, this ratio is corrected by an additional factor $\sim (m_1^{-s_{m_1}-\nu\,p}/m_2^{-s_{m_2}-\nu\,p})$, where $s_{m_1}$ and $s_{m_2}$ refer to the Sommerfeld mass indices of the halo of mass $m_1$ and the halo of mass $m_2$, respectively [see \citeeq{eq:sm_values}]---indeed, the two halos can be in different Sommerfeld regimes. We can still verify from the top panels of \citefig{fig:comparison_JSomm_subhalos} that, for instance, when halos are in the Coulomb regime (asymptotic values on the very left parts of the panels), then successive curves should be asymptotically split by a factor of $\sim 35$ for both the $s$- and $p$-wave cases, according to \citeeq{eq:J_ratio}. This is again close to the accurate numerical evaluation.

For the $p$-wave case illustrated in the top right panel of \citefig{fig:comparison_JSomm_subhalos}, the main differences with the $s$-wave case are the following. (i) In the saturation regime, the baseline enhancement is $\propto 1/\epsilon_\phi^3$ (instead of $1/\epsilon_\phi$). (ii) The amplitude of the resonance peak scales like $n^2 \propto 1/\epsilon_\phi$ (instead of $1/n^2 \propto \epsilon_\phi$), and therefore increases with the order of the resonance (linearly with $1/\epsilon_\phi$, as $\epsilon_\phi$ decreases). (iii) The overall $\propto \overline{v}^2$ suppression factor in the cross section (effectively captured in our ansatz for the Sommerfeld boost factor above), is compensated for by the Sommerfeld enhancement, except on the baseline of the saturation regime where it contributes an additional splitting factor $\propto m^{2\nu}$ (very right part of the plot), which then disappears in the Coulomb regime (very left part of the plot). \new{A full description of resonance properties can be found around \citeeq{eq:S_ansatz_res_v}}.

\begin{figure}[t!]
\centering
\includegraphics[width=0.99\linewidth]{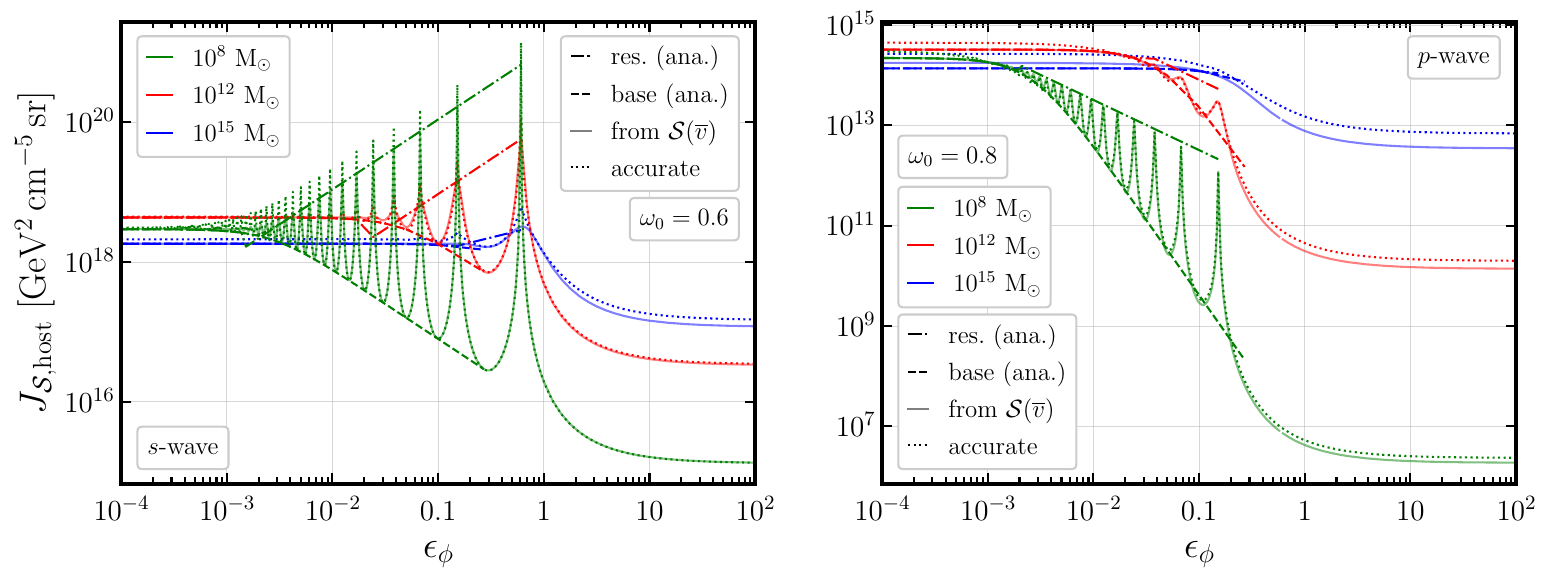}
\includegraphics[width=0.99\linewidth]{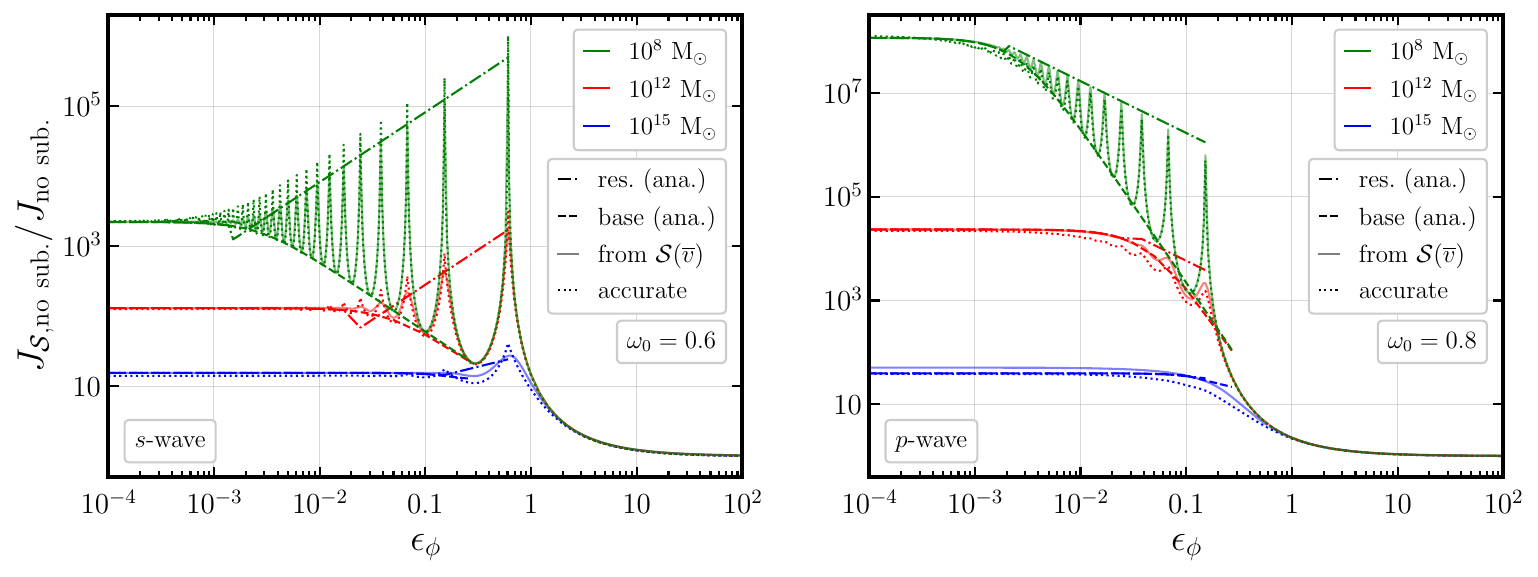}
\caption{{\bf Top left panel:} Sommerfeld-enhanced $J_{\cal S}$ factors over entire halos obtained for an $s$-wave annihilation process calculated (i) from the full phase-space numerical integral of exact expressions (dotted curve), (ii) from the analytical approximation of the $J$-factor in \citeeq{eq:jm} times the Sommerfeld factor of \citeeq{eq:Sommerfeld_enhancement_s_wave}  and \citeeq{eq:Sommerfeld_enhancement_p_wave} evaluated at speeds $\overline{v}(m)$, and (iii) from the full analytical approximation of \citeeq{eq:JS_halo} (for the Coulomb and saturation baseline, and for the peaks envelope) -- we consider three halos of masses $m=10^8$, $10^{12}$, and $10^{15}\,\Msun$. {\bf Bottom left panel:} Corresponding effective Sommerfeld enhancement factors over entire halos ratio expressed as $J_{\cal S}/J$, neglecting subhalos. {\bf Right panels:} Same as left panels for a $p$-wave annihilation process.}
\label{fig:ansatz_halo}
\end{figure}

Finally, before moving to the study of the global contribution of a subhalo population to the Sommerfeld enhancement, it is interesting to compare the accurate numerical results at the level of single halos with those derived from our approximate ansatz of \citeeq{eq:S_ansatz_tot_m}. We report such a comparison in \citefig{fig:ansatz_halo} in terms of both the Sommerfeld-enhanced $J_{\cal S}$ factors (top panels) and the ratios $J_{\cal S}/J$ (bottom panels) for halos of masses $10^8$, $10^{12}$, and $10^{15}\,\Msun$, typical of dwarf galaxies, spiral galaxies, and galaxy clusters, placed at distances 0.1, 1, 100~Mpc, respectively. An averaged Sommerfeld enhancement factor at the level of an entire halo can be formulated from the ratio $J_{\cal S}/J$ -- $J$ and $J_{\cal S}$ are calculated from a full phase-space integration in the accurate numerical results shown in the plots. We actually compare these accurate results (dotted curves) with our analytical approximations (dashed and dot-dashed curves) --- see \citeeq{eq:jm} for the $J$-factor, and \citeeq{eq:JS_halo} for the Sommerfeld-enhanced $J_{\cal S}$-factor. The left (right, respectively) panels show the comparisons for an $s$-($p$-)wave annihilation. We have also reported solid curves that correspond to the exact Sommerfeld factors of \citeeq{eq:Sommerfeld_enhancement_s_wave} and \citeeq{eq:Sommerfeld_enhancement_p_wave} evaluated at a single characteristic velocity $\overline{v}(m)$ for each halo. We used \citeeq{eq:v_to_m_detail} for the latter, and tuned the constant $\omega_0$ to 0.6 (0.8, respectively) for $s$-($p$-)wave annihilation. The baselines and the peaks envelopes (dashed and dot-dashed curves) are instead calculated from our analytical ansatz of \citeeq{eq:S_ansatz_tot_m}, evaluated at the same characteristic velocities. We see that the analytical approximations capture the exact behaviors reasonably well. The peaks amplitudes are slightly underestimated because $\overline{v}(m)$ overestimates the typical speed at the very center of objects. As expected from our analytical approximations, the Sommerfeld enhancement at the level of a full halo is quite similar to the local velocity-dependent effective Sommerfeld enhancement depicted in \citefig{fig:ansatz} (except for the $v^2$ $p$-wave correction absorbed in the definition of ${\cal S}$ in the latter case, which does not change the scaling in $\epsilon_\phi$ but rescales the Sommerfeld enhancement by a factor of $\mu^{2\nu}$: this benefits more massive halos but at the same time is more representative of the scaling of the true cross section).

\subsubsection{Sommerfeld enhancement for a population of subhalos}
\label{sssec:S_for_all}
To understand the global Sommerfeld enhancement arising from a population of subhalos, it is convenient to combine the results obtained in the previous paragraph, where we have defined an ansatz for the Sommerfeld enhancement in terms of the (sub)halo virial mass $m$, with the analytical results obtained for the subhalo boost factor in \citesec{sssec:boost_noSomm}. We warn the reader that the analytical results derived from now on will be much less precise when compared to the numerical results (generically much more precise for $s$-wave than for $p$-wave processes). Still, they turn very useful to really understand the different features of the numerical results.

Given a (sub)halo of virial mass $m$, we can define a $J_{\cal S}$-factor corrected for the overall Sommerfeld effect according to \citeeq{eq:JS_halo}, which scales like $\propto \mu^{1-3\varepsilon-s_m}$. Exponent $s_m$ is the effective power-law mass index introduced in \citeeq{eq:S_ansatz_tot_m}, which takes different values for the different Sommerfeld regimes. From this, assuming $\mtunit<\mmin<\mtsat(\epsilon_\phi)<\mmax<\mtmax$ (which is not always the case\footnote{Among the possible departures from this assumed mass hierarchy, we indicate three generic variants: (i) $\mmin<\mtunit$, which is actually the case for the template parameters used in this paper; (ii) $\mtsat<\mmin$, which is generic in the Coulomb massless-mediator limit $\epsilon_\phi\to 0$, in which case there is no saturation regime; (iii) $\mmax<\mtsat$, which can happen for small host halos (typically dwarf galaxies) and moderate values of $\epsilon_\phi$. In all of those cases, one needs to recast the splitting of the mass integral accordingly, which leads to different mass boundaries in the asymptotic regimes.}), it is easy to express the total $J$-factor for a population of subhalos:
\ben
\label{eq:JS_decompose}
J_{{\cal S},{\rm sub}}(\epsilon_\phi) &=& \int_{\mmin}^{\mmax} \dd m\,\frac{\dd N_{\rm sub}}{\dd m}\,J_{\cal S}(m,\epsilon_\phi)\\
&=& \underbrace{\int_{\mmin}^{\mtsat(\epsilon_\phi)} \dd m\,\frac{\dd N_{\rm sub}}{\dd m}\,J_{\cal S}(m,\epsilon_\phi)}_\text{saturation+resonances}+\underbrace{\int_{\mtsat(\epsilon_\phi)}^{\mmax} \dd m\,\frac{\dd N_{\rm sub}}{\dd m}\,J_{\cal S}(m,\epsilon_\phi)}_\text{Coulomb}\nn\\
&\overset{\sim}{\propto}& \sum_{s\in\text{Somm. regimes}}\frac{1}{\alpha_s}\,\mu^{-\alpha_s}\Big|^\text{lower relevant mass bound}_\text{upper relevant mass bound}\nn\,.
\een
\change{Our sign convention for the Sommerfeld-enhanced subhalo boost factor mass index $\alpha_s$ is such that a positive value gives large values of the total subhalo $J$-factor, hence of the boost factor. This occurs when lighter subhalos contribute the most to the annihilation rate, hence when the integral above is dominated by contributions at the lower mass boundary [see discussion around \citeeq{eq:alpha_boost_sign}].} The important features of this total luminosity are therefore (i) the mass boundaries of the integral, and (ii) the effective subhalo mass index $\alpha_s$ (and its sign), which depends on the mass and mass-concentration indices $\alpha$ and $\varepsilon$, as well as on the Sommerfeld-enhancement mass index $s_m$ introduced earlier for different regimes. This effective index can easily be derived by integrating Eqs.~\eqref{eq:JS_Coulomb}-\eqref{eq:JS_Resonances} over the subhalo mass function. It generically reads:
\ben
\label{eq:alphaS}
\alpha_s = \alpha-2+3\varepsilon+s_m\,,
\een
where $s_m$ is the Sommerfeld mass index that depends on the Sommerfeld regime---it is given in \citeeq{eq:sm_values}.

In fact, both $\alpha_s$ and the mass boundaries depend on the relevant Sommerfeld regime, which is itself fixed by $\epsilon_\phi$ or, equivalently, by $\mtsat(\epsilon_\phi)$ [hence the splitting of the integral as a sum of different pieces in \citeeq{eq:JS_decompose}]. Consequently, for a given $\epsilon_\phi$, there can be two different contributions, assuming $\mmin<\mtsat(\epsilon_\phi)<\mmax<\mtmax$: one from the Coulomb regime, involving subhalo masses between $\mtsat(\epsilon_\phi)$ and $\mmax$, and another one from the saturation regime, involving subhalo masses lighter than $\mtsat(\epsilon_\phi)$. For resonant values of $\epsilon_\phi$ (\ie~still in the saturation regime), subhalos lighter than $\mtsat(\epsilon_\phi^{{\rm res},n})$, for any order $n$, are also the ones most involved in the enhancement. Since the minimal subhalo mass $m_{\rm min}$ is fixed for all host halos (it depends on the DM particle scenario itself), and the maximal subhalo mass $m_{\rm max}\simeq M_{\rm host}/100$ is always smaller than $\tilde{m}_{\rm max}$ in the configurations studied in this paper, we can advantageously split the subhalo population yield to the $J$-factor by taking the asymptotic form of the Sommerfeld enhancement factor relevant to each part of the integral of \citeeq{eq:JS_decompose}. This provides us with a one of our main fully analytical results:
\begin{tcolorbox}
\ben
\label{eq:jtot_somm}
J_{{\cal S},{\rm sub}}(\epsilon_\phi) & = & 
\theta(m_{\rm max}-\mtsat(\epsilon_\phi)) \, J_{{\cal S},{\rm sub}}^\text{Coul}(\epsilon_\phi) \\
&& +\theta(\mtsat(\epsilon_\phi)-m_{\rm min})\,J_{{\cal S},{\rm sub}}^\text{sat}(\epsilon_\phi)\nn\\
&& + \sum_{n=1+\frac{p}{2}}\delta_{\epsilon_\phi/\{\epsilon_\phi^{{\rm res},n}\}} \,\theta(\mtsat(\epsilon_\phi)-m_{\rm min})\,J_{{\cal S},{\rm sub}}^{{\rm res}}(\epsilon_\phi)\,.\nn
\een
\end{tcolorbox}
\noindent
We have deliberately separated the different Sommerfeld regimes for clarity. We have respectively for the Coulomb, saturation, and resonance regimes:
\begin{tcolorbox}
\bseq
\label{eq:JSsub}
\ben
\label{eq:JsubCoul}
J_{{\cal S},{\rm sub}}^\text{Coul}(\epsilon_\phi) &=& \frac{N_0(\muhost)\,J_0}{\alphacoul}\,(2\,v_0)^p\,S_0\,\mutmax^{\nu(1+p)}\,\mu^{-\alphacoul}\Big|^{{\rm max}(\mutsat,\mumin)}_{\mumax}\\
\label{eq:JsubSat}
J_{{\cal S},{\rm sub}}^\text{sat}(\epsilon_\phi) &=& \frac{N_0(\muhost)\,J_0}{\alphasat}\,(2\,v_0)^p\,S_0\,S_1\,\left\{\frac{\mutmax}{\mutsat}\right\}^{\nu(1+p)}\,\mu^{-\alphasat}\Big|^{\mumin}_{{\rm min}(\mutsat,\mumax)}\\
\label{eq:JsubRes}
J_{{\cal S},{\rm sub}}^{{\rm res}}(\epsilon_\phi) &=& \frac{N_0(\muhost)\,J_0}{\alphares}\,(2\,v_0)^p\,S_0^{\rm res}\,\left\{\frac{\mutmax}{\mutsat}\right\}^{\nu(1+p)}\,\mutsat^{2\,\nu}\\
&&\times\left\{\mu^{-\alphares}\Big|^{{\rm max}(\mumin,\mutunit)}_{{\rm min}(\mutsat,\mumax)} + \frac{\theta(\mutunit-\mumin)}{\mutunit^{2\nu}}\frac{\alphares}{\alphares^{\rm unit}}\mu^{-\alphares^{\rm unit}}\Big|^{\mumin}_{\mutunit}\right\}\,,\nn
\een
\eseq
\end{tcolorbox}
\noindent
where $\mumin$ and $\mumax$ are the subhalo reduced minimal and maximal masses, given a host halo of mass $M_{\rm host}$ and a DM particle scenario. The boost mass index \citeeq{eq:alphaS} allows us to determine the different indices, $\alphacoul$, $\alphasat$, and $\alphares$, from the Sommerfeld mass index $s_m$ of \citeeq{eq:sm_values} given for the three Sommerfeld regimes. This separation is rather artificial though, because the Sommerfeld enhancement factor smoothly transits between these regimes. That, together with the fact that we approximate phase-space integrals by evaluating the relevant functions at characteristic velocities, which induces nonphysical thresholds between the saturation/resonant and Coulomb regimes, will be the main source of numerical errors with respect (i) to the full mass integral of the Sommerfeld factor, and (ii) a fortiori also to the exact numerical integration over both mass and phase space. However, this division has the virtue of providing fully analytical scaling relations and a fine understanding of parameter dependencies, despite the significant cost in precision.

As an additional detail, mind the last term of the result obtained for resonances, which features $\mutunit$ and includes the possibility of having $\mumin<\mutunit$, in which case we have to account for the unitarity saturation of resonances. In this small corner of the parameter space, the Sommerfeld-corrected index $\alphares$ changes, which we write explicitly by using $\alphares^{\rm unit}$ (this term is not crucial, so we will mostly neglect it in forthcoming discussion). As the frames indicate, these are still very insightful results which allow us to fully understand how the Sommerfeld enhancement propagates over a full population of subhalos.

Assuming a value for $\epsilon_\phi$ such that $\mmin<\mtsat(\epsilon_\phi)<\mmax$, the upper subhalo mass range $\in[\mtsat(\epsilon_\phi),m_{\rm max}]$ lies in the Coulomb regime, while the lower one $\in[m_{\rm min},\mtsat(\epsilon_\phi)]$ lies in the saturation regime, for which the asymptotic mass slopes associated with $J_{{\cal S},{\rm sub}}$ take different values---resonances further show up in the saturation regime. Each regime is featured by its own index $\alpha_s$, whose generic form is given in \citeeq{eq:alphaS}. The dominant boundary term of each piece in \citeeq{eq:JSsub} will be selected according to the sign of each index: as explained above, a positive sign implies a dominant contribution from lighter and therefore more numerous and more concentrated subhalos. Let us inspect these indices in more detail, by combining \citeeq{eq:sm_values} and \citeeq{eq:alphaS}:
\ben
\label{eq:alpha_boost_somm}
\alpha_s =
\begin{cases}
\alphacoul &\equiv \alpha-2+3\varepsilon+\nu \\
\alphasat & \equiv \alpha-2+3\varepsilon-\nu\,p \\
\alphares & \equiv \alpha-2+3\varepsilon+ \nu\,(2-p)\\
\alphares^{\rm unit} & \equiv \alpha-2+3\varepsilon- \nu\,p=\alphasat
\end{cases}
\overset{\text{numerical}}{\underset{\text{evaluation}}{\longrightarrow}}
\begin{cases}
\alphacoul  \approx 0.43\\
\alphasat  \approx
\begin{cases}
  0.10\;\text{(for $p=0$)}\\
  -0.57 \;\text{(for $p=2$)}
\end{cases}\\
\alphares  \approx
\begin{cases}
  0.76 \;\text{(for $p=0$)}\\
  0.10 \; \text{(for $p=2$)}
\end{cases}
\end{cases}
\een
From \citeeq{eq:nu_approx}, we have the characteristic-speed-to-mass index $\nu\simeq 1/3$. Parameter $p=0/2$ for an $s/p$-wave annihilation. Since $\alpha\approx 1.95$ and $\varepsilon\approx 0.05$, we see that $\alphacoul>0$ quite generically. Therefore, the Coulomb regime is $\mtsat$-dominated (provided $\mtsat<\mmax$, which is not always the case in particular if the host halo is a dwarf galaxy). On the other hand, in the saturation regime, $\alphasat>0$ for the $s$-wave case ($p=0$), while it gets negative for the $p$-wave case ($p=2$). Therefore, the saturation regime is either $\mmin$-dominated ($s$-wave) or $\mtsat$-dominated ($p$-wave). In the latter case, this means that mostly subhalo masses down to $\mtsat(\epsilon_\phi)$ participate in an extra-Sommerfeld enhancement, whereas the lower part of the mass function does not add up a significant yield---this actually comes from the $v^2$ $p$-wave suppression factor absorbed in our effective Sommerfeld ansatz, which remains in the saturation regime of $p$-wave annihilation. In contrast, on resonances, we see that $\alphares$ is positive for both $s$-wave and $p$-wave annihilation processes. Therefore, all subhalos down to the cutoff mass $m_{\rm min}$ participate in the extra-enhancement on resonances in both cases. There is still a fundamental different between $s$- and $p$-wave resonances that needs to be emphasized: there is formally a velocity dependence in the $s$-wave case, which can be seen from the $2\nu$ contribution to $\alpha_{\rm res}$, while $p$-wave resonance amplitudes do not depend on velocity---see detailed discussion around \citeeq{eq:S_ansatz_res_v}. \change{Finally, it is important to stress that values of $\alpha_s$ close to 0 are subject to uncertainties. A small change in the mass function slope $\alpha$, for instance, could change the hierarchy in the contributing masses, hence in the global enhancement. This concerns mostly the saturation regime of the $s$-wave annihilation and resonances of the $p$-wave annihilation.}

\new{The previous discussion is illustrated in \citefig{fig:N_times_JS_vs_m}, where we have actually calculated a dimensionless quantity proportional to the product of the integrated number of subhalos more massive than $m$, $N(>m)\propto m^{1-\alpha}$, with the $J_{\cal S}(m)$ factor for a single (sub)halo of mass $m$ (divided by $J(1\,\Msun)$ to get a dimensionless quantity). This is meant to capture the dominant scaling of the global $J_{{\cal S},{\rm sub}}$ factor given in \citeeq{eq:jtot_somm} with the lower mass bound $m$, which also gives insight on the most contributing mass range in $J_{{\cal S},{\rm sub}}$ [\citeeq{eq:jtot_somm}]. Again, we take the three different Sommerfeld configurations used before: $\epsilon_\phi=0.1$ (moderate enhancement), $10^{-3}$ (significant enhancement), and the $n=8$ resonance ($\epsilon_\phi\sim 10^{-2}$, strong enhancement). The corresponding saturating masses $\mtsat(\epsilon_\phi)$ are shown as vertical dashed lines, marking the transition between subhalos mostly in the Coulomb regime ($m>\mtsat$) or mostly in the saturation regime ($m<\mtsat$). For the $s$-wave case (left panel), we see that the lower bound is always the most contributing one (curves increase as the mass boundary $m$ decreases in all Sommerfeld regimes), consistently with the positive values of $\alpha_{s}$ found in \citeeq{eq:alpha_boost_somm}. In contrast, the $p$-wave curves (right panel) only increase down to the saturation mass, below which contributions become negligible; except of course on the resonance, where the contribution increases as the boundary mass $m$ decreases down to the unitary limit. This is again consistent with the fact that $\alphasat<0$ while $\alphares>0$ with our choice of parameters for the $p$-wave case.}

\begin{figure}[t!]
\centering
\includegraphics[width=0.99\linewidth]{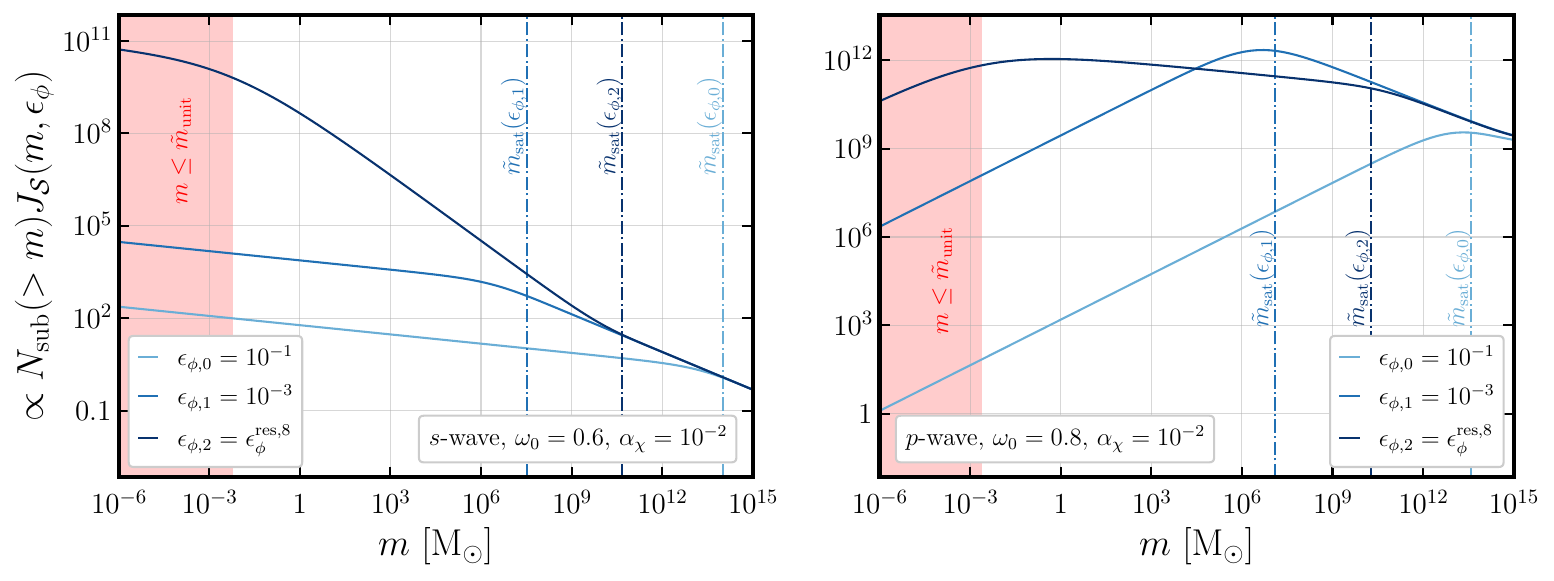}
\caption{\new{Estimate of the contribution of the subhalo population to the Sommerfeld-enhanced $J_{{\cal S},{\rm sub}}$ factor above some mass $m$, for different values of the reduced Bohr radius $\epsilon_\phi$: a large value 0.1 (moderate effect), a small value of $10^{-3}$ (significant effect), and an intermediate value of $\sim 10^{-2}$ sitting on the $n=8$ resonance (strong effect). A maximum in the curves show the subhalo mass range that contributes the most to the annihilation signal. This recasts most of the information already included in \citefig{fig:v_dependence} (effective Sommerfeld enhancement as a function of velocity), and in \citefig{fig:JS_vs_m}. Transition from Coulomb to saturation regimes occurs around $\mtsat(\epsilon_\phi)$, reported as vertical dash-dotted lines. {\bf Left panel:} $s$-wave case. {\bf Right panel:} $p$-wave case.}}
\label{fig:N_times_JS_vs_m}
\end{figure}

\begin{figure}[t!]
\centering
\includegraphics[width=0.99\linewidth]{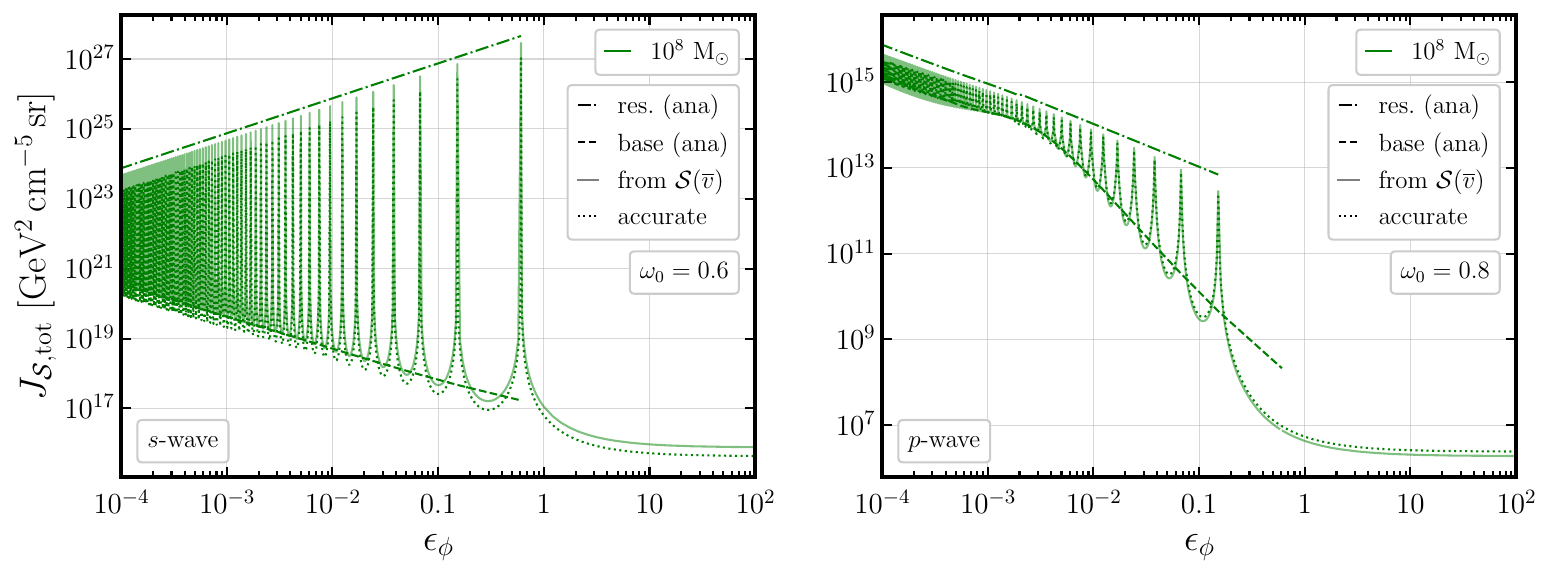} \hfill
\includegraphics[width=0.99\linewidth]{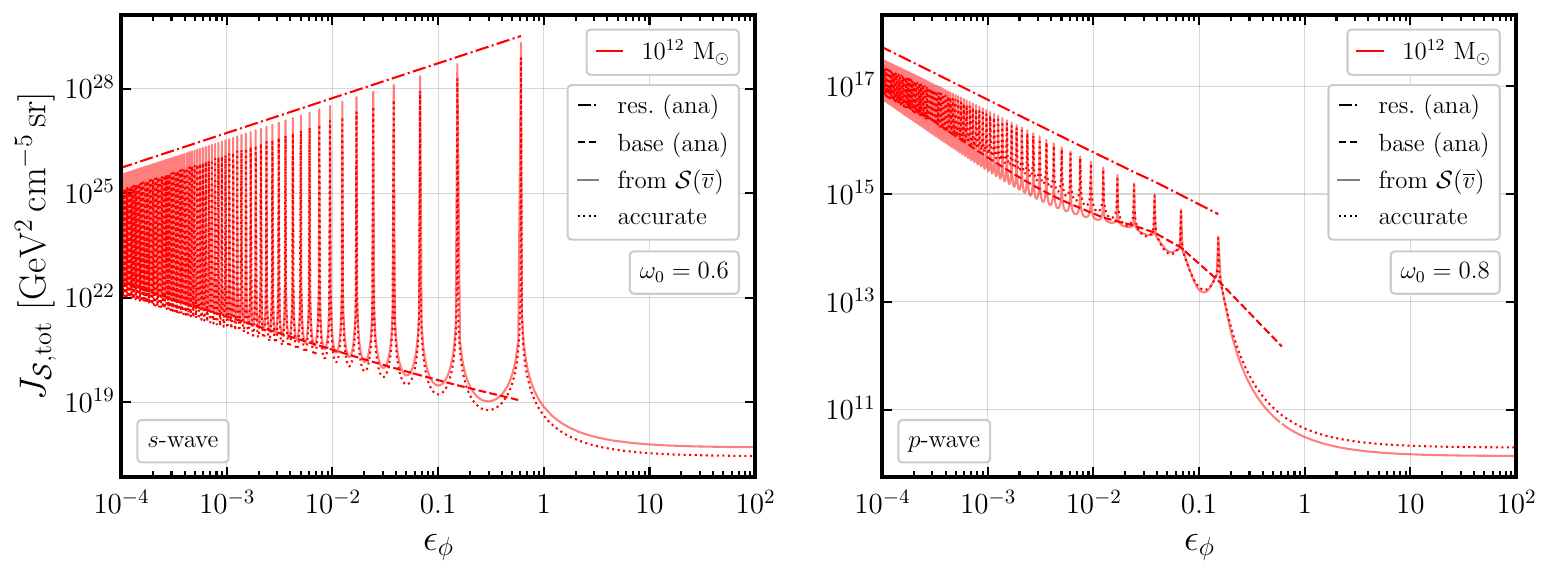} \hfill
\includegraphics[width=0.99\linewidth]{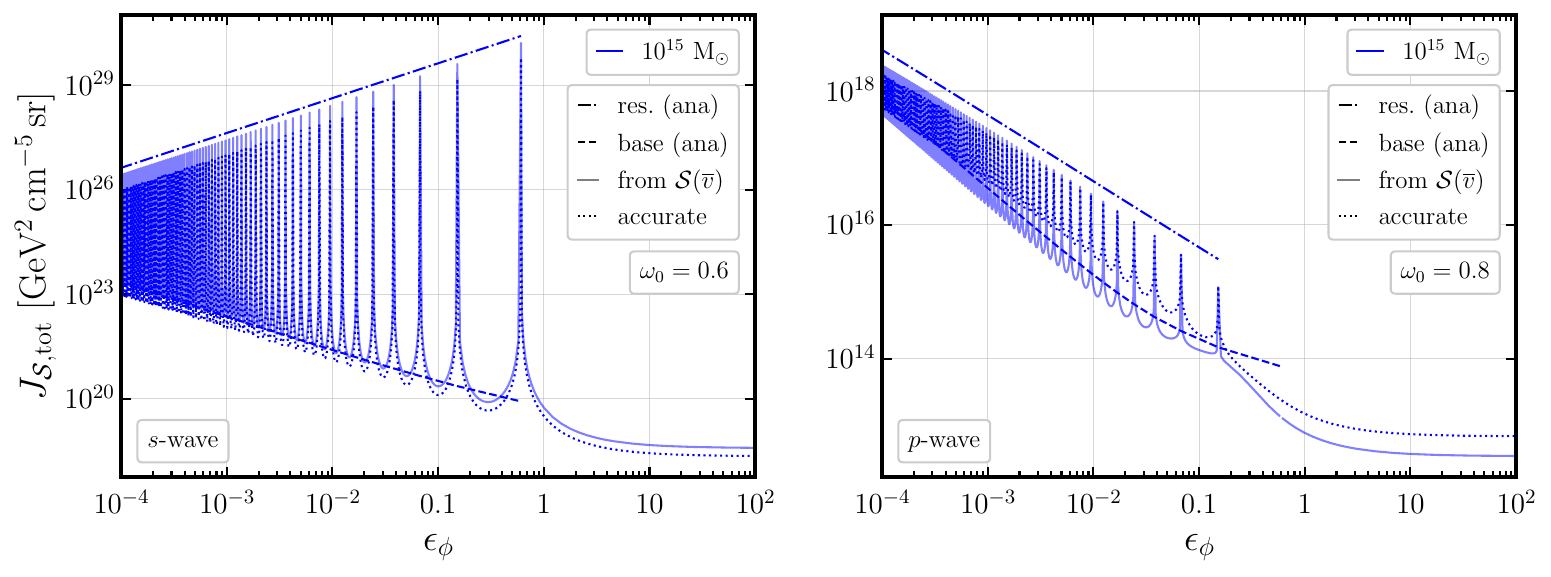}
\caption{{\bf Left column:} Global Sommerfeld-enhanced subhalo $J$-factors for $s$-wave annihilation, after integration of the whole subhalo population for three different host halo masses, $10^8$, $10^{12}$, and $10^{15}\,\Msun$, from top to bottom panels. {\bf Right column:} Same for $p$-wave annihilation.}
\label{fig:sub_Jsub_only}
\end{figure}

\begin{figure}[t!]
\centering
\includegraphics[width=0.99\linewidth]{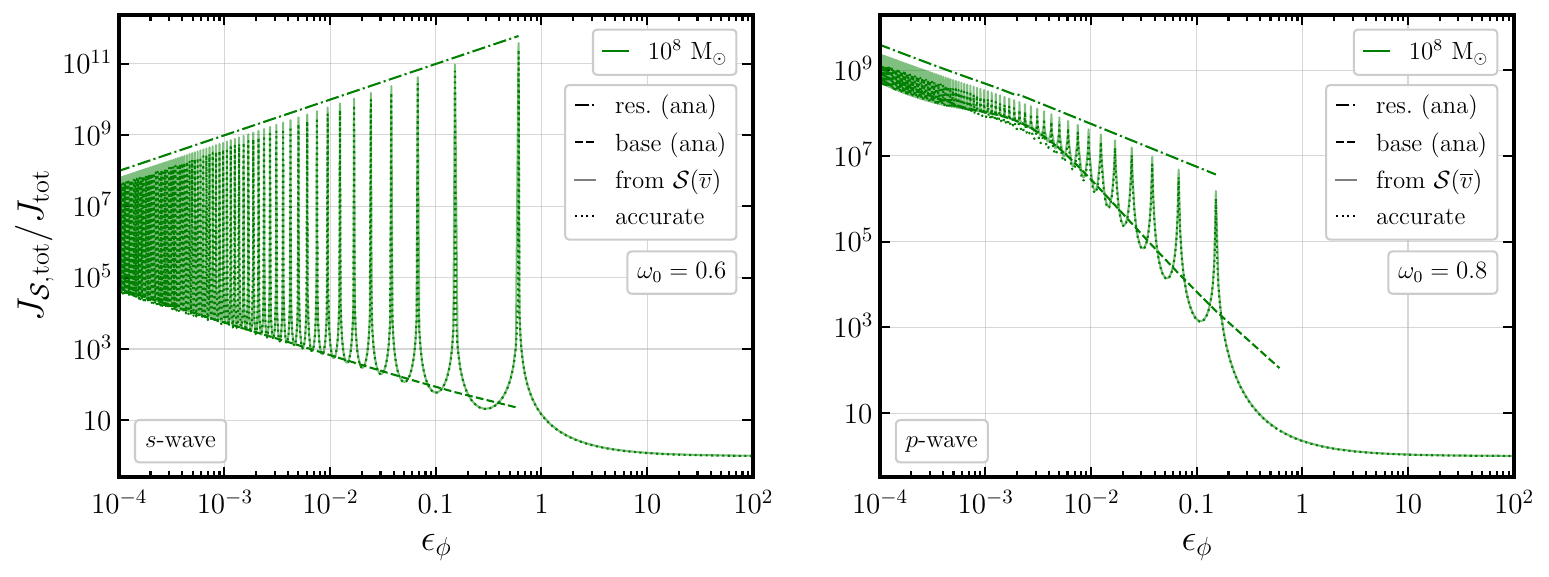} \hfill
\includegraphics[width=0.99\linewidth]{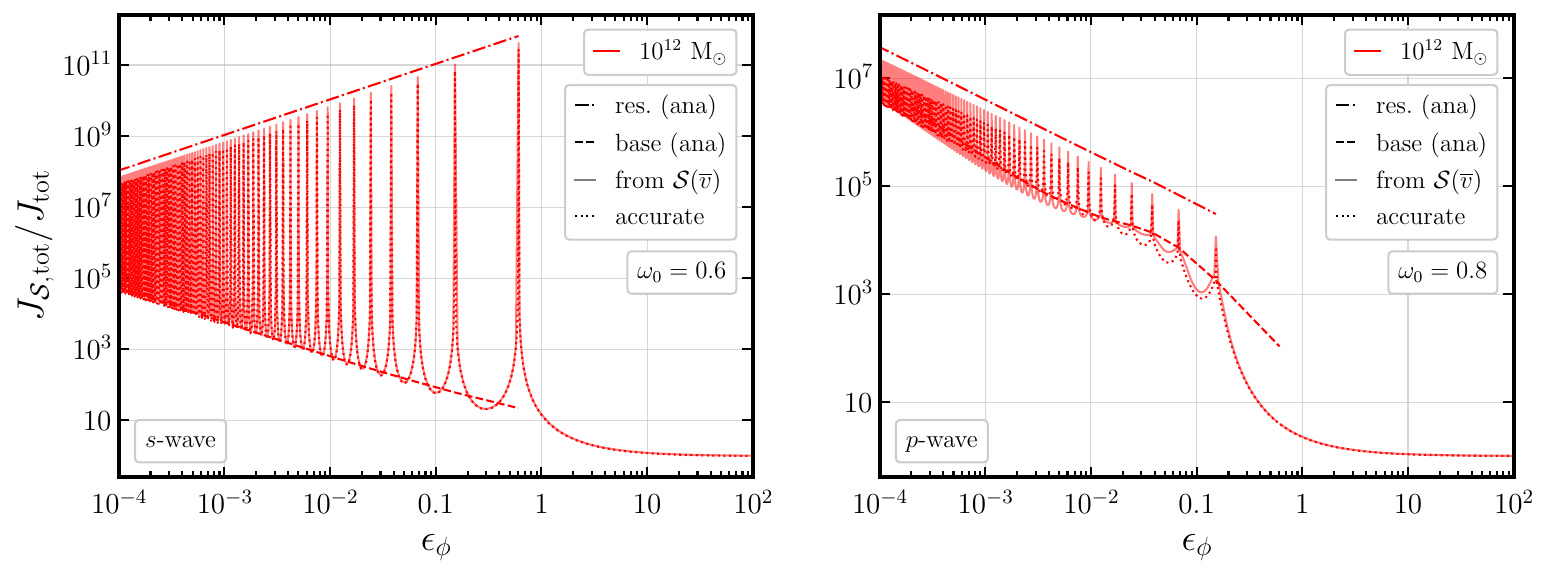} \hfill
\includegraphics[width=0.99\linewidth]{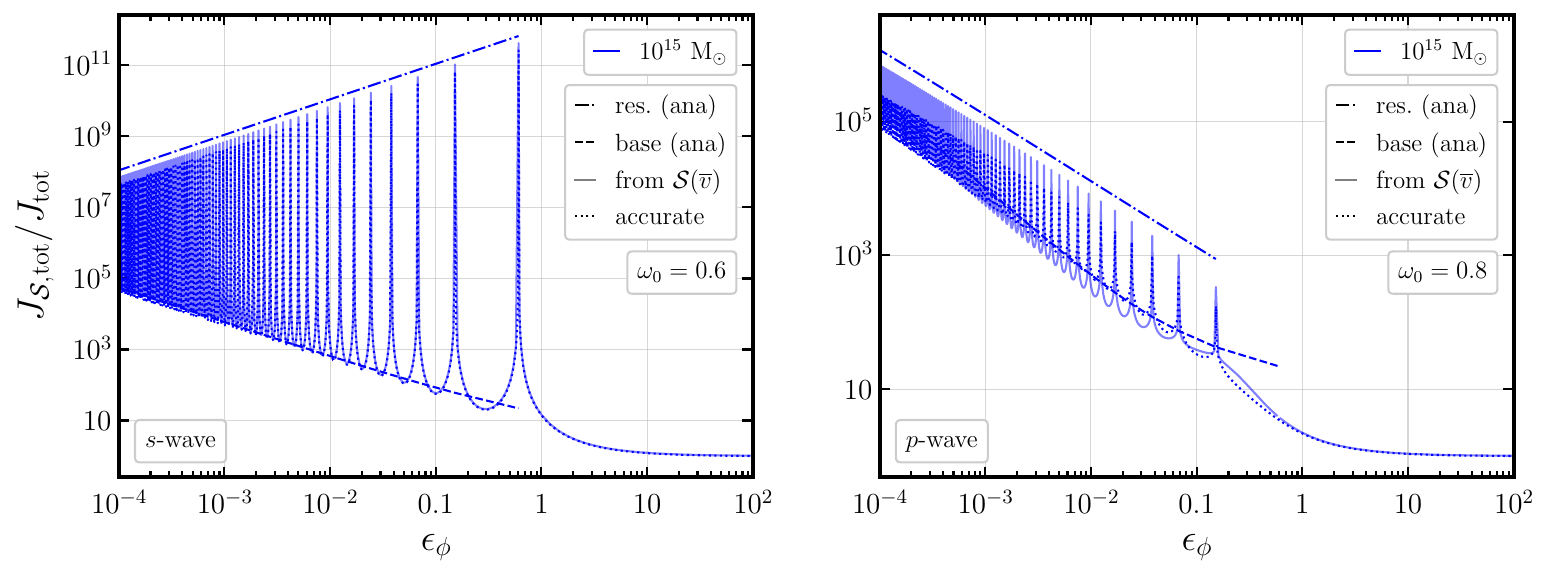}
\caption{{\bf Left column:} Global Sommerfeld enhancement for $s$-wave annihilation after integration of the whole subhalo population for three different host halo masses, $10^8$, $10^{12}$, and $10^{15}\,\Msun$, from top to bottom panels, expressed in terms of the ratio of the Sommerfeld-enhanced-to-Sommerfeld-free $J$-factors. {\bf Right column:} Same for $p$-wave annihilation.}
\label{fig:sub_Jsub}
\end{figure}

The peak-to-baseline ratio in the saturation regime for a full subhalo population, $\overline{\cal R}_{\rm sub}$, is given by:
\ben
\label{eq:sub_peak_to_base}
\overline{\cal R}_{\rm sub}(\epsilon_\phi,\mumin,\alpha) &=& \frac{S_0^{\rm res}}{S_0 S_1} \,\mutsat^{2\nu+\textcolor{blue}{\frac{p}{2}\alphasat}}\,\textcolor{blue}{\mumin}^{-\alphares}\mumin^{\textcolor{blue}{\frac{(2-p)}{2}\alphasat}}\\
&=& \frac{\pi^2}{6}\,\mutsat^{2\nu+\textcolor{blue}{\frac{p}{2}\alphasat}}\,
\textcolor{blue}{\mumin}^{-\alphares}\mumin^{\textcolor{blue}{\frac{(2-p)}{2}\alphasat}}\nn\\
&\propto& \epsilon_\phi^{2+\textcolor{blue}{\frac{p}{2}\frac{\alphasat}{\nu}}}\,
\textcolor{blue}{\mumin}^{-\alphares}\mumin^{\textcolor{blue}{\frac{(2-p)}{2}\alphasat}}\,.\nn
\een
For $s$-wave processes, we have $\overline{\cal R}_{\rm sub}\propto \epsilon_\phi^2$, while for $p$-wave processes,\footnote{The \textcolor{blue}{indices or parameters in blue} featuring factors of $p$ are tricks to account for the change of sign of $\alphasat$ here between the $s$- and $p$-wave processes. \change{Indeed, the sign of the index decides whether one picks only the lower or the upper bound of the integral, as generically illustrated in \citeeq{eq:JS_decompose}, so as to write a simplified approximate results in the limit $m_{\rm lower}\ll m_{\rm upper}$. It turns out that with our choice of parameters, the sign of the saturation mass index $\alphasat$ changes from the $s$- to the $p$-wave case, hence the associated final results scale with different boundary masses (for instance $\propto m_{\rm lower}^{-\alphasat}$ in one case, and $\propto m_{\rm upper}^{-\alphasat}$). Therefore, while \citeeq{eq:JSsub} is generic, \citeeq{eq:sub_peak_to_base} is not and is only valid for our choice of reference parameters. In the same vein, other equations with blue indices are not generic. All generic results can formally be expressed from \citeeq{eq:JSsub}, but would lead to rather tedious expressions. Parameter $\mumin$ also appears in blue whenever it could be traded for $\mutunit$, \ie, when $\mumin<\mutunit$.}} $\overline{\cal R}_{\rm sub}\propto \epsilon_\phi^{0.3}$. This result predicts that the peak-to-baseline ratio should decrease much faster as $\epsilon_\phi$ decreases in the $s$-wave case than in the $p$-wave case. Note that the above ratio assumes $\mutsat<\mumax$, which is not always verified (notably for light host halos). If instead $\mutsat>\mumax$, then the dependence in $\epsilon_\phi$ becomes $\propto \epsilon_\phi^2$ in both cases, and the ratio decreases fast with $\epsilon_\phi$ (the baseline increases fast) until $\mutsat$ enters the subhalo mass range, whence the dependence becomes much weaker.

We can further determine the effective Sommerfeld enhancement at the level of a subhalo population, which helps understand how the Sommerfeld effect manifests itself on top of the subhalo boost factor. We define this global Sommerfeld enhancement as
\ben
\label{eq:global_Ssub}
\widetilde{\cal S}(M_{\rm host},\mmin,\epsilon_\phi)\equiv \frac{J_{{\cal S},{\rm sub}}(\epsilon_\phi)}{J_{\rm sub}}\,,
\een
where the Sommerfeld-enhanced subhalo contribution $J_{{\cal S},{\rm sub}}$ is given in Eqs.~\eqref{eq:JsubCoul}-\eqref{eq:JsubRes} for the different regimes, while the Sommerfeld-free subhalo contribution $J_{\rm sub}$ is given in \citeeq{eq:JSfreesub} for the $s$- and $p$-wave cases.

In \citefig{fig:sub_Jsub_only} and \citefig{fig:sub_Jsub}, we compare the exact numerical calculations with the analytical approximations of respectively the total Sommerfeld-enhanced $J$-factors $J_{{\cal S},{\rm tot}}\equiv J_{{\cal S},{\rm host}}+J_{{\cal S},{\rm sub}}\simeq J_{{\cal S},{\rm sub}}$, and their ratios to the Sommerfeld-free cases, $J_{{\cal S},{\rm tot}}/J_{{\rm tot}}$, for the three template host halos introduced before. Note the resemblance with \citefig{fig:ansatz_halo} (top and bottom panels, respectively), which compares calculations of the Sommerfeld enhancement at the level of a single halo---this is indicative of the fact that a few specific masses drive the overall enhancement in each of the Sommerfeld regimes. Again, the dotted curves represent the full integrated results, while the solid curves represent the subhalo mass integral performed over the exact Sommerfeld factor evaluated at subhalo mass-dependent characteristic velocities. The analytical baselines and the peaks envelopes (dot-dashed curves) are instead calculated from the ansatz of \citeeq{eq:S_ansatz_tot_m}, integrated over the subhalo mass function. Panels in the left (respectively right) column display our results for the $s$-wave ($p$-wave) case. While the results for the $J$-factors can be understood from \citeeq{eq:JSsub}, we can better interpret the ratio $J_{{\cal S},{\rm tot}}/J_{{\rm tot}}\approx J_{{\cal S},{\rm sub}}/J_{{\rm sub}}\equiv \widetilde{\cal S}$, \ie~the overall effective Sommerfeld enhancement of the subhalo population, by inspecting the analytical expression of each asymptotic regime of the mass-integrated Sommerfeld enhancement $\widetilde{\cal S}$:\footnote{Note that for $p$-wave annihilation, the Sommerfeld-free $J$-factor $J_{{\rm tot}}\gg J_{{\rm sub}}$, so that $J_{{\cal S},{\rm tot}}/J_{{\rm tot}}$ as reported in \citefig{fig:sub_Jsub} does not strictly measure the amplitude of the overall Sommerfeld boost factor $\widetilde{\cal S}\equiv J_{{\cal S},{\rm sub}}/J_{{\rm sub}}$ in that case, but rather the full combined boost factor and its scaling with $\epsilon_\phi$.}
\bseq
\ben
\widetilde{\cal S}_{\rm Coul} &=& \frac{\alpha_{\rm boost}}{\alphacoul}\,S_0\,\mutmax^{(1+p)\,\nu}\,
\frac{\left({\rm max}(\mutsat,\mumin)\right)^{-\alphacoul}}{\mu^{-\alpha_{\rm boost}}\Big|^{\mumin}_{\mumax}}\,,\\
\widetilde{\cal S}_{\rm sat} &= & \frac{\alpha_{\rm boost}}{\alphasat}\,S_0 \,S_1 \,\mutmax^\nu \,\mutsat^{-\nu(p+1)}\,\frac{\mu^{-\alphasat}\Big|^{\mumin}_{{\min}(\mumax,\mutsat)}}{\mu^{-\alpha_{\rm boost}}\Big|^{\mumin}_{\mumax}}\,,\\ 
\widetilde{\cal S}_{\rm res} &= & \frac{\alpha_{\rm boost}}{\alphares}\,S_0^{\rm res} \,\mutmax^\nu\,\mutsat^{-\nu(p-1)}\,\frac{\mu^{-\alphares}\Big|^{{\rm max}(\mumin,\mutunit)}_{{\rm min}(\mumax,\mutsat)}}{\mu^{-\alpha_{\rm boost}}\Big|^{\mumin}_{\mumax}}\,.
\een
\eseq
We have removed the relevant step functions assuming $\mmin<\mtsat<\mmax$ for simplicity, but the general result can easily be derived from Eqs.~\eqref{eq:jtot_somm}-\eqref{eq:jtot_somm} and \citeeq{eq:JSfreesub}---a couple of footprints of the general result are still indicated with the min() and max() functions. The only parameter that depends on $\epsilon_\phi$ is $\mutsat$, while the only parameter that depends on $M_{\rm host}$ is $\mumax\simeq \muhost/100$. The other masses, including the minimal dimensionless subhalo mass $\mumin$, are taken universal.\footnote{Strictly speaking, $\mumin$ depends on the full underlying particle physics scenario, and may therefore depend on $\epsilon_\phi$ \cite{vandenAarssenEtAl2012}. Here, we assume that self-interactions play no role in setting the kinetic decoupling of DM particles in the early universe, and thereby that $\mumin$ does not depend on $\epsilon_\phi$.} Since $\mumin\ll\mutsat(\epsilon_\phi)$ over a wide range in $\epsilon_\phi$, we can assume that most subhalos are in the saturation regime. Therefore, they contribute both to the baseline and to the resonance peaks of the overall Sommerfeld factor (we can disregard the Coulomb regime). Looking each term of the initial ratio expression, we see that the denominator $J_{{\rm sub}}$ is $\propto \mumin^{-\alpha_{\rm boost}}$ in the $s$-wave case, and therefore can be assumed constant for any host halo. This explains why all plots in the left column of \citefig{fig:sub_Jsub}, which are associated with different host halo masses, look the same. On the other hand, in the $p$-wave case, the denominator is $\propto \mumax^{-\alpha_{\rm boost}} \propto (\muhost/100)^{-\alpha_{\rm boost}}$, and therefore $\widetilde{\cal S}$ indirectly depends on the host halo mass, which is readily verified in the right panels.

It is now straightforward to further predict the scaling in $\epsilon_\phi\propto \mutsat^{\nu}$ [see \citeeq{eq:msat}] and $\muhost$ from the previous analytical expressions, by accounting for the signs of the $\alpha$'s indices given in \citeeq{eq:alpha_boost_somm}:
\bseq
\label{eq:Stilde_approx}
\ben
\widetilde{\cal S}_{\rm Coul} &\propto& \muhost^{\textcolor{blue}{\frac{p}{2}\alpha_{\rm boost}}}\,\mumin^{\textcolor{blue}{\frac{(2-p)}{2}\alpha_{\rm boost}}}
\,  \epsilon_\phi^{-\frac{\alphacoul}{\nu}}\\
\widetilde{\cal S}_{\rm sat} &\propto& \muhost^{\textcolor{blue}{\frac{p}{2}\alpha_{\rm boost}}}\,\mumin^{\textcolor{blue}{\frac{(2-p)}{2}(\alpha_{\rm boost}-\alphasat)}}\,\epsilon_\phi^{-(p+1)}\times
\begin{cases}
\muhost^{-\textcolor{blue}{\frac{p}{2}\alphasat}} \,\,{\rm if}\,\mumax<\mutsat\\
\epsilon_\phi^{-\textcolor{blue}{\frac{p}{2}\frac{\alphasat}{\nu}}} \,\,{\rm else}
\end{cases}\\
\widetilde{\cal S}_{\rm res} &\propto& \muhost^{\textcolor{blue}{\frac{p}{2}\alpha_{\rm boost}}}\,\mumin^{\textcolor{blue}{\frac{(2-p)}{2}\alpha_{\rm boost}}} \left\{{\rm max\left( \mumin,\mutunit\right)}\right\}^{-\alphares}\,\epsilon_\phi^{-(p-1)}\,.
\een
\eseq
We recall that $p=0$ (2) for an $s$-($p$-)wave annihilation. From these expressions, we can understand why in the $s$-wave case (left-column panels of \citefig{fig:sub_Jsub_only} and \citefig{fig:sub_Jsub}) the baseline of the saturation regime goes $\propto \epsilon_\phi^{-1}$, while the curve following the amplitude of the peaks is instead $\propto \epsilon_\phi$. We also understand why in the $p$-wave case (right-column panels), the baseline of the saturation regime experiences a change in the slope in $\epsilon_\phi$ when $\mutsat<\mumax\sim\muhost/100$. This is due to the fact that the exponent $\alphasat$ is negative in the $p$-wave case, which implies that it is the upper bound of the subhalo mass integral ${\rm min(\mumax,\mutsat)}$ that matters. Indeed, as shown in the second equation above, the scaling in $\epsilon_\phi$ goes from $\propto \epsilon_\phi^{-3}$ when $\mumax<\mutsat$ to a much more moderate $\propto \epsilon_\phi^{-3-\frac{\alphasat}{\nu}}\sim \epsilon_\phi^{-1.3}$ when $\mumax\geqslant\mutsat$, which explains why the increase of the ratio is first very steep as $\epsilon_\phi$ decreases from large values, and then changes of slope. This is particularly visible for the lightest host halo with a mass of $10^8\,\Msun$ because then $\mmax\sim 10^{6}\Msun$, which corresponds to a scaling transition around $\epsilon_\phi^{\rm sat}(\mmax)\sim 6\times 10^{-3}$ (see bottom right panel of \citefig{fig:comparison_JSomm_subhalos}), above which no subhalo can participate in the saturation regime. The same transition is slightly less visible for the host halo of $10^{12}\,\Msun$. On the other hand, the scaling of the resonance peaks does not feature any such transition, as expected from the analytical results. The peak-to-baseline ratio can be fully understood from \citeeq{eq:sub_peak_to_base}, and associated discussion. We emphasize that in our template calculations, the peak amplitudes are fixed by $\mtunit \sim 8\times 10^{-4}\,\Msun$, not by $\mmin=10^{-6}\,\Msun<\mtunit$.

From these plots, we see that a quick integration of our simplified ansatz describes reasonably well the more accurate numerical results, slightly degrading from the $s$-wave to the $p$-wave case. This departure from the numerical results comes from the error made by changing the phase-space integral by an evaluation through a characteristic speed $\overline{v}(m)$, which needs to be adjusted by playing with the value of $\omega_0$ in \citeeq{eq:v_to_m} (the tuning values are given in the plots, and are fixed once and for all for a given configuration). Further splitting the subhalo mass integral into analytical asymptotic pieces as done just above to get analytical approximations and insight on the full result would induce bigger numerical errors (a factor of a few for $s$-wave processes, up to an order of magnitude for $p$-wave processes), because the actual Sommerfeld enhancement factor transits smoothly between regimes over the available mass range. Still, the full analytical prediction gets the scaling relations correct, which strongly helps in the interpretation. We also see from \citefig{fig:sub_Jsub_only} that a very simple expression, like that in \citeeq{eq:jtot_somm}, can be used for quick signal predictions to a reasonable precision, without resorting to a complex numerical machinery.

To summarize, independently of the scaling relations, we see that the overall Sommerfeld effect induced by subhalos does not change the host target hierarchy in the $s$-wave annihilation case, because it is driven by the minimal subhalo mass $\mumin$, taken the same for all host halos and all values of $\epsilon_\phi$. Decreasing $\mumin$ would simply enhance the signal by the same amount for all host halos (though one should keep in mind the unitarity limit on resonances, set by $\mutunit$ if $\mumin<\mutunit$). In contrast, in the $p$-wave case, we see that the subhalo contribution to the global Sommerfeld enhancement is relatively stronger for lighter host halos (see the right panels of \citefig{fig:sub_Jsub}), and could therefore potentially invert the hierarchy of the Sommerfeld-free $p$-wave signal set by the (squared) dispersion velocities of the most massive subhalos. This is due to the fact that while further suppressing the overall Sommerfeld-free signal, subhalos now act as extra enhancement factors due to their smaller dispersion velocities, making the Sommerfeld-enhanced to Sommerfeld-free ratio much more contrasted than in the $s$-wave case.

\subsubsection{Sommerfeld-enhanced subhalo boost factor}
\label{sssec:S_boost}

\begin{figure}[t!]
\centering
\includegraphics[width=0.99\linewidth]{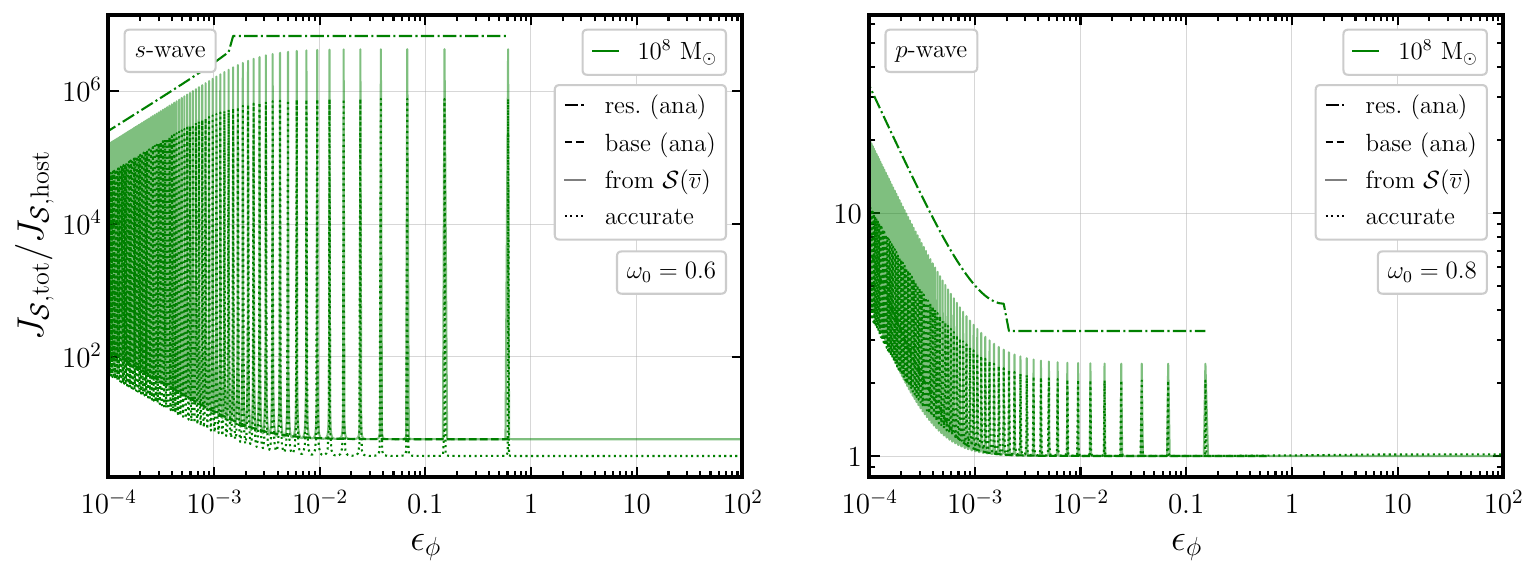} \hfill
\includegraphics[width=0.99\linewidth]{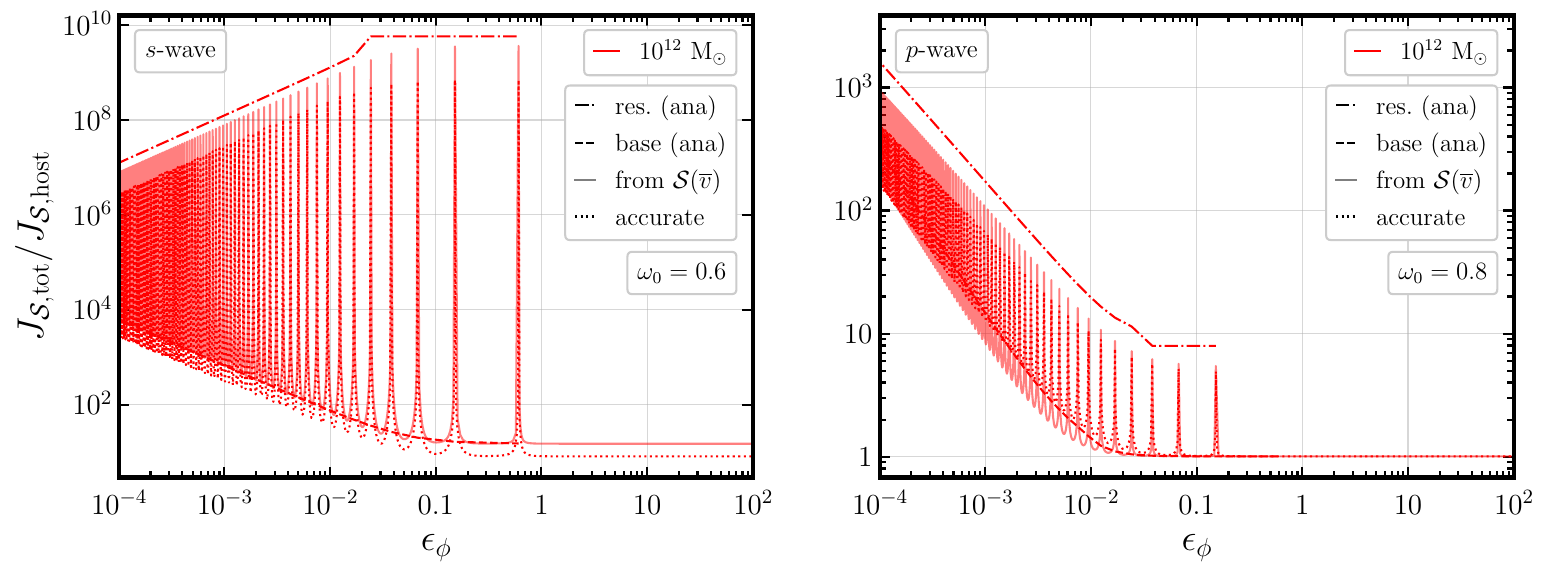} \hfill
\includegraphics[width=0.99\linewidth]{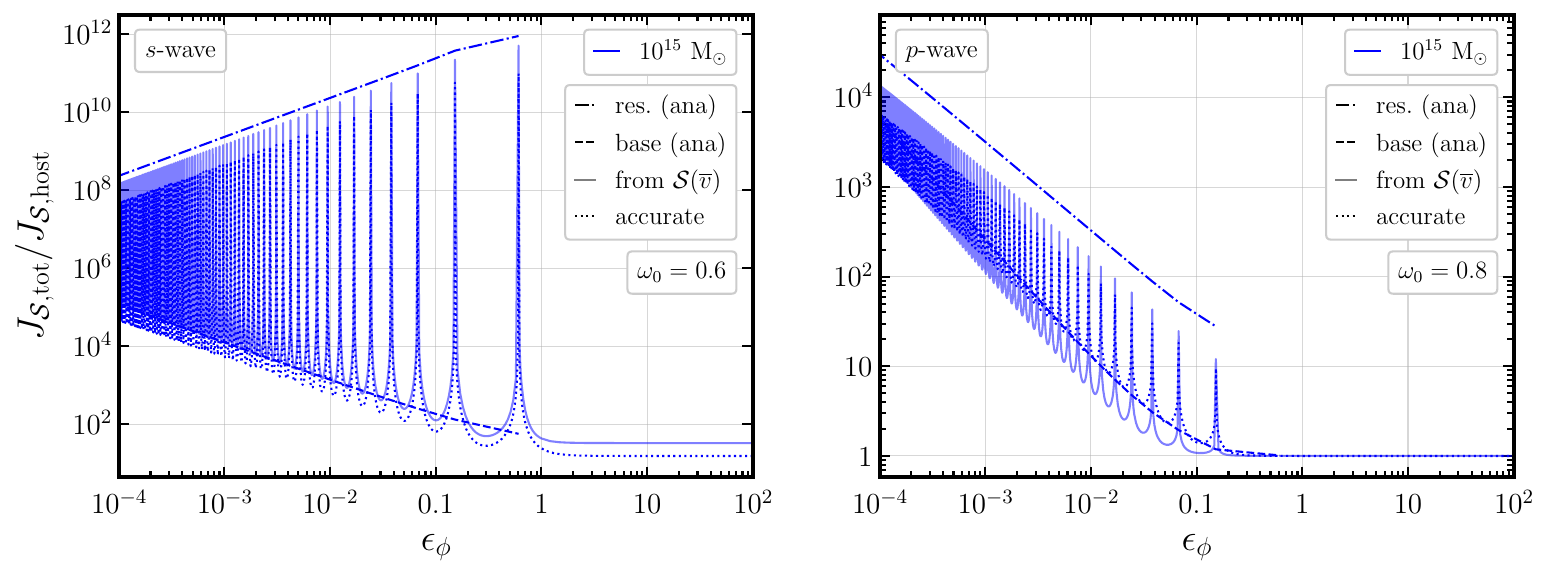}
\caption{{\bf Left column:} Subhalo boost factor for $s$-wave Sommerfeld-enhanced DM annihilation, calculated for three typical host halos of masses $10^8$, $10^{12}$, and $10^{15}~\Msun$, from top to bottom panels. The different curves show (i) the exact numerical results (dotted lines), (ii) an approximation in which we calculate the mass integral numerically by taking the exact Sommerfeld formula but evaluated at the characteristic velocity of (sub)halos (plain curves), and (iii) the integrated analytical ansatz, which is reported for both the baseline and the peak amplitude (dot-dashed curves). {\bf Right column:} Same for $p$-wave annihilation.}
\label{fig:sub_boost}
\end{figure}

We can now determine the overall Sommerfeld-corrected subhalo boost factor for a host of mass $M_{\rm host}$. It may be written as [see \citeeq{eq:boost_Sfree}]
\ben
{\cal B}_{\cal S} \simeq 1 +  \frac{J_{{\cal S},{\rm sub}}(\epsilon_\phi)}{J_{{\cal S},{\rm host}}(\epsilon_\phi)}\,,
\een
where the Sommerfeld-enhanced contribution of the host in the denominator is given in \citeeq{eq:JS_halo}, while the Sommerfeld-enhanced contribution of the subhalo population is given in \citeeq{eq:jtot_somm}. Here, the difficulty comes from the fact that different Sommerfeld regimes come about at different values of $\epsilon_\phi$ depending on the (sub)halo masses (including the host halo). The clearest way to understand the net impact of subhalos in Sommerfeld-enhanced scenarios is to separate the discussion in terms of the different Sommerfeld regimes for the host halo.

We order the different Sommerfeld regimes of the host halo by varying the reduced Bohr radius $\epsilon_\phi$ from large to small values. Therefore, we first discuss the saturation regime, and then the Coulomb regime. Note that for host halo masses of $10^{15}$, $10^{12}$, and $10^{8}~\Msun$, the transition between these regimes occurs around $\epsilon_\phi^{\rm sat}\sim 0.5$, $5\times 10^{-2}$, $3\times 10^{-3}$, respectively. A concrete illustration is given in \citefig{fig:sub_boost}, where we see the Sommerfeld-corrected subhalo boost factors computed for the $s$-wave ($p$-wave) annihilation case in the left-column (respectively right-column) panels, and for different host halo masses.

\paragraph{Saturation regime of the host halo}

The saturation regime of the host halo corresponds to values of $\epsilon_\phi>\epsilon_\phi^{\rm sat}(M_{\rm host})$, or equivalently $\mtsat(\epsilon_\phi)>M_{\rm host}$---see Eqs.~\eqref{eq:def_eps_star} and \eqref{eq:msat}. Since subhalos are all lighter than the host halo, they are also all in the saturation regime. The boost factor can then be written as, assuming that $\mmin<\mmax<M_{\rm host}<\mtsat(\epsilon_\phi)$:
\begin{tcolorbox}
\ben
\label{eq:boost_sat_sub}
{\cal B}_{\cal S}^{\rm sat} -1 &\simeq&
\frac{J_{{\cal S},{\rm sub}}^\text{sat}(\epsilon_\phi)}{J_{{\cal S},{\rm host}}^\text{sat}(\epsilon_\phi) } =
\frac{\gamma}{\textcolor{blue}{(1-p)\alphasat}} \frac{A_{0,{\rm host}}^{2}}{A_{0,{\rm sub}}^{2}}\,\left\{\frac{\mumin}{\muhost}\right\}^{-\textcolor{blue}{\frac{(2-p)}{2}\alphasat}} \,
\left\{\frac{\mumax}{\muhost}\right\}^{-\textcolor{blue}{\frac{p}{2}\alphasat}} \,,
\een
\end{tcolorbox}
\noindent
where the multiple appearance of $p$ here is simply a trick to account for the change of sign in the boost mass spectral index $\alphasat$ between the $s$- and $p$-wave cases in our specific choice of parameters, which makes either $\mumin$ or ${\rm min}(\mumax,\mutsat)=\mumax\propto\muhost$ dominate the mass function integral. From this equation, we clearly understand why in the saturation regime of the host halo (right parts of panels in \citefig{fig:sub_boost}), the baselines of both the $s$-wave and $p$-wave boost factors remain constant: this is due to the fact that they are independent of $\epsilon_\phi$. The $s$-wave one has its amplitude $\propto \muhost/\mumin$, though hindered by a small power index $\alphasat\sim 0.1$. In contrast, the boost factor amplitude for $p$-wave annihilation is vanishingly small because $(\mumax/\muhost)^{-\alphasat}<1$, as a consequence of $\alphasat<0$ in that case (asymptotically similar to the Sommerfeld-free case).

On resonances, this becomes
\begin{tcolorbox}
\ben
\label{eq:boost_res_sub}
{\cal B}_{\cal S}^{\rm res} -1 \simeq  \frac{J_{{\cal S},{\rm sub}}^\text{res}(\epsilon_\phi)}{J_{{\cal S},{\rm host}}^\text{res}(\epsilon_\phi) }=\frac{\gamma}{\alphares} \frac{A_{0,{\rm host}}^{2}}{A_{0,{\rm sub}}^{2}}\,\left\{\frac{\textcolor{blue}{\mumin}}{\muhost}\right\}^{-\alphares} \,.
\een
\end{tcolorbox}
\noindent
The minimal subhalo mass $\mumin$ is featured in blue to keep in mind that it should be replaced by $\mutunit$ when $\mumin<\mutunit$ (this is the case in our template examples, but this is not generic).
Interestingly, when both the host halo and its subhalos sit on resonances ($m<\mtsat$), the boost factor does not depend on Sommerfeld parameters, and remains flat as a function of $\epsilon_\phi$. This can be seen from all panels of \citefig{fig:sub_boost} by inspecting the right-hand-side peaks (more peaks are concerned as the host halo mass decreases, as the latter remains longer in the saturation regime). Not visible in this formula but also theoretically important, there is a formal difference between predictions of the $s$- and $p$-wave boost factors on resonant peaks. In the latter case, the Sommerfeld-enhanced annihilation cross section does not depend on DM velocity (the $v^2$ suppression is canceled out by the $1/v^2$ enhancement), which in principle reduces the potential error associated with the approximation of trading the phase-space average of the effective Sommerfeld factor for a its local expression evaluated at an average characteristic velocity.

\paragraph{Coulomb regime of the host halo}

When the host halo is in the Coulomb regime, then subhalos can themselves be either in the Coulomb regime or in the saturation regime (which includes resonances). Therefore, we need to combine all possibilities, which depend on whether $\mtsat$ lies within the subhalo mass range $[\mmin,\mmax]$ or not. The corresponding expression for the boost factor is slightly more involved:
\ben
\label{eq:boost_coul_sub}
    {\cal B}_{\cal S}^{\rm Coul} -1 &\simeq& \frac{J_{{\cal S},{\rm sub}}(\epsilon_\phi)}{J_{{\cal S},{\rm host}}^\text{Coul}(\epsilon_\phi) } = \gamma\,\mu_{\rm host}^{\alphacoul}\,\Bigg\{
    \frac{\theta\left(\mumax-\mutsat \right)}{\alphacoul} \left[{\rm max}\left(\textcolor{blue}{\mutsat,\mumin}\right)\right]^{-\alphacoul} \\
    &+ & 
    \theta(\mutsat-\mumin)\frac{S_1}{\textcolor{blue}{(1-p)\alphasat}}\,\left\{ \frac{\mutmax}{\mutsat} \right\}^{\nu(p+1)}\,\mumin^{-\textcolor{blue}{\frac{(2-p)}{2}\alphasat}}\,\left[{\rm min}\left(\mutsat,\mumax\right)\right]^{-\textcolor{blue}{\frac{p}{2}\alphasat}}  \nn\\
    &+ & 
    \sum_{n=1+\frac{p}{2}}\delta_{\epsilon_\phi/\{\epsilon_\phi^{{\rm res},n}\}}\,\theta(\mutsat-\mumin)\,\frac{S_0^{\rm res}}{\alphares S_0}\,\left\{ \frac{\mutmax}{\mutsat} \right\}^{\nu(p+1)}\,\textcolor{blue}{\mumin}^{-\alphares}\,\mutsat^{2\nu}
    \Bigg\}\,.\nn
\een
Assuming $\mmin<\mtsat<\mmax$, this expression simplifies to:
\begin{tcolorbox}
\ben
\label{eq:boost_coul_sub_approx}
    {\cal B}_{\cal S}^{\rm Coul} -1 &\simeq& \gamma\,\mu_{\rm host}^{\alphacoul}\,\Bigg\{
    \frac{\textcolor{blue}{\mutsat}^{-\alphacoul}}{\alphacoul}  \,\textcolor{red}{\propto} \,\epsilon_\phi^{-\frac{\alphacoul}{\nu}}\\
    +& &
    \frac{S_1}{\textcolor{blue}{(1-p)\alphasat}}\,\left\{ \frac{\mutmax}{\mutsat} \right\}^{\nu(p+1)}\,\mumin^{-\textcolor{blue}{\frac{(2-p)}{2}\alphasat}}\,\mutsat^{-\textcolor{blue}{\frac{p}{2}\alphasat}} \,\textcolor{red}{\propto}\, \epsilon_\phi^{-(p+1)-\textcolor{blue}{\frac{\alphasat p}{2\nu}}}\,\mumin^{-\textcolor{blue}{\frac{(2-p)}{2}\alphasat}} \nn\\
    +& &
    \sum_{n=1+\frac{p}{2}}\delta_{\epsilon_\phi/\{\epsilon_\phi^{{\rm res},n}\}}\,\frac{S_0^{\rm res}}{\alphares S_0}\,\left\{ \frac{\mutmax}{\mutsat} \right\}^{\nu(p+1)}\,\textcolor{blue}{\mumin}^{-\alphares}\,\mutsat^{2\nu}\,\textcolor{red}{\propto}\, \epsilon_\phi^{1-p}\,\textcolor{blue}{\mumin}^{-\alphares}
    \Bigg\}\,.\nn
\een
\end{tcolorbox}
\noindent
From the top to bottom lines, we have the Coulomb/Coulomb, saturation/Coulomb, and resonant peaks/Coulomb terms. The scaling in $\epsilon_\phi$ and $\mumin$ is made explicit at the end of each line, for convenience.

Quite generically, we first see that when the host halo lies in the Coulomb regime, the subhalo boost factor is $\propto\muhost^{\alphacoul}$, with $\alphacoul\sim 0.57 >0$ here. Consequently, the boost factor increases with the host halo mass for both $s$- and $p$-wave Sommerfeld-enhanced annihilation processes, a result similar to the Sommerfeld-free result for the $s$-wave annihilation [see \citeeq{eq:boost_swave_approx}].

The Coulomb/Coulomb term is not visible \citefig{fig:sub_boost}, and would asymptotically take over in the extreme left parts of the panels at lower values of $\epsilon_\phi<\epsilon_\phi^{\rm sat}(\mumin)$. It would then freeze in as $\mutsat\sim\mumin$, and remain constant, $\propto \mumin^{-\alphacoul}$, similar to the Sommerfeld-free boost factor for $s$-wave annihilation processes.

The saturation/Coulomb term characterizes the baseline of the boost factor over a large range of $\epsilon_\phi\in[\epsilon_\phi^{\rm sat}(\mumin),\epsilon_\phi^{\rm sat}(\muhost)]$. In the $s$-wave case, it scales like $\propto \muhost^{\alphacoul}\,\epsilon_\phi^{-1}\,\mumin^{-\alphasat}$. Except for the explicit host halo mass dependence, which sets an absolute hierarchy, we see that the scaling in $\mumin$ and $\epsilon_\phi$ is the same for all halos. The only implicit difference is that the onset of the subhalo saturation regime at $\epsilon_\phi^{\rm sat}(\muhost)$ shifts to lower values as $\muhost$ decreases. We can therefore understand why the boost factor behaves the same, \ie~it increases linearly $\propto \epsilon_\phi^{-1}$ as $\epsilon_\phi$ decreases, while with some increasing delay as $\muhost$ decreases. Besides, the minimal subhalo mass $\mumin$ participates in setting the overall amplitude of the saturation baseline of the subhalo boost factor, which increases as $\mumin$ decreases. In the $p$-wave case, the baseline scales like $\propto \muhost^{\alphacoul}\, \epsilon_\phi^{-3-\frac{\alphasat}{\nu}}\sim \muhost^{\alphacoul}\, \epsilon_\phi^{-1.3}$, which is independent of $\mumin$ (as long as $\mumin<\mutsat$). The slope in $\epsilon_\phi$ is therefore slightly steeper than in the $s$-wave case, but the delay in the onset of the saturation regime as $\epsilon_\phi$ decreases is the same.

Finally, the resonant peaks/Coulomb term, which characterizes the amplitude of the subhalo boost factor on resonant peaks, scales like $\propto \muhost^{\alphacoul}\, \epsilon_\phi^{1-p}\,{\rm max}(\mumin,\mutunit)^{-\alphares}$. In addition to the host-halo mass hierarchy set by $\muhost^{\alphacoul}$, we first see that the amplitude of the boost is also affected by $\mumin$ (or $\mutunit$), but more for $s$-wave ($\alphares\sim 0.76$) than for $p$-wave processes ($\alphares\sim 0.1$). In contrast, we also see that the dependence in $\epsilon_\phi$ is inverted from the $s$- to $p$-wave case, with a scaling $\propto \epsilon_\phi$ in the former case, but $\propto \epsilon_\phi^{-1}$ in the latter case. This explains why the relative amplitude of peaks with respect to the baseline of the subhalo boost decreases faster, $\propto \epsilon_\phi^{2}$, in the $s$-wave configuration than in the $p$-wave one, for which the relative decrease is accordingly $\propto \epsilon_\phi^{0.3}$. This behavior matches exactly the peak-to-baseline ratio of the subhalo signal derived in \citeeq{eq:sub_peak_to_base}, which means that the signal itself is completely driven by subhalos.

From the plots of \citefig{fig:sub_boost}, we see that our semi-analytical approximations (numerical mass integrals of analytical expressions) come with significant errors, but get the scaling relations and the orders of magnitude correct. Note that the numerical errors are more exacerbated in the subhalo boost factor than in individual signals because it is a ratio that combines quite different mass scales ($\muhost\gg \mumin$).

\subsubsection{Absolute Sommerfeld-enhanced subhalo boost factor}

As a last useful result which may help re-evaluating the hierarchy of targets, we calculate the Sommerfeld-enhanced subhalo boost factor with respect to the Sommerfeld-free signal of the host halo. We recall that we keep the subhalo mass slope $\alpha$ ``fixed by theory'', which determines the signs of the Sommerfeld mass slopes $\alphacoul$, $\alphasat$, and $\alphares$. In that case, we get:
\ben
\label{eq:boost_sub_S_noS}
{\cal B}_{{\cal S}/\text{no-}{\cal S}}-1 &\simeq& \frac{J_{{\cal S},{\rm sub}}(\epsilon_\phi)}{J_{\rm host}(\muhost) }\\
&=&
\gamma\,\muhost^{\alphasat} \Bigg\{ \theta(\mumax-\mutsat)\,\frac{S_0}{\alphacoul}\mutmax^{\nu(p+1)}\left[{\rm max}(\textcolor{blue}{\mutsat,\mumin})\right]^{-\alphacoul}\nn\\
&+& \theta(\mutsat-\mumin)\,\left\{\frac{\mutmax}{\mutsat} \right\}^{\nu(p+1)}\times\nn\\
&& \Big\{ \frac{S_0S_1}{\textcolor{blue}{(1-p)\alphasat}}\,\mumin^{-\textcolor{blue}{\frac{(2-p)}{2}\alphasat}}\,\left[ {\rm min}(\mutsat,\mumax)\right]^{-\textcolor{blue}{\frac{p}{2}\alphasat}}\nn\\
&&+\sum_{n=1+\frac{p}{2}}\delta_{\epsilon_\phi/\{\epsilon_\phi^{{\rm res},n}\}} \,\frac{S_0^{\rm res}}{\alphares}\,\mutsat^{2\,\nu}\,\textcolor{blue}{\mumin}^{-\alphares} \Big\}
\Bigg\}\,.\nn
\een

Assuming $\mmin<\mtsat<\mmax$, this simplifies as follows:
\begin{tcolorbox}
\ben
\label{eq:boost_sub_S_noS_approx}
{\cal B}_{{\cal S}/\text{no-}{\cal S}}-1 &\simeq& \gamma\,\muhost^{\alphasat} \Bigg\{ \frac{S_0}{\alphacoul}\mutmax^{\nu(p+1)}\,\textcolor{blue}{\mutsat}^{-\alphacoul}\\
&+& \left\{\frac{\mutmax}{\mutsat} \right\}^{\nu(p+1)}\times
\Big\{ \frac{S_0S_1}{\textcolor{blue}{(1-p)\alphasat}}\,\mumin^{-\textcolor{blue}{\frac{(2-p)}{2}\alphasat}}\,\mutsat^{-\textcolor{blue}{\frac{p}{2}\alphasat}}\nn\\
&& +\sum_{n=1+\frac{p}{2}}\delta_{\epsilon_\phi/\{\epsilon_\phi^{{\rm res},n}\}} \,\frac{S_0^{\rm res}}{\alphares}\,\mutsat^{2\,\nu}\,\textcolor{blue}{\mumin}^{-\alphares} \Big\}
\Bigg\}\,.\nn
\een
\end{tcolorbox}
\noindent
From this equation, we see the crucial roles played by both $\mmin$ and $\epsilon_\phi$ in the $s$-wave case ($p=0$, $\alphasat\sim 0.1$) to set the boost amplitude. We also see that for $s$-wave processes, the global factor of $\muhost^{\alphasat}$ makes the boost factor increase as $\muhost$ increases, which exacerbates the signal hierarchy between targets as function of their mass. In contrast, the $p$-wave boost factor ($p=2$, $\alphasat<0$) is almost entirely fixed by $\epsilon_\phi$ through $\mutsat$, as long as $\mutsat>\mumin$. Besides, the global factor of $\muhost^{\alphasat}$ makes the boost factor decrease as $\muhost$ increases for $p$-wave processes ($\alphasat\sim -0.57$), which now tends to invert the signal hierarchy between targets as function of their masses. On resonance peaks, it is again $\mumin$ (or $\mutunit$ if $\mumin<\mutunit$) that sets the amplitude, with a stronger impact in the $s$-wave ($\alphares\sim 0.57$) than in the $p$-wave case ($\alphares\sim 0.1$).

\begin{figure}[t!]
	\centering
	\includegraphics[width=0.99\linewidth]{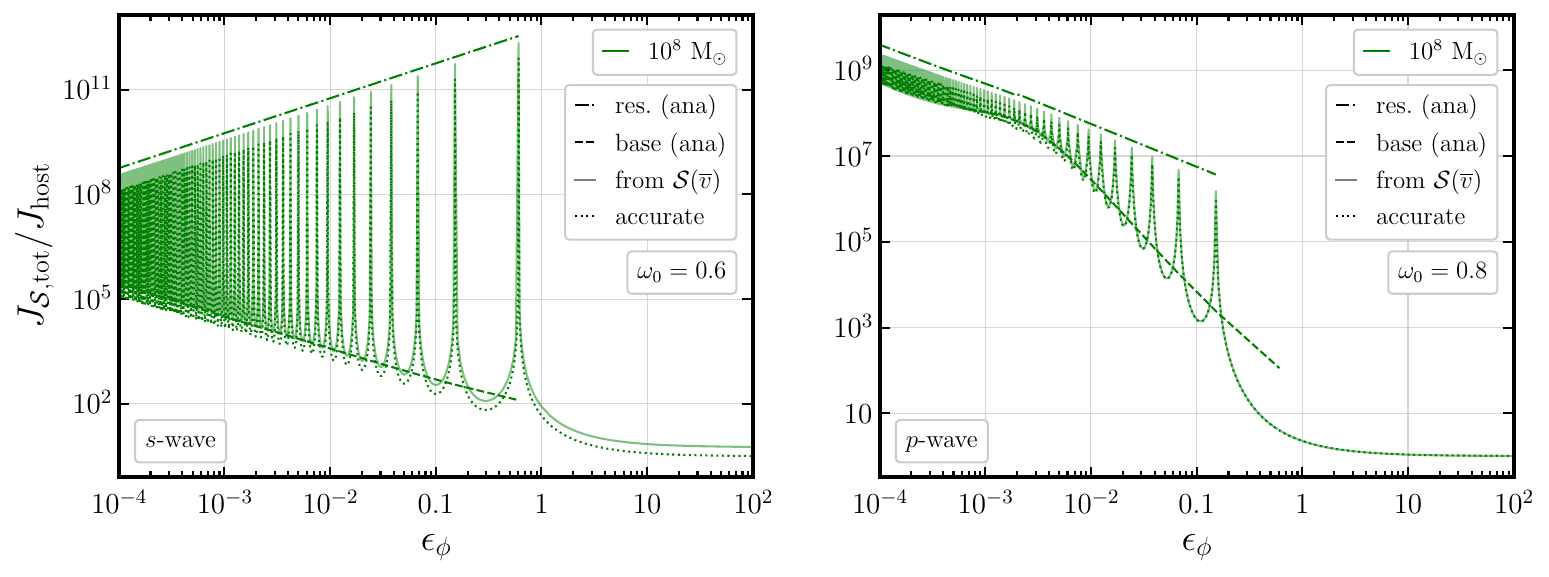} \hfill
	\includegraphics[width=0.99\linewidth]{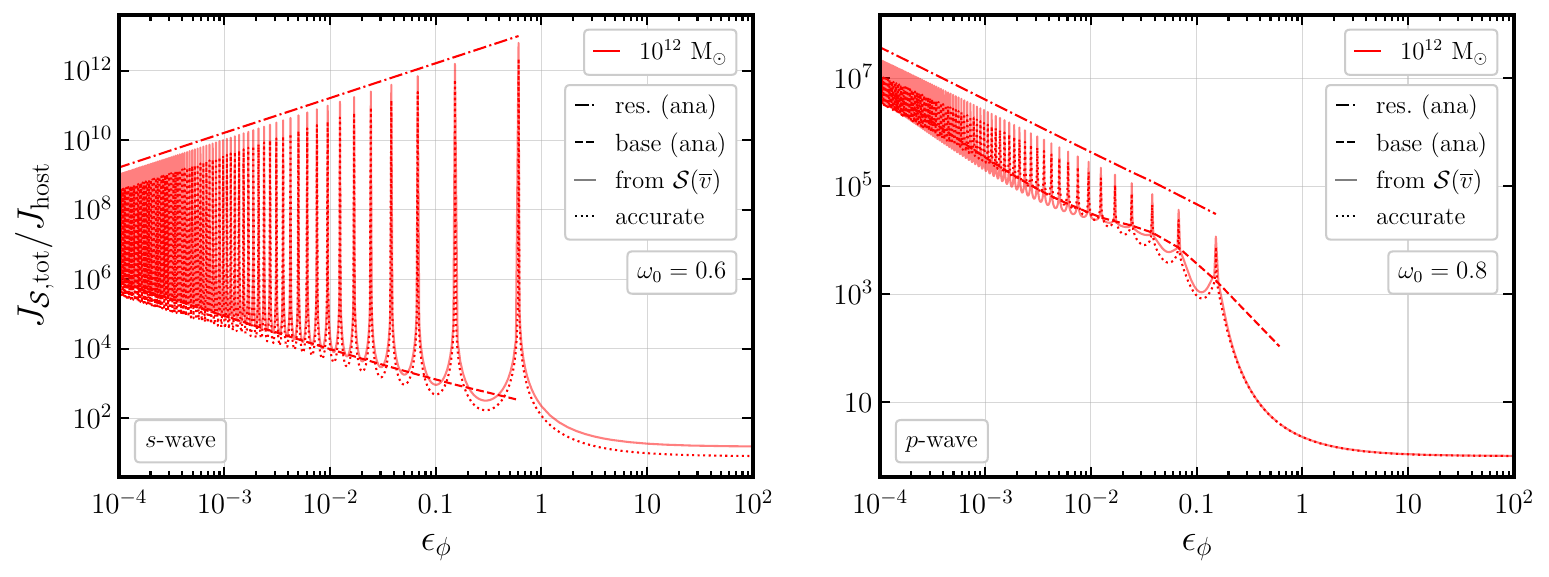} \hfill
	\includegraphics[width=0.99\linewidth]{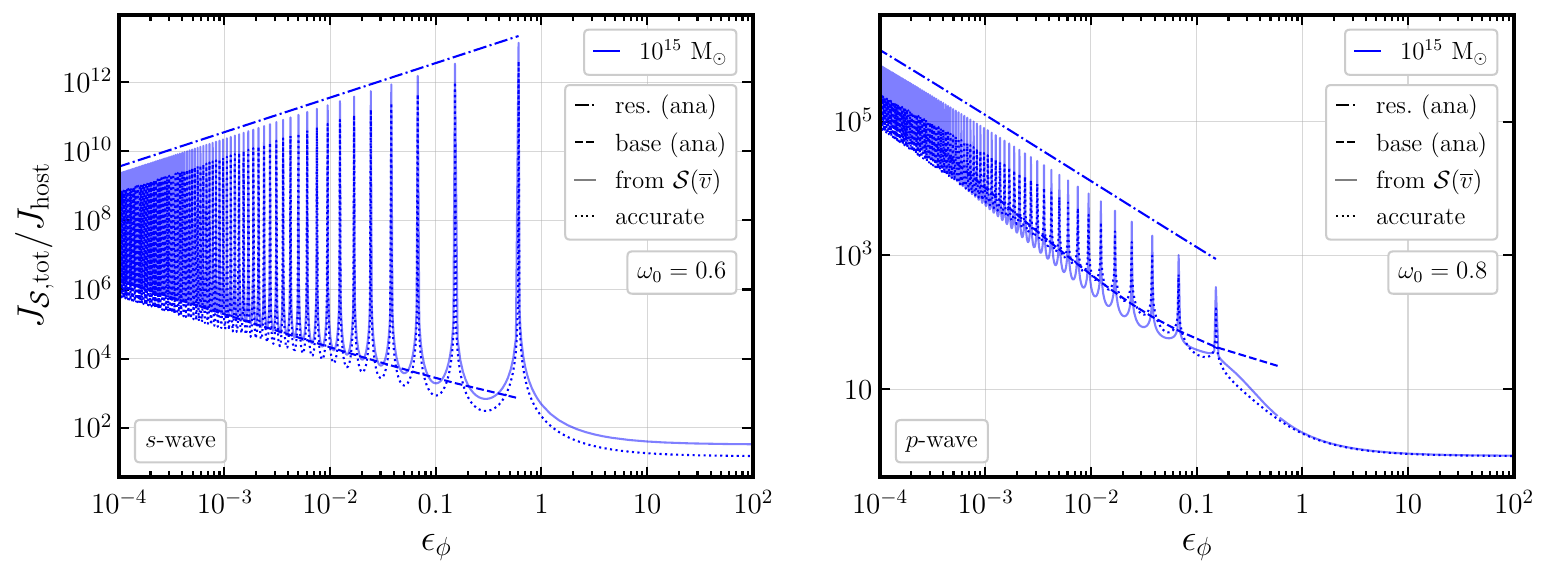}
	\caption{{\bf Left column:} Absolute subhalo boost factor for $s$-wave Sommerfeld-enhanced DM annihilation, calculated for three typical host halos of masses $10^8$, $10^{12}$, and $10^{15}~\Msun$, from top to bottom panels. The different curves show (i) the exact numerical results (dotted lines), (ii) an approximation in which we calculate the mass integral numerically by taking the exact Sommerfeld formula but evaluated at the characteristic velocity of (sub)halos (plain curves), and (iii) the integrated analytical ansatz, which is reported for both the baseline and the peak amplitude (dot-dashed curves). {\bf Right column:} Same for $p$-wave annihilation.}
	\label{fig:abs_boost}
\end{figure}

In \citefig{fig:abs_boost}, we display our different results for the absolute boost factor introduced just above, for the different reference host halo masses. The full numerical calculation results appear as dotted curves, the mass integral performed over the exact Sommerfeld enhancement factor evaluated at the characteristic speeds of subhalos as solid curves, and the integrated ansatz envelope as dot-dashed curves. There is a reasonable agreement between the analytical approximation and the full numerical results. The left-column panels show the results for $s$-wave annihilation, while the right-column panels show our results for $p$-wave annihilation (the latter are quite similar to the right panels of \citefig{fig:sub_Jsub}, because $J_{{\cal S},{\rm sub}}(\epsilon_\phi)/J_{\rm host}\simeq J_{{\cal S},{\rm sub}}(\epsilon_\phi)/J_{\rm tot}$ in the $p$-wave case). We do not discuss longer the former, which exhibit no surprise, but we emphasize the inverted hierarchy now occurring in the latter, where it is evident that the absolute boost factor can then be much larger for less massive host halos (from bottom to top right panels). We have already explained why above, but these plots allow us to be slightly more quantitative. Let us for instance compare a dwarf galaxy-like host halo of mass $m_1\sim10^8\,\Msun$ and a galaxy cluster-like host halo of mass $m_2\sim 10^{15}\,\Msun$, both located at the same distance. The $p$-wave suppression factor induces an extra relative reduction of $\sim (m_1/m_2)^{2\nu}\sim (m_1/m_2)^{2/3}\sim 10^{-3.7}$, not favorable to the lightest host halo. As soon as both host halos have entered the Coulomb regime and have their subhalos contributing in the saturation regime, then a boost factor applies with an inverse balance, giving a boost factor ratio of $\sim (m_1/m_2)^{\alpha_{\rm sat}}\sim (m_1/m_2)^{-0.57}\sim 10^{4}$ (in perfect agreement with the numerical results in the plots), which fully compensates for the initial Sommerfeld-free $p$-wave penalty. Such a compensation might actually change the initial hierarchy between targets of different masses, depending on their respective distances to the observer.

\subsubsection{Caveats}
Before summarizing and concluding, it is useful to mention a few caveats.
\bi
\item[(i)] The calculations of the $J$-factors (enhanced or not) presented in this paper assume the integration of a subhalo population over an entire target object, \ie,~within its virial radius. If the real target halo has its tidal radius significantly smaller than its virial radius, or if the angular size used to perform the signal analysis is significantly smaller than the angular extension of the target halo, then although the host $J$-factor may not change significantly, the subhalo contribution (hence the subhalo boost factor) can be more strongly affected because subhalos dominate the overall mass profile in the outskirts of their host halo (they are subject to gravitational tides in the central regions, where they can experience strong mass losses and even disruption). Our results likely overestimate the contribution of the subhalo population to the signal in that case. A possible way out is to rescale our results for subhalos by the tidal-to-virial (or contained-to-virial) mass ratio of the host halo, assuming the missing or lost mass is mostly made of subhalos. This may be particularly relevant to dwarf satellite galaxies that orbit within our Milky Way, and also to targets with large angular extensions in the sky such as galaxy clusters.
\item[(ii)] We have used fixed values for the subhalo mass slope $\alpha$ and the free-streaming cutoff mass $\mmin$. We have motivated the former from theoretical arguments in the framework of concordance cosmology (see \citeapp{app:subhalo_model}), and the latter by uncorrelating DM self-interactions from DM-baryons interactions, but all this does not come without uncertainties. If the primordial power spectrum departs from almost scale invariant and exhibits extra features on small scales, then our predictions would be strongly affected. It is a priori possible to adapt our analysis by starting from another subhalo mass function inferred from a modified primordial spectrum, but explaining what would happen in different specific scenarios goes beyond the scope of this paper. One can easily guess, though, the impact if the only change is in the mass slope $\alpha$, or even if the mass function exhibits spectral breaks or bumps. On the other hand, if $\mmin$ becomes related to the intrinsic Sommerfeld parameters, then it is a priori easy to determine the consequences from our results.
\item[(iii)] We have limited our study to NFW profiles for both the host halo and the entire subhalo population. Changing the shape of profiles either for the host halo or its subhalos would mostly change our analytical approximation for the $J$-factor of \citeeq{eq:jm}, which would propagate in subhalo-to-host ratios and then affect our predictions. However, we do not expect a significant change in the overall subhalo signal, because subhalos are mostly characterized by cuspy profiles and are essentially not subject to baryonic feedback. Anyway, modifications of inner shapes are in principle not difficult to account for, even analytically. For changes to Einasto profiles \cite{Einasto1965,NavarroEtAl2004,MerrittEtAl2006}, see, \eg, ~SL17. For other changes, one simply needs to feed our results with another $J$-factor-to-mass relation [see \citeeq{eq:jm}]. 
\item[(iv)] We do not include sub-subhalos (nor any subsequent sublayers), which would a priori tend to further increase the Sommerfeld enhancement. Indeed, our subhalo mass function introduced in \citeapp{app:subhalo_model} only contains the first generation of objects (those accreted directly into the host halo). It is actually not difficult to incorporate all layers in the model, which would tend to sharpen the mass function by increasing the effective mass index $\alpha$ \cite{Facchinetti2021}, and further proceed with the hard-sphere approximation discussed above or in the appendix. This goes beyond the scope of this paper, and would certainly add extra theoretical uncertainties related to tidal effects internal to the different layers of subhalos. A more detailed semi-analytical subhalo population study is in preparation \cite{FacchinettiEtAl2022a}, where it is shown that these additional layers mostly shape the lower part of the subhalo mass range (see also ref.~\cite{JiangEtAl2016} for another merger-tree inferred subhalo population example).
\ei

\section{Summary and conclusion}
\label{sec:concl}

In this paper, we have reviewed quite in detail how the presence of DM subhalos affects the gamma-ray signal amplitude predictions in a scenario in which DM self-interacts through long-range interactions, leading to the Sommerfeld enhancement of the annihilation cross section. We have proposed a simplifying analytical ansatz in \citeeq{eq:S_ansatz_tot_v} to incorporate the rather complex Sommerfeld enhancement factor in the signal predictions, showing that calculations can then be performed fully analytically. This helps better understand the dependencies of signal predictions in terms of the main physical parameters. These parameters are the Sommerfeld-enhancement parameters on the one hand, dictated by particle physics only, and the subhalo parameters on the other hand, dictated both by cosmology (DM power spectrum and structure formation) and particle physics (minimal subhalo mass). We adopted a simplifying description for the former by means of a DM fine-structure constant $\alpha_\chi$, that we have kept fixed to $0.01$ throughout the paper, and of the reduced Bohr radius $\epsilon_\phi$, which then characterizes the mediator-to-DM mass ratio (the Sommerfeld-enhancement regime typically corresponds to $0<\epsilon_\phi<1$, triggered at velocities $v<\pi\alpha_\chi$). Although it is formally DM velocity-dependent and related to local interactions, we have shown that an averaged Sommerfeld factor could be expressed at the level of a full DM halo of virial mass $m$ by means of the corresponding characteristic dispersion velocity $\overline{v}(m)$, and that $\epsilon_\phi$ could be turned into a transition velocity $\vsat(\epsilon_\phi)$ below which halos transit from the Coulomb regime to the saturation regime of the Sommerfeld enhancement. This global expression of the Sommerfeld effect allowed us to perform the whole chain of calculation fully analytically up to the gamma-ray signal amplitude in terms of $J$-factors. Our main results for the subhalo population signal, given a host halo mass, are summarized in Eqs.~\eqref{eq:jtot_somm}-\eqref{eq:JSsub}, to be compared with the Sommerfeld-free results in \citeeq{eq:JSfreesub}. They can also be expressed as a Sommerfeld-enhanced subhalo boost factor with respect to the Sommerfeld-enhanced smooth-halo approximation [see Eqs.~\eqref{eq:boost_sat_sub}-\eqref{eq:boost_coul_sub_approx}], or with respect to the Sommerfeld-free smooth-halo approximation [see Eqs.~\eqref{eq:boost_sub_S_noS}-\eqref{eq:boost_sub_S_noS_approx}]. We have shown that our analytical results are in reasonable agreement with the more accurate numerical calculations (but still in excellent qualitative agreement), by a factor of a few (with respect to amplitudes of several orders of magnitude), and can therefore be used for quick estimates associated with specific targets.

As a general conclusion, we see that the Sommerfeld enhancement exacerbates the subhalo boost factor, and vice versa. This is true not only for $s$-wave processes, for which a subhalo boost factor was already present in the Sommerfeld-free case, but also and more dramatically so for $p$-wave processes, for which subhalos tended to further suppress the signal in the Sommerfeld-free case. In the latter case, this comes from a full compensation of the $v^2$ $p$-wave suppression factor by the velocity-dependent Sommerfeld factor. This may lead to changes in the hierarchy of targets as a function of mass (assuming the same distance to the observer), initially more favorable to bigger host halos in the Sommerfeld-free case, but then conversely to less massive halos in the Sommerfeld-enhanced case. For both $s$- and $p$-wave processes, the enhancement at resonances is phenomenal. For $s$-wave processes, we have boost factors ranging from $\sim 10^8$ for dwarf-like host halos ($10^8\Msun$) up to $\sim 10^{13}$ for galaxy cluster-like host halos ($10^{15}\Msun$), for the first peaks, decreasing like $1/n^2\propto \epsilon_\phi$, where $n$ is the order of the resonance. They are more moderate for $p$-wave annihilation, ranging from a few for a $10^8\Msun$ host halo up to a few tens for a $10^{15}\Msun$ host halo, but increasing like $n^2\propto 1/\epsilon_\phi$ with the order of resonance -- -this still shows that subhalos provide the dominant contribution to the overall signal for Sommerfeld-enhanced $p$-wave annihilation processes, and therefore must be included in the predictions.

There has been lots of studies considering the Sommerfeld enhancement induced by subhalos, \eg~\cite{KuhlenEtAl2009,Bovy2009,LattanziEtAl2009,PieriEtAl2009,SlatyerEtAl2012,ZavalaEtAl2014a,BoddyEtAl2019,RunburgEtAl2021} (see also refs.~\cite{BoddyEtAl2017,BoddyEtAl2018,BoucherEtAl2021,AndoEtAl2021,BoardEtAl2021}). Most of them address the $s$-wave case, and overall, our results are in qualitative agreement with these. There are quantitative differences coming from the different theoretical assumptions or parameters used, but our analytical results can be applied to a wide range of model configurations, and should allow to recover (or complete in overlooked regimes) those of past studies. We are not aware of such a full analytical derivation, especially for $p$-wave annihilation, thus we hope that our study will allow the reader to grasp the very details of the Sommerfeld-enhanced subhalo contributions to gamma-ray signals. These questions are further explored in a companion study \cite{LacroixEtAl2022}, dedicated to a thorough analysis of the combined Sommerfeld and subhalo enhancement effects on concrete target examples.

\acknowledgments
\change{We would like to thank the anonymous referee for her/his constructive and precise comments which helped us improve the presentation of our results}. This work has been supported by funding from the ANR project ANR-18-CE31-0006 ({\em GaDaMa}), from the national CNRS-INSU programs PNHE and PNCG, and from European Union's Horizon 2020 research and innovation program under the Marie Sk\l{}odowska-Curie grant agreement N$^{\rm o}$ 860881-HIDDeN. JPR work is supported by grant SEV-2016-0597-17-2 funded by MCIN/AEI/10.13039/501100011033 and ``ESF Investing in your future''. MASC was also supported by the {\it Atracci\'on de Talento} contracts no. 2016-T1/TIC-1542 and 2020-5A/TIC-19725 granted by the Comunidad de Madrid in Spain. The work of JPR and MASC was additionally supported by the grants PGC2018-095161-B-I00 and CEX2020-001007-S, both funded by MCIN/AEI/10.13039/501100011033 and by ``ERDF A way of making Europe''. GF acknowledges support of the ARC program of the Federation Wallonie-Bruxelles and of the Excellence of Science (EoS) project No. 30820817 - be.h “The H boson gateway to physics beyond the Standard Model”.

\appendix

\section{Short review of the Sommerfeld enhancement}
\label{app:sommerfeld}
In this appendix section, we shortly review the impact of DM self-interaction on DM self-annihilation, which leads to the Sommerfeld enhancement. We consider a phenomenological scenario in which DM particles self-interact through the exchange of a (light) mediator $\phi$ of mass $m_{\phi}$ with coupling $g_{\chi} = \sqrt{4\pi\alpha_{\chi}}$, where $\alpha_{\chi}$ plays the role of a dark fine structure constant. In this approach, attractive self-interactions between non-relativistic DM particles are described by an attractive Yukawa potential,
\ben
V_{\rm Y}(r) = -\alpha_{\chi} \dfrac{\mathrm{e}^{-m_{\phi}r}}{r}\,,
\een
with $r$ the relative distance between two annihilating DM particles. In the absence of self-interaction, \ie~for $\alpha_{\chi} = 0$, the annihilation cross section times relative velocity $(\sigma v_{\rm rel})_{0}$ is computed perturbatively from the short-range annihilation process. However, a long-range Yukawa potential, which encodes multiple exchanges of the light mediator between the two incoming DM particles, can distort the wave function of the corresponding two-body system in a non-perturbative way, leading to Sommerfeld enhancement of the annihilation cross section.\footnote{We restrict ourselves to symmetric DM with attractive interactions, for which the Sommerfeld factor is effectively an enhancement factor.} The Sommerfeld-enhanced cross section is then expressed as in \citeeq{eq:enhanced_cs}, which we simply repeat here \cite{Sommerfeld1931}:
\ben
\sigma v_{\rm rel} = (\sigma v_{\rm rel})_{0} \times {\cal S}_{\ell} \,,
\een
where $v_{\rm rel}$ is the relative speed of DM particles, and the enhancement factor ${\cal S}_{\ell}$ is computed by solving the Schr\"{o}dinger equation for the radial part of the wave function $R_{\ell}(r)$ for the partial wave with angular momentum $\ell$ (e.g.~\cite{Cassel2010,Iengo2009,Slatyer2010}),
\ben
\label{eq:radial_Schroedinger}
\left( -\dfrac{\hbar^{2}}{m_{\chi}} \partial_{r}^{2} - m_{\chi}\dfrac{v^{2}}{c^{2}} + V_{\rm Y}(r) + \dfrac{\hbar^{2}\ell(\ell+1)}{m_{\chi}r^{2}} \right) \chi_{\ell}(r) = 0\,,
\een
where $\chi_{\ell}(r) = r R_{\ell}(r)$ and $v = v_{\rm rel}/2$ the velocity of the incoming DM particles in the center-of-mass frame. \citeeq{eq:radial_Schroedinger} is solved with the boundary conditions that the interaction only leads to outgoing spherical plane waves at infinity, and with $R_{\ell}(r) \propto r^{\ell}$ as $r \rightarrow 0$. Then the Sommerfeld enhancement factor for partial wave $\ell$ reads \cite{Cassel2010,Iengo2009,Slatyer2010}
\ben
{\cal S}_{\ell} = \left| \dfrac{(2 \ell + 1)!!\, \chi_{\ell}^{\ell+1}(0)}{(\ell+1)!\, k^{\ell+1}} \right|^{2}\,,
\een
where $k = m_{\chi}v/\hbar$, and $(2 \ell + 1)!! \equiv (2 \ell + 1)!/(2^{\ell} \ell!)$. The radial function $\chi_{\ell}$ can only be obtained numerically when assuming a Yukawa potential, but a good approximation of the latter is given by the Hulth\'{e}n potential,
\ben
V_{\rm H}(r) = -\dfrac{\alpha_{\chi} m_{\ast} \mathrm{e}^{-m_{\ast}r}}{1-\mathrm{e}^{-m_{\ast}r}}\,,
\een
for $m_{\ast} = (\pi^{2}/6)m_{\phi}$. \new{Note that strictly speaking, the above result is only valid for $s$-wave annihilation ($\ell=0$), as an extra centrifugal term must be added to derive analytical expressions for larger partial-wave expansion modes \cite{Cassel2010}. Accounting for this generalization to $\ell\neq 0$, the radial Schr\"{o}dinger equation can be solved analytically for the Hulth\'{e}n potential, leading to a closed form of the Sommerfeld enhancement factor $S_{\ell}$}:
\ben
\label{eq:Sommerfeld_analytic_Cassel}
{\cal S}_{\ell} = \left| \dfrac{\Gamma(a^{-}) \Gamma(a^{+})}{\Gamma (1 + \ell + 2 {\rm i} \epsilon_{v}/\epsilon_{\phi}^{*})} \dfrac{1}{\ell !} \right|^{2}\,,
\een
where $\epsilon_\phi$ and $\epsilon_v$ have been defined in \citeeq{eq:def_epsilon}. Other parameters are:  $\epsilon_{\phi}^{*} \equiv \pi^{2}\epsilon_{\phi}/6$, $\Gamma$ is the Gamma function, and $a^{\pm} = 1 + \ell + {\rm i} \epsilon_{v}/\epsilon_{\phi}^{*} \left(1 \pm \sqrt{1-\epsilon_{\phi}^{*}/\epsilon_{v}^{2}}\right)$, with a square root to be understood as a complex number.

From this equation, we may derive the relevant expressions for the $s$-wave and $p$-wave annihilation processes, corresponding to $\ell = 0$ and $\ell = 1$, respectively. They are given in \citeeq{eq:Sommerfeld_enhancement_s_wave} and \citeeq{eq:Sommerfeld_enhancement_p_wave}. This already covers a broad variety of underlying particle-physics models.

\section{Building up a semi-analytical subhalo population model}
\label{app:subhalo_model}
Here we provide more technical details as for the modeling of subhalo populations in host halos. This theoretical modeling is improved from ref.~\cite{StrefEtAl2017} (SL17), to which we add a subhalo mass fraction normalization based on first-principle arguments rather than calibrated from cosmological simulation results. We start by rewriting \citeeq{eq:nsub} that describes the differential number density of subhalos,
\ben
\frac{\dd n_{\rm sub}(m,R)}{\dd m} = \frac{\dd^2 N_{\rm sub}}{\dd m\,\dd V} = N_{\rm tot} \,\frac{1}{K_{\rm tidal}}\frac{\dd \overline{\cal P}_V(R)}{\dd V} \int \dd c\, \frac{\dd^2{\cal P}_{c,m}(c,m,R)}{\dd c\,\dd m}\,.
\een
As we shall see below, in the above formulation, the concentration and mass pdfs are actually intricate as a result of tidal effects. Therefore, in contrast to many works, we see an explicit dependence of the mass-concentration pdf (consequently also of the mass function) on the position $R$, which makes the phase space fully intricate. This spatial dependence is induced by tidal stripping effects, which depend on the position of subhalos in the host halo and on its detailed gravitational potential (including all components, DM and baryons). We shall discuss tidal effects in more detail below. Note that in order to interpret $N_{\rm tot}$ as the total number of subhalos in the host, one must have the volume integral of the above equation over the host halo normalized to $N_{\rm tot}$, which constrains the full phase-space integral of the pdfs to be equal to the constant $K_{\rm tidal}$. All this will be more clearly defined below.

Before specifying the pdfs, we can already provide the link between the subhalo number density of \citeeq{eq:nsub} and the associated averaged density profile of \citeeq{eq:rhosub}, which is the tidal mass:
\ben
\label{eq:tidal_mass}
m_{\rm t}(m,c,R) = \frac{4\pi}{3} r_{\rm s}^3 \rho_{0} \left\{ \frac{1}{3}\int_0^{x_{\rm t}}\dd x\,x^2\,f(x) \right\}\,,
\een
where we define the subhalo profile shape $f(x)$ in terms of the dimensionless radius $x\equiv r/r_{\rm s}$ and subhalo scale density $\rho_{0}$ as follows:
\ben
f(x) \equiv \frac{\rho(r)}{\rho_{0}}\,.
\een
The dependence of the tidal mass $m_{\rm t}$ on the virial mass $m$ and concentration $c$ appears indirectly as a dependence on the $r_{\rm s}$ and $\rho_{0}$. The dependence on the radial position $R$ within the host halo is further hidden in the upper bound of the volume integral over $f(x)$, the dimensionless tidal radius $x_{\rm t}=x_{\rm t}(m,c,R)\equiv r_{\rm t}(m,c,R)/r_{\rm s}$---we impose $x_{\rm t}= {\rm min} (x_{\rm t},x_{200})$, such that $m_{\rm t}(m,c,R)\leq m$. Here, $r_{\rm t}$ is the subhalo tidal radius and $r_{\rm s}$ its scale radius, given an inner density profile shape $f(x)$. In the following, we will only consider an NFW profile for subhalos,\footnote{See \cite{StrefEtAl2017} for discussion on the impact on changing the inner subhalo profile).} such that
\ben
\label{eq:f_of_x_NFW}
f(x)=f_{\rm nfw}(x) = x^{-1}(1+x)^{-2}\,.
\een
The tidal radius further depends on the virial mass, concentration, position (somewhat related to accretion time), and can be predicted. Our model actually provides such a prediction, based on a detailed description of both the baryonic and global DM components within the host halo \cite{StrefEtAl2017,HuettenEtAl2019,Facchinetti2021}. A simplification of the model is to consider that the density profile within $x_{\rm t}$ is not significantly affected by gravitational tides, which is a reasonable approximation \cite{StenDelos2019,ErraniEtAl2020,ErraniEtAl2020a} and can further be justified in some cases from adiabatic invariance arguments \cite{Weinberg1994,GnedinEtAl1999}. Trying to describe more precisely the evolution of the inner profile would lead to very little change in our predictions, but would be prohibitive in terms of numerical convergence, since DM subhalos may cover up to $\sim$20 orders of magnitude in mass for galaxy clusters.

A related important ingredient of our subhalo population model is the {\em tidal disruption threshold}, $\epsilon_{\rm t}\geq 0$, which basically allows us to disrupt subhalos with $x_{\rm t}\leq \epsilon_{\rm t}$. This tidal disruption criterion is inspired from studies of tidal disruption performed with dedicated numerical simulations \cite{HayashiEtAl2003}, but might be an oversimplified description of this complex process. Still, it allows us to effectively implement tidal disruption in a very efficient way, and study the impact of either aggressive disruption ($\epsilon_{\rm t}\sim 1$), or subhalos strongly resilient to tidal disruption ($\epsilon_{\rm t}\ll 1$). The recent literature tends to suggest that the latter case is more likely \cite{vandenBoschEtAl2018}. The calculation of the tidal radius $x_{\rm t}$ and the value taken for the disruption threshold $\epsilon_{\rm t}$ are actually key parameters at the origin of the spatial dependence of the mass and concentration pdfs introduced in \citeeq{eq:nsub}.

We now specify the pdfs introduced in \citeeq{eq:nsub}. For the initial spatial distribution, we adopt the hard-sphere approximation and simply assume that should subhalos be hard spheres with a negligible encounter rate, they would simply track the global host gravitational potential (like the bodies of $N$-body simulations), such that:
\ben
\frac{\dd\overline{\cal P}_V(R)}{\dd V} = \frac{\rho_{\rm host}(R)}{M_{\rm host}} \theta(R_{\rm host}-R)\,,
\een
where $R_{\rm host}$ is the radial extent of the host halo, and $M_{\rm host}$ is the total DM mass within $R_{\rm host}$ and allows for normalization to unity over the volume of the host halo. We emphasize that this spatial pdf is {\em not} the actual spatial distribution of the subhalo population, which accounts for tidal stripping and can formally simply be inferred from \citeeq{eq:nsub} as:
\ben
\frac{\dd{\cal P}_V^{\rm actual}(R)}{\dd V} = \frac{n_{\rm sub}}{N_{\rm tot}}\neq \frac{\dd\overline{\cal P}_V(R)}{\dd V} \,.
\een
The difference between the ``initial'' and ``final'' spatial pdf will become more striking after the impact of tidal stripping on the concentration and mass pdfs is discussed. This explains the term ``{\em driving} pdf'' used earlier. Note that if our tidal disruption parameter $\epsilon_{\rm t}\longrightarrow 0$, then the actual spatial distribution tends to the initial one (\ie~the host profile), a trend confirmed by recent work on the resilience of subhalos to tidal effects \cite{GreenEtAl2021}. On the other hand, non-zero values of $\epsilon_{\rm t}$ up to $\sim 0.1$ allow us to recover antibiased spatial distributions found in several past analyses of cosmological simulations \cite{DiemandEtAl2004,DiemandEtAl2007a,SpringelEtAl2008,HanEtAl2016b}, though very likely affected by spurious numerical effects \cite{vandenBoschEtAl2018a,vandenBoschEtAl2018}---we will shortly come back to that below. For completeness, in our numerical study, we will use the following two values:
\ben
\epsilon_{\rm t} =
\begin{cases}
1\;\;\;\text{(fragile subhalos)}\\
0.01\;\;\;\text{(resilient subhalos)}\,,
\end{cases}
\een
with the former very conservatively limiting the number of subhalos, and the latter being more realistic according to recent literature.

We resort to the mass-concentration relation as fitted in ref.~\cite{Sanchez-CondeEtAl2014}, to which we further assign a log-normal pdf of constant width $\sigma_c=0.14\,\log(10)$ (in natural logarithm basis), which stems from analyses of cosmological simulations and associated interpretations \cite{MaccioEtAl2008,PradaEtAl2012,DuttonEtAl2014,Sanchez-CondeEtAl2014}. This pdf, hidden in the mass pdf in \citeeq{eq:nsub} (this will appear explicitly below), is initially universal. We denote this universal initial pdf $\dd\overline{\cal P}_c(c,c_0(m))/\dd c$, where $c_0(m)$ carries the mass dependence and refers to the mass-concentration relation proposed in ref.~\cite{Sanchez-CondeEtAl2014} (this pdf is taken log-normal, normalized to unity within $1\leq c <\infty$---see SL17 for details). The spatial dependence of the evolved concentration pdf is then fully induced by our tidal disruption criterion, according to:
\ben
\frac{\dd{\cal P}_c(c,c_0(m),R)}{\dd c} = \frac{\dd\overline{\cal P}_c(c,c_0(m))}{\dd c}
\times  \theta(x_{\rm t}(m,c,R) - \epsilon_t)\,.
\label{eq:c_ev_pdf}
\een
The main technical difficulty here concerns the calculation of the dimensionless tidal radius, which is detailed in ref.~\cite{StrefEtAl2017}. Note that in addition to being spatial-dependent, ${\cal P}_c$ is no longer normalized to unity because of tidal disruption (unless $\epsilon_{\rm t}=0$), which will actually allow us to predict the total number of surviving subhalos after tidal disruption.

Finally, for the subhalo mass function, we significantly improve over the initial version of the subhalo population model of ref.~\cite{StrefEtAl2017}, which was previously used either with power-law mass functions \cite{StrefEtAl2019,HuettenEtAl2019,CaloreEtAl2019} or with the Sheth-Tormen mass function \cite{FacchinettiEtAl2020}. Here, instead, we fully resort to merger-tree techniques. This semi-analytical approach is still based on the extended Press-Schechter formalism \cite{PressEtAl1974,BondEtAl1991a,LaceyEtAl1993}, which allows us not only to self-consistently incorporate relevant cosmological information,\footnote{We use the most recent Planck cosmological parameters \cite{PlanckCollab2020}.} but also to predict the subhalo mass fraction in host halos of different sizes (from dwarf galaxies to galaxy clusters)---in ref.~\cite{StrefEtAl2017}, the subhalo mass fraction was a tunable free parameter of the model. We perform a calculation similar to the one presented in refs.~\cite{JiangEtAl2016,vandenBoschEtAl2016}, which compares very well with cosmological simulations when artificial tidal disruption is included \cite{vandenBoschEtAl2016,GreenEtAl2021}. We have used the merger-tree algorithm introduced in ref.~\cite{ColeEtAl2000} on purpose, because it only depends on cosmological parameters and is not tuned on cosmological simulations---more details on the model upgrade will be given in subsequent papers \cite{Facchinetti2021,FacchinettiEtAl2022a}. We find that the {\em unevolved} subhalo mass functions for several realizations of merger trees and for different host halos can be very well fitted by the parametric function proposed in ref.~\cite{JiangEtAl2016}:
\ben
\label{eq:dNdm}
\frac{ \dd N(m,M_{\rm host})}{\dd m} = \frac{1}{M_{\rm host}} \left[ \gamma_1\left(\frac{m}{M_{\rm host}}\right)^{-\alpha_1} + \gamma_2\left(\frac{m}{M_{\rm host}}\right)^{-\alpha_2} \right] \exp\left\{ -\beta \left(\frac{m}{M_{\rm host}}\right)^{\zeta} \right\}\,,
\een
with the best-fit parameters $\gamma_1  = 0.014, \gamma_2 = 0.41, \alpha_1 = 1.965, \alpha_2 = 1.57, \beta = 20, \zeta = 3.4$. These parameters very slightly differ from the parameters found in ref.~\cite{JiangEtAl2016}, because they derive from different cosmological inputs and normalization procedure. However, this only leads to order percent differences in terms of global subhalo mass fraction. The mass function just above counts the average number of subhalos accumulated along the history of the host halo per ``bin'' of mass (here assumed to be hard spheres,  i.e. keeping their virial masses after accretion). As in other studies, this leads to an {\em unevolved} effective subhalo mass fraction of $\sim 10$\% in a mass range $m/M_{\rm host}\in[10^{-5},10^{-3}]$ \cite{DiemandEtAl2007a,SpringelEtAl2008,GiocoliEtAl2008,PieriEtAl2011,vandenBoschEtAl2016,AndoEtAl2019,IshiyamaEtAl2020a} (this fraction is calculated by taking subhalos with their virial masses, not their actual tidal masses). Like for the concentration, the {\em unevolved} mass function is universal here, prior to any tidal stripping effect.

Note that despite the rather complex form of \citeeq{eq:dNdm}, the unevolved mass function remains rather close to a single power-law function $\propto (m/m_0)^{-\alpha}$, with $m_0$ an arbitrary normalization and $\alpha\approx \alpha_1 \simeq 1.96$. Therefore, introducing $\mu\equiv m/m_0$ and $\mu_{\rm host}\equiv M_{\rm host}/m_0$, a useful approximation is the following:
\ben
\label{eq:dNdm_approx}
\frac{ \dd N(m,M_{\rm host})}{\dd m} &\simeq& \frac{N_0}{m_0}\,\mu^{-\alpha}\\
\text{with} \;\; N_0 &\equiv& \gamma \,\mu_{\rm host}^{\alpha-1}
= 1.4\times 10^{12(\alpha-1)-2}\,\left\{\frac{\mu_{\rm host}}{10^{12}}\right\}^{\alpha-1}
\approx 4.67\times 10^{9}\,\left\{\frac{\mu_{\rm host}}{10^{12}}\right\}^{0.96}
\,,\nn
\een
where we have used $\gamma=\gamma_1$. This is the approximation we use to get analytical understanding of our numerical results.

For completeness, it is useful to define the total number of subhalos prior to tidal stripping, $N_{\rm tot}^0$, from the considered subhalo mass range $[m_{\rm min},m_{\rm max}]$:
\ben
N_{\rm tot}^0 = N_{\rm tot}^0(m_{\rm min},m_{\rm max},M_{\rm host}) = \int_{m_{\rm min}}^{m_{\rm max}}\dd m \frac{ \dd N(m,M_{\rm host})}{\dd m}\,,
\een
such that we can now fully define the unevolved mass pdf prior to tidal effects:
\ben
\label{eq:dpdm}
\frac{\dd \overline{\cal P}_m(m)}{\dd m} = \frac{1}{N_{\rm tot}^0}\frac{ \dd N(m,M_{\rm host})}{\dd m}\,.
\een
This unevolved pdf is then normalized to unity within the considered subhalo mass range.

The {\em evolved} subhalo mass function, in contrast, accounts for tidal stripping and as a consequence, as emphasized in ref.~\cite{StrefEtAl2017}, becomes spatially dependent. This is again induced by the disruption parameter $\epsilon_{\rm t}$, that depletes the subhalo population according to position, mass, and concentration. In our model, the spatially dependent evolved mass pdf is somewhat entangled with the concentration pdf, but can be formally derived from (in terms of the {\em virial} mass $m$):
\ben
\label{eq:ev_mass_func}
\frac{\dd^2 {\cal P}_{c,m}(c,m,R)}{\dd c\,\dd m} &=& \frac{\dd \overline{\cal P}_m(m)}{\dd m}  \times \frac{\dd{\cal P}_c(c,c_0(m),R)}{\dd c} \\
\Rightarrow
\frac{\dd {\cal P}_m(m,R)}{\dd m} &=& \frac{\dd \overline{\cal P}_m(m)}{\dd m}  \times \int \dd c \frac{\dd{\cal P}_c(c,c_0(m),R)}{\dd c} \,,\nn
\een
where ${\cal P}_c$ is the evolved concentration pdf given in \citeeq{eq:c_ev_pdf}. Since ${\cal P}_c$ is not normalized to unity because of tidal disruption, neither is ${\cal P}_m$ (except if $\epsilon_{\rm t}=0$). As a result, we see explicitly here how the concentration pdf is entangled with the mass function as a consequence of tidal disruption, in the formulation of \citeeq{eq:nsub}.

We have now all the necessary ingredients to determine the total number of subhalos $N_{\rm tot}$ and the normalization constant $K_{\rm tidal}$ introduced in \citeeq{eq:nsub}. Indeed, we have
\ben
K_{\rm tidal} = \int_{V_{\rm host}} \dd V\int \frac{\dd\overline{\cal P}_V(R)}{\dd V} \dd m \frac{\dd {\cal P}_m(m,R)}{\dd m} \leq 1\,,
\een
and
\ben
N_{\rm tot} = N_{\rm tot}^0\times K_{\rm tidal} \leq N_{\rm tot}^0\,.
\een
Note that we have $K_{\rm tidal}\to 1$ and $N_{\rm tot} \to N_{\rm tot}^0$ in the limit $\epsilon_{\rm t}\to 0$, \ie~in the absence of tidal disruption (which does not mean absence of tidal stripping).

For more physical insight on the {\em real} subhalo mass function, it might prove useful to have access to the tidal mass distribution instead of the virial mass distribution. Indeed, virial masses have no physical meaning for subhalos, for which the only true masses are the tidal ones. The actual spatial-dependent tidal mass function is simply given by:
\ben
\label{eq:dpdmt}
\frac{\dd {\cal P}_{m_{\rm t}}(m_{\rm t},R)}{\dd m_{\rm t}} 
 &=&  \int\dd m  \int\dd c \frac{\dd^2 {\cal P}_{c,m}(c,m,R)}{\dd c\,\dd m} \delta(m_{\rm t}-m_{\rm t}^\star(m,c,R))\\
&\neq& \frac{\dd {\cal P}_m(m,R)}{\dd m} \,.\nn
\een
In this equation $m_{\rm t}$ is a free variable and $m_{\rm t}^\star(m,c,R)$ is the tidal mass calculated from the model, given $m$, $c$, and $R$; the cross pdf $\dd^2 {\cal P}_{c,m}/\dd c\,\dd m$ was introduced in \citeeq{eq:ev_mass_func}. An important remark to make here is that even in the absence of tidal disruption (\ie~$\epsilon_{\rm t}= 0$), the real tidal mass function still differs from the nonphysical virial mass function, simply as a consequence of tidal stripping. Furthermore, as already mentioned above, tidal effects are typically much stronger in the central region of the host, and much less important beyond the scale radius. The model effectively leads to a selection in concentration: more concentrated (\ie~denser) subhalos are more resilient to tidal effects, which explains why originally lighter subhalos (which formed earlier and are denser) survive more efficiently in the central regions of the host. Consequently, the mass function is strongly altered in the central parts of the host halo, and becomes much steeper than the {\em unevolved} one. In contrast, the mass function is very close to its initial shape in the outskirts of the host halo. All this is consistent with other works showing a spatial evolution of the mass-concentration relation \cite{KuhlenEtAl2008,PieriEtAl2011,MolineEtAl2017}. Therefore, so long as we are not concerned with the inner subhalo population of the host halo, the relevant mass function should remain close to \citeeq{eq:dNdm}. This is actually the case for all targets considered in this paper, for which total luminosity of subhalos dominates over that of the host beyond the scale radius, typically.

As we have just seen, tidal stripping and tidal disruption slightly degrade the effective global unevolved subhalo mass fraction, but more importantly, they can strongly flatten the spatial distribution of subhalos in the host center. A damping of the subhalo population is even predicted in the very inner parts of the host for most of the model parameters used in our study (which is only moderately reflected by the mass fraction, which integrates subhalos over the whole host volume). The level of this flattening is obviously driven by the value taken for $\epsilon_{\rm t}$, the disruption efficiency parameter, with a population damping more severe for larger values. This is consistent with other analytical studies based on different approaches (see \eg~\cite{GreenEtAl2021}).

\bibliographystyle{JHEP.bst}
\bibliography{somm_sub_th.bib}

\end{document}